\documentclass[fleqn,usenatbib]{mnras}

\usepackage{newtxtext,newtxmath}

\usepackage[T1]{fontenc}
\usepackage{comment}
\usepackage{booktabs}
\usepackage{lscape}
\usepackage{rotating}
\usepackage{longtable} 
\usepackage{pdflscape} 
\usepackage{caption}
\usepackage{threeparttable}         
\usepackage{threeparttablex} 
\usepackage[dvipsnames]{xcolor}
\usepackage{multirow}
\usepackage{soul}

\DeclareRobustCommand{\VAN}[3]{#2}
\let\VANthebibliography\thebibliography
\def\thebibliography{\DeclareRobustCommand{\VAN}[3]{##3}\VANthebibliography}

\usepackage{graphicx}	
\usepackage{amsmath}	

\usepackage{etoolbox}

\defcitealias{Martins2005}{M05}
\defcitealias{MartinsPlez2006}{MP06}
\defcitealias{Lorenzo2022}{L22}
\defcitealias{Mathys1988}{M88}
\defcitealias{Sota2011}{S11}

\newcommand{\ha}{$\rm H\alpha$}

\newcommand{\HeI}{He~{\sc i}}
\newcommand{\HeII}{He~{\sc ii}}

\newcommand{\NIII}{N~{\sc iii}}
\newcommand{\NIV}{N~{\sc iv}}

\newcommand{\HI}{H~{\sc i}}
\newcommand{\HII}{H~{\sc ii}}
\newcommand{\Htwo}{H$_{\mathrm{2}}$}

\newcommand{\EBV}{$\mathrm{E}(B-V)$}

\newcommand{\Teff}{\mbox{$T_{\mathrm{eff}}$}}
\newcommand{\logg}{log~$g$}
\newcommand{\logQ}{log~$Q_{\mathrm{wind}}$}
\newcommand{\vsini}{$ v \sin i$}
\newcommand{\logL}{log~$L$}
\newcommand{\Mdot}{$\dot{M}$}
\newcommand{\Qphot}{$Q_{\mathrm{phot}}$}

\newcommand{\logqH}{log~$q_\mathrm{H I}$}
\newcommand{\logqHeI}{log~$q_{\mathrm{He I}}$}
\newcommand{\logqHeII}{log~$q_{\mathrm{He II}}$}
\newcommand{\logQH}{log~{$Q_\mathrm{H I}$}}
\newcommand{\logQHeI}{log~$Q_{\mathrm{He I}}$}
\newcommand{\logQHeII}{log~$Q_{\mathrm{He II}}$}

\newcommand{\qH}{$q_H$}
\newcommand{\qHeI}{$q_{\mathrm{He I}}$}
\newcommand{\qHeII}{$q_{\mathrm{He II}}$}
\newcommand{\QH}{$Q_H$}

\newcommand{\QHeII}{$Q_{\mathrm{He II}}$}

\newcommand{\kms}{\mbox{km~s$^{-1}$}}
\newcommand{\Msun}{\mbox{$M_{\odot}$}}
\newcommand{\Lsun}{\mbox{$L_{\odot}$}}
\newcommand{\Rsun}{\mbox{$R_{\odot}$}}
\newcommand{\Zsun}{\mbox{$Z_{\odot}$}}
\newcommand{\Osun}{\mbox{$O_{\odot}$}}

\newcommand{\NCNOgrid}{13700}

\newcommand{\NOBfilter}{5150}
\newcommand{\TEFFCNOmin}{15}
\newcommand{\TEFFCNOmax}{57}

\newcommand{\GRAVCNOmin}{1.6}
\newcommand{\GRAVCNOmax}{4.4}

\title[A reference framework for XMP OB star studies]{A reference framework for extremely metal-poor OB star studies: calibrations for stellar parameters and intrinsic colours}

\author[]{
M. Lorenzo,$^{1,2}$\thanks{E-mail: mlorenzo@cab.inta-csic.es}
M. Garcia,$^{1}$
N. Castro,$^{3,4}$
F. Najarro$^{1}$
M. Cerviño,$^{1}$
A. Herrero,$^{5, 6}$
S. Simón-Díaz$^{5, 6}$
\\
$^{1}$Centro de Astrobiología, CSIC-INTA. Crtra. de Torrejón a Ajalvir km 4. 28850 Torrejón de Ardoz (Madrid), Spain\\
$^{2}$Departamento de Física Teórica, Universidad Autónoma de Madrid (UAM), Campus de Cantoblanco, E-28049 Madrid, Spain\\
$^{3}$Institut für Astrophysik, Georg-August-Universität, Friedrich-Hund-Platz 1, 37077 Göttingen, Germany\\
$^{4}$Leibniz-Institut für Astrophysik Potsdam (AIP), An der Sternwarte 16, 14482, Potsdam, Germany\\
$^{5}$Instituto de Astrof\'isica de Canarias, E-38200 La Laguna, Tenerife, Spain\\
$^{6}$Departamento de Astrof\'isica, Universidad de La Laguna, E-38205 La Laguna, Tenerife, Spain\\
}

\date{Accepted XXX. Received YYY; in original form ZZZ}

\pubyear{2024}

\begin{document}
\label{firstpage}
\pagerange{\pageref{firstpage}--\pageref{lastpage}}
\maketitle

\begin{abstract}
We provide the first reference framework for  extremely metal-poor (XMP)~OB-type stars.
We parsed a grid of 0.10~\Zsun~\textsc{fastwind} models,
covering the parameter space of O stars and early-B supergiants,
through contemporary spectral classification criteria 
to deliver
a calibration of key stellar properties as a function of spectral type,
and tabulated colours for the most common photometric systems.
By using an extensive grid of models, we account for the different combinations of stellar parameters that result in the same spectral morphology and provide a range of parameters and colours compatible with each spectral subtype and luminosity class.
We supply updated photometric criteria to optimize candidate selection of OB stars in XMP environments.
We find 0.10~\Zsun~OB stars are 1-6~kK hotter and produce higher ionizing fluxes than their Galactic analogues. 
In addition, we find a bimodal distribution of the \HeII-ionizing flux with spectral type; because of its known dependence on effective temperature and the wind, \logqHeII~for individual XMP late-O type stars could be underestimated by up to 4~orders of magnitude by other calibrations, some of them used by population synthesis codes.
Finally, we used our calibrated colours to map the extinction of the 0.10~\Zsun~galaxy Sextans~A finding that reddening is non-negligible and uneven.
\end{abstract}





\section{Introduction}
\label{sec:introduction}

In a Universe whose average metal content has been ever-growing since the Big Bang, the study of extremely metal-poor \citep[XMP, defined as $\leq$~0.10~\Zsun~by, e.g., ][]{KunthOstlin2000} massive stars is critical to understanding earlier cosmic epochs \citep{MadauDickinson2014}.
Their characterization is the missing link towards the first, metal-free stars of the Universe, the early chemical enrichment detected at redshifts up to $z$~$\sim$~10 \citep[e.g.,][]{Bunker2023}, and the reionization epoch, all processes in which XMP massive stars may have played a significant role according to recent observations from the James Webb Space Telescope \citep[e.g.,][]{Endsley2022, Topping2022, Atek2023}.


Our Local Group (LG) and vicinity host star-forming dwarf galaxies with an extreme paucity of metals and resolved stellar populations, offering a unique opportunity to determine the stellar properties of XMP massive stars.
In the future, these properties will be entered into population synthesis and photoionization codes in order to better interpret the observations at $z >$~2 \citep{MadauDickinson2014}.
However, despite the current efforts to characterize local XMP massive stars \citep[e.g.,][]{Camacho2016, Ramachandran2021,Gull2022, Telford2023}, their study is still in its infancy because of i) the small number of high-quality observations and ii) the lack of reference material to guide the studies.

In particular, there are no calibrations of stellar parameters or magnitudes and colours for XMP massive stars as a function of spectral type. 
In the absence of such scales, authors usually resort to calibrations for the Milky Way (MW) or the Magellanic Clouds (MC) \citep[such as][]{Martins2005, MartinsPlez2006, Ramachandran2019}. However, both theory and observations have shown the strong dependency between some stellar properties and metallicity.

Effective temperature (\Teff) is one of the most affected parameters due to \textit{line-blanketing}.
A fraction of the flux produced in the stellar interior is blocked by absorption lines in the photosphere, resulting in a steeper 
temperature gradient in the inner atmosphere to preserve the total stellar luminosity \citep[e.g. ][]{Pauldrach2001, LanzHubeny2003}.
Since most absorption lines are produced by metals due to their more intricate atomic structures, metal-poor stars experience less \textit{line-blanketing} and consequently, need higher \Teff~to achieve the same ionization state than their metal-rich analogues, impacting the relation between \Teff~and spectral type (SpT).
This shift in the \Teff~-~SpT relation is already measurable in regimes with a slightly decreased metallicity.
The O stars of the Large Magellanic Cloud (LMC, 0.50~\Zsun) are 1$\sim$kK hotter than their Galactic counterparts \citep[e.g., ][]{Massey2004, Mokiem2007a, Ramirez-Agudelo2017, Sabin-Sanjulian2017}, and the difference increases to 3-4~kK \citep[e.g.,][]{ Massey2004, Mokiem2006,Trundle2007,Ramachandran2019} in the Small Magellanic Cloud (SMC, 0.20~\Zsun).
The hotter temperatures impact on their spectral energy distributions (SED), making their colours bluer and their $V$-magnitudes fainter.
\citet{Evans2019} estimated that the absolute visual magnitudes of XMP~O stars at the Zero Age Main Sequence (ZAMS) could be 0.5~mag fainter than their metal-rich analogues with the same mass.
Metal-poor massive stars are also more compact, resulting in higher surface gravities (\logg) \citep{Yoon2006, Ekstrom2008}.
Lastly, because absorption and re-emission of photons by metal lines are the main drivers of radiation-driven winds, metal-poor massive stars are expected to experience weaker mass and angular momentum loss during their evolution \citep{Kudritzki1987,Vink2001,Mokiem2007b}. 
The different wind behaviour can affect the luminosity of the stars and their SED.

The aims of this work are to construct the first calibration of stellar parameters and colours for XMP O stars and B supergiants. 
Because the current sample of known XMP OB stars is reduced, it does not uniformly cover all spectral subtypes \citep[e.g.][hereafter L22]{Lorenzo2022}, and the spectral quality of the data is low, we based this work on a grid of 0.10~\Zsun$,$ stellar atmosphere models.

The paper is structured as follows.
In Section~\ref{sec:GRID_ranges}, we present our grid of \textsc{fastwind} stellar atmosphere models,
and assign a spectral type to each of them.
In Sect.~\ref{sec:params}, we obtain calibrations of \Teff~and ionizing fluxes as a function of spectral type, and discuss the results. Likewise, we tabulate magnitudes and colours in Sect.~\ref{sec:phot_results}.
As an example of scientific exploitation, we build an extinction map of the 0.10~\Zsun~dwarf galaxy Sextans~A in Sect.~\ref{sec:ExtMap}, using our calibration and the spectroscopic sample of XMP massive stars from \citetalias{Lorenzo2022}.
Finally, a summary and general conclusions are given in Sect. \ref{sec:conclusions}.


\section{A grid of \textsc{fastwind} models to calibrate the stellar parameters of XMP OB stars}
\label{sec:GRID_ranges}

\begin{figure}
\centering
    \includegraphics[width=\hsize]{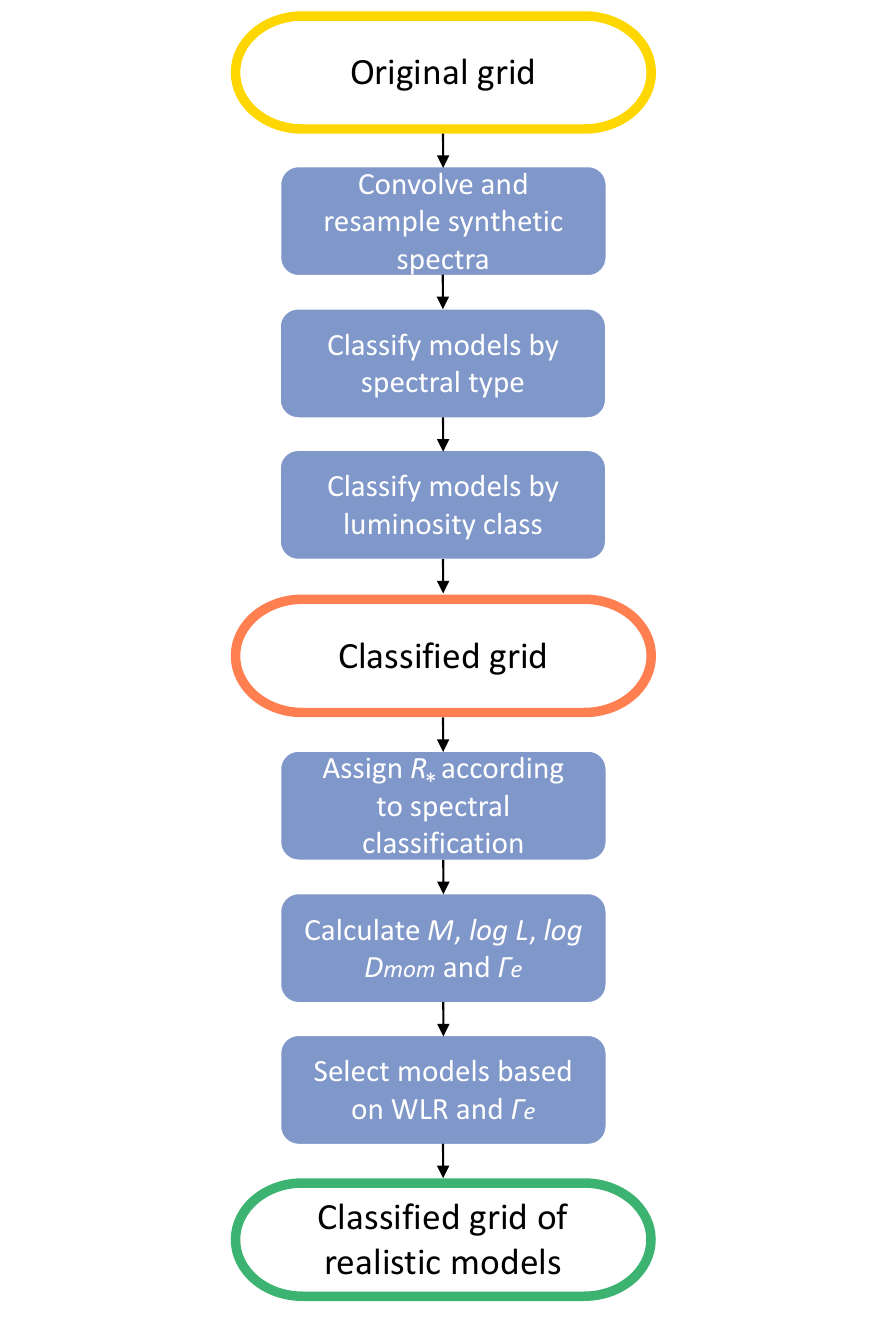}
    \caption{Flowchart illustrating the strategy followed to assign spectral types to our grid of \textsc{fastwind} synthetic spectra and distil the subsample of models that hold realistic physical properties.
    }
    \label{fig:FlowChart}
\end{figure}

To build the theoretical calibrations of stellar parameters and photometry of XMP OB stars, we computed an extensive grid of stellar
atmosphere models described in Sect.~\ref{sec:GRID_desc}. 
We processed the synthetic spectra to match the observations of spectral standards and classified them (Sect.~\ref{sec:GRID_proc}),
and then we discarded unrealistic models from the grid (Sect.~\ref{sec:GRID_filters}). 
The complete process is outlined in Figure~\ref{fig:FlowChart}. 

\subsection{Grid description}
\label{sec:GRID_desc}

The grid covers the parameter space of XMP OB-type stars with \NCNOgrid~models.
We calculated them using \textsc{fastwind} (version 10.6), a stellar atmosphere code that accounts for line-blanketing and NLTE effects in "unified", extended and spherically-symmetric atmospheres \citep{Santolaya-Rey1997, Puls2005, Rivero-Gonzalez2012}. 
\textsc{fastwind} can provide high volumes of synthetic spectra in reduced periods of time, an essential feature to building our comprehensive grid. In addition, this code provides the continuum of the spectral energy distribution from 20 to 10$^7$~\AA, allowing us to calculate the synthetic photometry and production of ionizing flux of the models.

The grid was computed using detailed model atoms for hydrogen and helium, and treating carbon, nitrogen, oxygen, silicon and magnesium explicitly in the formal solution. All other elements \citep[those included in the full comprehensive line list of][]{Pauldrach1998, Pauldrach2001} were treated as background elements (i.e., allowing them to contribute to line-blocking/blanketing but not calculating their spectral line profiles).

\begin{figure}
    \includegraphics[width=\hsize]{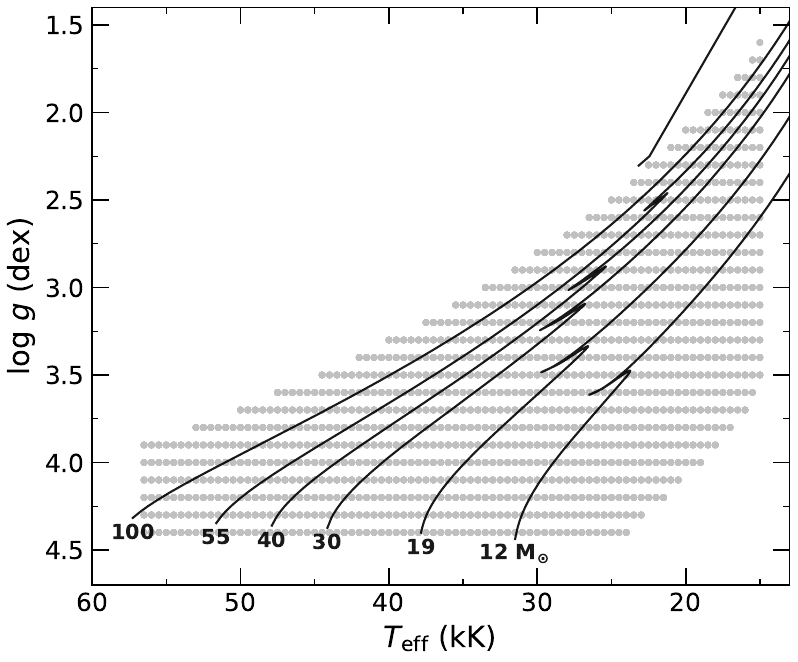}
    \caption{Parameter space covered by our grid of \textsc{fastwind} models in the Kiel diagram.
     \mbox{$Z$ = 0.10~\Zsun}, \mbox{\vsini~= 100 \kms} evolutionary tracks by \citet{Szecsi2022} are also included. We note that each point represents ten models corresponding to the ten different values of \logQ~we covered in the grid, between -15 and -11.7~dex.}
    \label{fig:GRIDranges}
\end{figure}

The models have 0.10~\Zsun~ metallicity \citep[scaled using the solar abundances from][]{Asplund2009}. 
We do not consider any departures from the chemical abundance pattern of the Sun, nor any element enrichment due to stellar evolution. 
The grid covers effective temperatures from \TEFFCNOmin~to \TEFFCNOmax~kK and surface gravities from \GRAVCNOmin~to \GRAVCNOmax~dex (depending on \Teff) in steps of 0.5~kK and 0.1~dex respectively. 
The ranges were tailored to cover objects with spectral types between $\sim$O2-B3 and the main luminosity classes (I, III, V, in the case of O~stars and class~I for the B-types). 
In Fig.~\ref{fig:GRIDranges}, we show the distribution of our grid in the Kiel diagram, along with the evolutionary tracks from \citet{Szecsi2022} for rotational velocity of 100~\kms~and metallicity of 0.10~\Zsun.

We initially allowed the wind strength parameter,
defined as $Q_{\mathrm{wind}} = \dot{M}/(R_* v_{\infty})^{1.5}$ \citep{Puls1996}
where $\dot{M}$ represents the mass loss rate (\Msun yr$^{-1}$), $R_*$ the stellar radius (\Rsun) and $v_{\infty}$ the terminal velocity (\kms), to range from \mbox{\logQ~=~-15~dex} to \mbox{\logQ~=~-11.7~dex}. 
Although we expect weak winds at the considered 0.10~\Zsun~metallicity, we kept this parameter free in the complete grid
and later selected models according to the expected physical properties of XMP OB stars (Sect.~\ref{sec:GRID_filters}). 
We also considered smooth, stationary winds with no clumping nor X-ray emission from wind embedded shocks.
The exponent of the wind-velocity law ($\beta$) and the helium abundance ($Y_{\mathrm{He}}$) were kept constant to values of 1.00 and 0.10, respectively. 
The microturbulence ($\xi$) for the atmosphere models was set according to \logg~and then was varied from 1~to 30~\kms~in the formal solution.




\subsection{Spectral classification of the grid models}
\label{sec:GRID_proc}

We assigned spectral type (SpT) and luminosity class (LC) to each model following standard spectral classification schemes for O and B stars 
\citep[e.g.][]{Sota2011,Negueruela2024},
adapted to 0.10~\Zsun\, metallicity following \citet{Lennon1997}'s strategy (see Lorenzo et al. in prep).
We previously processed the synthetic spectra to match the
observations of \citet{Sota2011}’s standard stars, with spectral resolution $R$~=~2500 and typical projected rotational velocities of \vsini~=~70~\kms \citep{Sota2011,Holgado2018}. No noise was introduced (but see below).



\begin{figure}
    \includegraphics[width=\hsize]{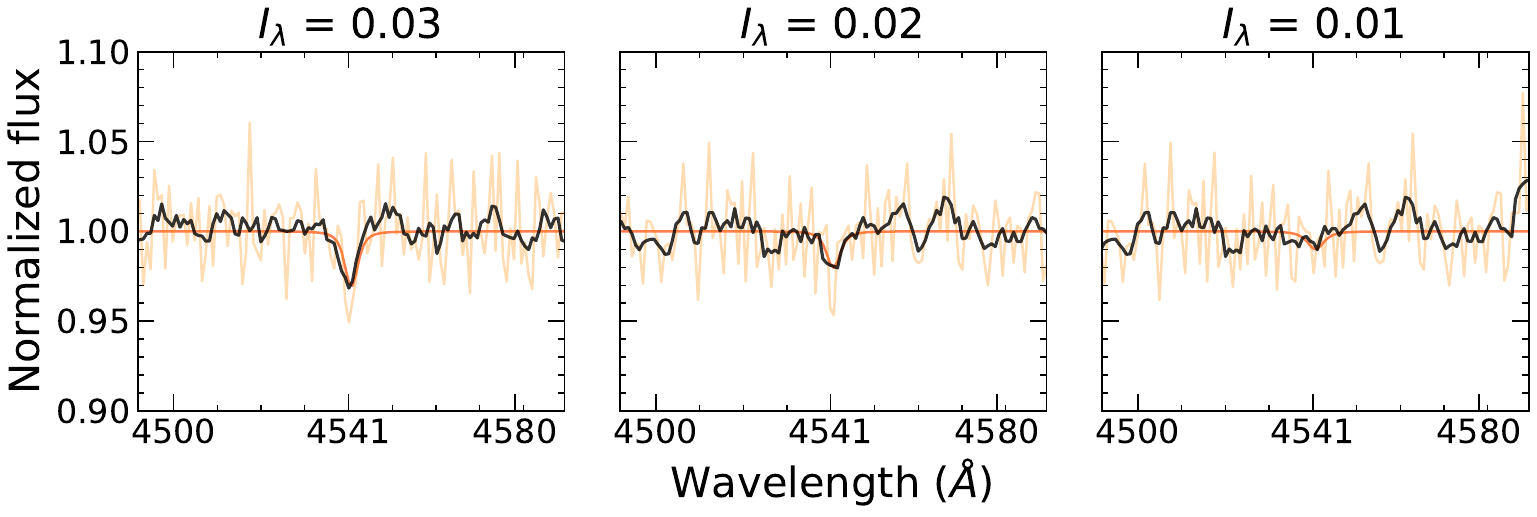}
    \caption{\HeII~4541 line with different line-intensities, as indicated above each panel. The profile was convolved 
    to $R$~=~2500 and \vsini~=~70~\kms (in red), including Gaussian noise with \mbox{$\sigma = 1/50$~sp.~pix.$^{-1}$} (in orange) and after applying a five-point smooth (in black). The line is no longer detected at intensities lower than 0.02.}
    \label{fig:DetLim}
\end{figure}

Assigning spectral subtypes to the grid models in an automatic way required to
first find criteria deciding if key diagnostic lines were detected. 
To define the detectability limit, we visually assessed at which intensity with respect to the local continuum (\mbox{$I_{\mathrm{\lambda}}$ = $F_{\mathrm{minimum}}$ - $F_{\mathrm{continuum}}$}) the lines would no longer be detected.
To this end, we plotted the \HeII~{4541} transition with different line-intensities, as shown in Fig.~\ref{fig:DetLim}.
We then convolved these profiles to $R$~=~2500 and \vsini~=~70~\kms
and introduced zero-mean Gaussian noise with a standard deviation ($\sigma$) equal to the inverse of a given signal-to-noise ratio (S/N).
Finally, we smoothed the data with a kernel width of five points to replicate the procedure followed in the spectral classification of poor S/N observations \citep[e.g.,][]{GarciaHerrero2013, Camacho2016, Lorenzo2022}. We applied the procedure for different line intensities and inspected visually whether the line was detected or not.

\begin{figure}
    \includegraphics[width=0.9\hsize]{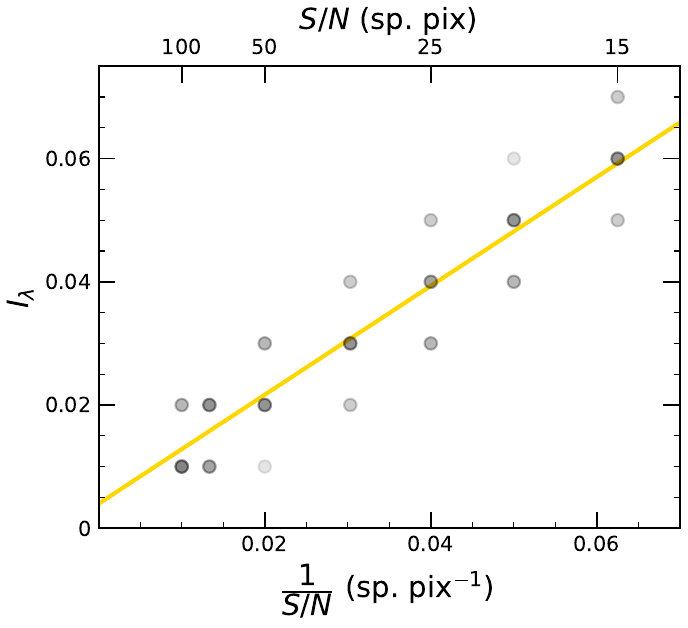}
    \caption{Minimum intensity at which spectral lines are detected for different levels of noise. We used various random seeds to simulate the noise. The points fit a linear function of slope 0.884 and intercept 0.004, shown in yellow in the Figure. We approximate this function to a 1:1 relation $I_{\mathrm{\lambda}}$~=~1/(S/N) throughout the text.}
    \label{fig:DetLim_SNR}
\end{figure}

We repeated this method for different values of S/N to determine its impact on the detectability limit. Figure~\ref{fig:DetLim_SNR} shows the minimum intensity at which lines are detected for different values of S/N using different random seeds to simulate noise. 
By fitting the trend, we approximated that the detectability limit roughly follows S/N as  $I_{\mathrm{\lambda}}$~=~1/(S/N) or, equivalently, $I_{\mathrm{\lambda}}$~=~$\sigma$, where $\sigma$ is the noise at the continuum.

During the process of classifying the grid,
we adopted the detectability limit corresponding to S/N~=~50~sp.~pix. for O~stars.
With this S/N, we can detect lines with intensities  $I_{\mathrm{\lambda}}$~$\gtrsim$~0.02. 
The extremely low metallicity of the grid severely hampers the classification of B-type stars, particularly at low S/N and resolution, as it relies on Si and Mg lines that are weaker than the He lines. Therefore, we assessed the detectability limit at S/N~=~100~sp.~pix., i.e. $I_{\mathrm{\lambda}}$~=~0.01, for them. 
Because B supergiants are usually brighter than O stars, it is reasonable to expect that higher S/N will be achieved with the same observational set-up.


Finally, we adopted one single value of $\xi$ per model, $\xi$~=~5~kms$^{-1}$ for \Teff~typical of O~stars
and $\xi$~=~10~kms$^{-1}$ for B~supergiants. In Appendix~\ref{sec:Appx_micro}, we discuss that the assigned spectral classification using a higher value of microturbulence for the same model would not differ by more than 0.5 spectral subtypes.

\subsection{Selection of models with realistic parameters} 
\label{sec:GRID_filters}

Our grid of \textsc{fastwind} models serves an additional purpose to perform automatic $\chi^2$ spectral fitting of observed spectra of XMP OB~stars in future works. 
In order to minimize edge effects (i.e. the $\chi^2$ distribution being truncated by the coverage limits), the parameter space is broader than observed from real stars. 
For this work, we needed to discard unrealistic models by filtering according to the expected physical properties of the stars.

First, we discarded models with flux conservation errors larger than 3\%.

Secondly, we selected models that follow the \textit{modified  wind momentum - luminosity relationship} (WLR),
\begin{equation}
    \log D_{\mathrm{mom}} = \log \left( \dot{M}v_{\infty}\sqrt{R/R_{\odot}} \right) = 
    \log D_{\mathrm{0}} + x \log L/L_{\odot}.
\end{equation}
This equation expresses that the mechanical wind momentum of massive stars ($D_{\mathrm{mom}}$) is mainly a function of photon momentum \citep[e.g.,][]{Kudritzki1995, KudritzkiPuls2000}, translating into a direct dependence on stellar luminosity. The slope ($x$) and intercept ($\log D_{\mathrm{0}}$) vary according to metallicity because metal atoms receive the energy and momentum from photons and carry the wind \citep{Kudritzki1999, Vink2000, Puls2000}.

We scaled the intercept of the empirical, unclumped WLR found by \citet{Mokiem2007b} for stars in the SMC to the metallicity of the grid (0.10~\Zsun) using the $Z$-dependence relations of mass-loss and terminal velocity given by \citet{Mokiem2007b} and \citet{Leitherer1992} \citep[see also][]{Hawcroft2024}, respectively. 
For luminosities larger than $\log L/L_{\odot}$ = 5.6~dex, we accepted all models that followed the WLR allowing for $\pm$~1~dex departures to reflect the typical scatter of observational points. 
For lower luminosities, we also allowed for models reflecting the weak wind problem.
Weak winds have been detected at $\log L/L_{\odot} \leq$~5.2~dex in the Milky Way \citep{Martins2005b, Marcolino2009} and at $\leq$~5.4~dex in the SMC \citep{Bouret2003}, thus we conservatively adapted to the metallicity of the grid and set the luminosity threshold for weak winds to $\leq$~5.6~dex.

To check for consistency, we alternatively selected models following the WLR and $\dot{M} \propto Z$ scale by \citet{Bjorklund2021}. 
The selection of models was equivalent to those obtained with \citet{Mokiem2007b}'s when allowing for $\pm$~0.5~dex in the intercept of the WLR. 
We note that the different allowed range with respect to \citet{Mokiem2007b}'s is caused by the different units of $D_{\mathrm{mom}}$ used in the two works.

We performed a third selection of models based on the Eddington factor $\Gamma_{\mathrm{e}}$, which evaluates the proximity to the Eddington limit calculated considering that opacity is only due to electron scattering. We used the expression given by \citet{Grafener2011},
\begin{equation}
    \log \Gamma_{\mathrm{e}} = -4.813 + \log(1+X_{\mathrm{H}}^\mathrm{s}) + \log(L/L_{\odot}) - \log(M/M_{\odot}),
\end{equation}
where $X_{\mathrm{H}}^\mathrm{s}$ corresponds to the hydrogen mass fraction at the stellar surface.
For the O stars, we defined an upper limit of $\Gamma_{\mathrm{e}}$~=~0.55, which is based on the maximum $\Gamma_{\mathrm{e}}$ measured in O4-6 supergiants with solar metallicity~by \citet{Grafener2011} and the results of \citet{Holgado2018}. 
For the B stars, we used the same value, 0.55, based on the results of spectroscopic analysis of B~supergiants in the MW and other galaxies of the LG \citep[][ etc]{Urbaneja2002, Evans2004, Crowther2006}.

To calculate the mass, luminosity, \Mdot, $v_{\infty}$ and $\log D_{\mathrm{mom}}$ of the models classified as O stars, 
we used the calibrated radii for each spectral subtype of \citet{Martins2005}. Radii were subsequently scaled to account for the metallicity difference by considering the calibrated \Teff~for the same spectral type at each $Z$ and assuming constant luminosity \citep[as discussed by][]{Evans2019}.
This procedure was done iteratively. 
We first adopted \citet{Martins2005}'s radii for the spectral types assigned to the models. Then, 
we accepted the models based on their $\Gamma_{\mathrm{e}}$ and WLR location as explained above,
and obtained a first value of typical \Teff~for each spectral type. 
On a second iteration, we accounted for metallicity by scaling the radii based on the \Teff~differences we found between \citet{Martins2005}'s MW calibration and our \Teff~scale from the first iteration and again selected the models that complied with the $\Gamma_{\mathrm{e}}$ and WLR constraints.
This procedure was repeated until the averaged \Teff~for each spectral type remained constant in consecutive iterations.
According to these calculations, a 0.10~\Zsun~O dwarf would be 20\% more compact than its Galactic analogue, whereas the radii of O giants and supergiants would be 10\% smaller.

Lacking calibrations for B supergiants, we performed a linear regression between the radii and the spectral subtypes determined in spectroscopic analysis of B supergiants \citep[][ etc]{Urbaneja2002, Evans2004, Crowther2006} and assigned those.

Finally, we accounted for the large uncertainty of the assigned radii by allowing variations of up to 20\% from the central value and propagated these variations to the quantities derived using this parameter. 
Therefore, each model with a given (\Teff, \logg) was assigned three different radii.
If none of the derived quantities for the three radii ($\Gamma_{\mathrm{e}}$ and WLR location) met our physical filters, we discarded the model.

\begin{figure}
    \includegraphics[width=0.9\hsize]{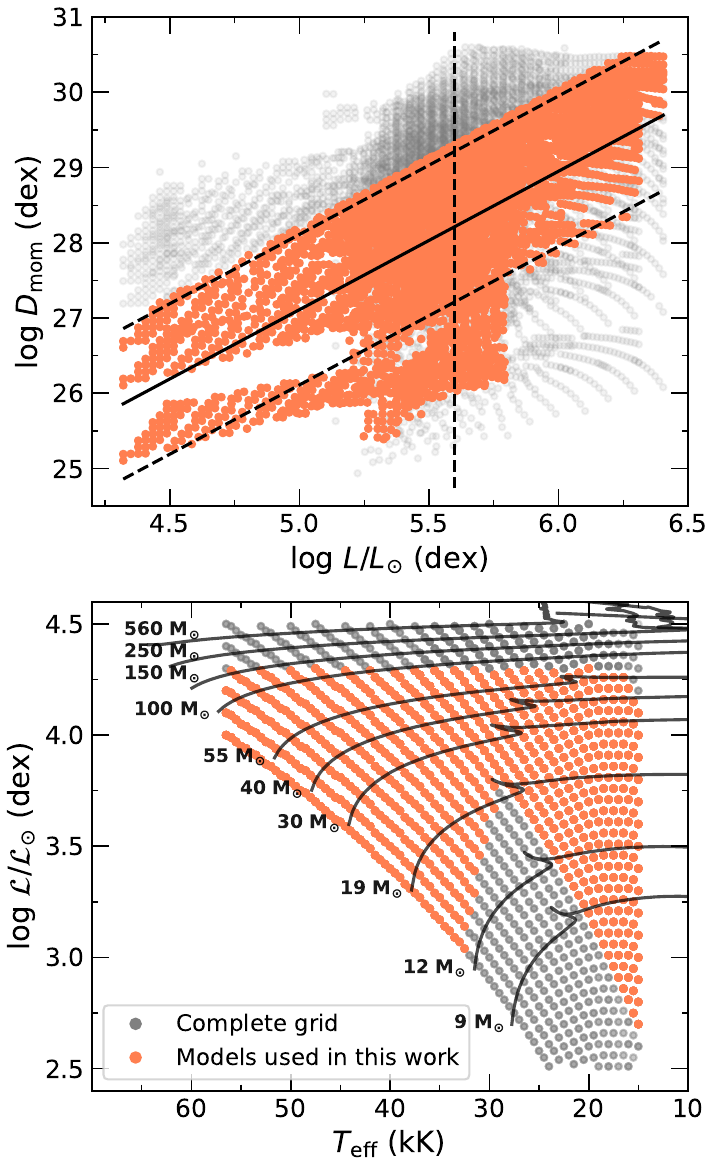}
    \caption{Complete grid of \textsc{fastwind} models (grey) and the selection of realistic models (orange) according to the observed properties of low-$Z$ OB-type stars (see Sect.~\ref{sec:GRID_filters}).
    In the top panel, we present the modified wind-momentum -- luminosity relation (WLR).
    As a reference, we include \citet{Mokiem2007b}'s WLR adapted to the metallicity of the grid (solid line). The parallel lines reflect the interval of $\log D_0$ values we allowed to reflect the scatter of the WLR. 
    We also allowed for all models under these lines and $\log L/L_{\odot} \leq$~5.6~dex (vertical line), to account for weak winds.
    Some of the selected models lie outside the mentioned limits. This is a consequence of the uncertainty implemented in the radii, which is carried over to $\log D_{\mathrm{mom}}$ and $\log L/L_{\odot}$.
    In the bottom panel, we present the grid in the sHRD. 
    The gap between 15 and 30~kK and $\log \mathcal{L}/\mathcal{L}_{\odot}~\leq$~3.75~dex in the final selection is caused by our rejection of the models with spectral morphologies compatible with B giants and dwarfs. In this panel, we also included the \mbox{$Z$ = 0.10~\Zsun}, \mbox{\vsini~= 100 \kms} evolutionary tracks of \citet{Szecsi2022} as reference.}
    \label{fig:GridFilters}
\end{figure}

Figure \ref{fig:GridFilters} shows the parameter space covered by the full grid and the selected models, in the WLR (top panel) and in the spectroscopic Hertzsprung-Russell diagram (sHRD, bottom panel). 
The down-selection process resulted in a set of \NOBfilter~models. 


\section{Calibration of stellar parameters with spectral type} 
\label{sec:params}

After assigning a spectral type to the distilled grid models,
we used them to produce calibrations of physical properties as a function of spectral type.
To that aim, we averaged the stellar parameters of all realistic models whose spectral morphology was compatible with a given spectral type.
By leaving all considered parameters free and not fixing them to a specific value in each (SpT, LC) pair, we better reflect the existing degeneracy in reproducing the same He ionization equilibrium (i.e. SpT) and width of the Balmer lines (i.e. LC) when combining different stellar parameters (\Teff, \logg~and \logQ).
Thus, this is the first theoretical calibration that considers that several combinations of stellar parameters can yield the same spectral morphology of a given (SpT, LC) pair.
Our methodology not only allowed us to provide a more realistic average of the stellar properties for each subtype but also the range of values compatible with a given morphology as seen, for example, in Figure~\ref{fig:TEFFscale} for \Teff.

Table~\ref{tab:StellarParams} collects the calibrated stellar parameters for each spectral subtype and luminosity class. 
For \Teff, \logg, \logqH~and \logqHeI~(defined below), we provide the mean, minimum and maximum values calculated over all the models classified with the same spectral subtype and luminosity class. 
For \logqHeII~(also defined below), we only include an upper limit (see discussion in Sect.~\ref{sec:qHeII}).
For all other parameters, such as radius, luminosity and ionizing photon production rates, we provide the average.

We excluded subtypes O3 and earlier from the calibration because
typical observations of XMP OB stars cannot detect \NIII~4640 and \NIV~4058 lines (Lorenzo et al. in prep), 
hence they cannot provide types earlier than O4.
Indeed, effective temperature is largely undetermined for these objects \citep[e.g. GHV-62024,][]{GarciaHerrero2013} unless 
observations
extend to the space ultraviolet (e.g. Furey et al. in prep).
In addition, since \textsc{fastwind}~does not provide reliable solutions below 15~kK (Puls, priv. comm), we only consider the spectral subtypes above this temperature (B3 or earlier).

We caution the reader that the physical properties that require the stellar radii were calculated using calibrated values (O~stars), or estimates for analogue stars by previous works (B~stars), as explained in Sect.~\ref{sec:GRID_filters}.
These quantities are provided in Table~\ref{tab:StellarParams}, along with the radii used. However, we strongly encourage the reader to use radii estimated from observed magnitudes when calculating e.g. luminosities and ionizing fluxes of individual stars.

\clearpage
\onecolumn
\begin{landscape}
\small
\setlength\LTleft{-15pt}
\setlength\LTright{-20pt}
\setlength\LTcapwidth{\linewidth}
\captionsetup[longtable]{labelfont=bf, font = footnotesize, labelsep=period}
\begin{longtable}{lccccc|ccccc}
\caption{Calibrated stellar parameters of O stars and early-B supergiants. For \Teff, \logg, \logqH~and \logqHeI, we provide the derived value, calculated by averaging (av.) over all models with a spectral morphology compatible with the considered spectral subtype, plus the minimum (min.) and maximum (max.) values of the distribution. For \logqHeII, we provide only the maximum since the \logqHeII~scale exhibits a bimodal distribution (see Sect. \ref{sec:qHeII}); this number must be considered as an upper limit. Column $R_*$ is the radii calibrated by \citet{Martins2005} and scaled to the metallicity of 0.10~\Zsun~for O stars, and the average of the radii provided in previous works for B supergiants. The luminosity (log $\dfrac{L}{\Lsun}$) and the total number of ionizing photons (\logQH, \logQHeI~and \logQHeII) were calculated with these radii. For these parameters, we provide only the mean value, except for \logQHeII~for which we provide only the maximum.}
\label{tab:StellarParams}\\
\toprule
\toprule
     SpT &        \Teff~(kK) & \logg~(cm s$^{-2}$) & \logqH~(cm$^{-2}$ s$^{-1}$) & \logqHeI~(cm$^{-2}$ s$^{-1}$) & \logqHeII~(cm$^{-2}$ s$^{-1}$) & $R_*$~(\Rsun) & log $\dfrac{L}{\Lsun}$ & \logQH~(s$^{-1}$) & \logQHeI~(s$^{-1}$) & \logQHeII~(s$^{-1}$) \\
 & av. (min., max.) & av. (min., max.) & av. (min., max.) & av. (min., max.) & max. & av. & av. & av. & av. & max. \\
\midrule
\endfirsthead
\caption[]{continued.} \\
\toprule
\toprule
     SpT &        \Teff~(kK) & \logg~(cm s$^{-2}$) & \logqH~(cm$^{-2}$ s$^{-1}$) & \logqHeI~(cm$^{-2}$ s$^{-1}$) & \logqHeII~(cm$^{-2}$ s$^{-1}$) & $R_*$~(\Rsun) & log $\dfrac{L}{\Lsun}$ & \logQH~(s$^{-1}$) & \logQHeI~(s$^{-1}$) & \logQHeII~(s$^{-1}$) \\
 & av. (min., max.) & av. (min., max.) & av. (min., max.) & av. (min., max.) & max. & av. & av. & av. & av. & max. \\
\midrule
\endhead
\midrule
\multicolumn{11}{r}{{\textit{continued on next page}}} \\

\endfoot

\bottomrule
\endlastfoot
    O4 V & 50.0 (48.0, 52.0) &      4.3 (4.1, 4.4) &        24.76 (24.68, 24.84) &          24.26 (24.14, 24.38) &                   $\leq~$19.99 &          9.85 &                   5.73 &             49.53 &               49.03 &         $\leq~$44.76 \\
  O4.5 V & 48.5 (46.5, 50.0) &      4.2 (4.1, 4.4) &        24.69 (24.61, 24.76) &          24.16 (24.04, 24.27) &                   $\leq~$19.80 &          9.35 &                   5.63 &             49.41 &               48.89 &         $\leq~$44.53 \\
    O5 V & 47.0 (45.0, 49.0) &      4.2 (4.0, 4.4) &        24.63 (24.54, 24.71) &          24.10 (23.95, 24.22) &                   $\leq~$19.65 &          8.86 &                   5.55 &             49.31 &               48.78 &         $\leq~$44.33 \\
  O5.5 V & 46.0 (44.0, 48.0) &      4.2 (4.0, 4.4) &        24.57 (24.48, 24.66) &          24.01 (23.87, 24.15) &                   $\leq~$19.50 &          8.49 &                   5.46 &             49.21 &               48.65 &         $\leq~$44.14 \\
    O6 V & 44.5 (42.0, 46.5) &      4.2 (4.0, 4.4) &        24.49 (24.36, 24.59) &          23.90 (23.70, 24.05) &                   $\leq~$19.34 &          8.18 &                   5.37 &             49.10 &               48.51 &         $\leq~$43.95 \\
  O6.5 V & 42.5 (39.5, 45.0) &      4.2 (3.9, 4.4) &        24.37 (24.19, 24.50) &          23.73 (23.45, 23.95) &                   $\leq~$19.15 &          7.83 &                   5.25 &             48.94 &               48.30 &         $\leq~$43.72 \\
    O7 V & 40.5 (37.5, 43.0) &      4.2 (3.8, 4.4) &        24.24 (24.04, 24.38) &          23.54 (23.18, 23.78) &                   $\leq~$18.99 &          7.50 &                   5.14 &             48.77 &               48.07 &         $\leq~$43.53 \\
  O7.5 V & 38.5 (36.5, 41.0) &      4.1 (3.8, 4.4) &        24.08 (23.94, 24.24) &          23.26 (22.98, 23.56) &                   $\leq~$18.47 &          7.15 &                   5.01 &             48.57 &               47.75 &         $\leq~$42.96 \\
    O8 V & 37.0 (35.0, 39.0) &      4.1 (3.7, 4.4) &        23.94 (23.80, 24.08) &          22.96 (22.61, 23.26) &                   $\leq~$17.75 &          6.82 &                   4.90 &             48.39 &               47.41 &         $\leq~$42.20 \\
  O8.5 V & 36.0 (34.0, 37.5) &      4.1 (3.7, 4.4) &        23.80 (23.66, 23.92) &          22.63 (22.24, 22.93) &                   $\leq~$17.20 &          6.49 &                   4.80 &             48.21 &               47.04 &         $\leq~$41.61 \\
    O9 V & 34.5 (32.5, 36.5) &      4.0 (3.6, 4.4) &        23.66 (23.47, 23.81) &          22.22 (21.63, 22.64) &                   $\leq~$16.36 &          6.18 &                   4.69 &             48.02 &               46.58 &         $\leq~$40.73 \\
  O9.5 V & 33.5 (31.5, 35.5) &      4.0 (3.6, 4.4) &        23.49 (23.27, 23.67) &          21.70 (21.02, 22.25) &                   $\leq~$16.11 &          5.91 &                   4.60 &             47.82 &               46.03 &         $\leq~$40.43 \\
  O9.7 V & 32.0 (29.5, 34.5) &      4.0 (3.5, 4.4) &        23.17 (22.75, 23.51) &          20.87 (19.91, 21.77) &                   $\leq~$15.73 &          5.54 &                   4.46 &             47.44 &               45.14 &         $\leq~$40.00 \\
\toprule
  O4 III & 46.5 (43.5, 49.5) &      3.9 (3.7, 4.2) &        24.63 (24.51, 24.76) &          24.07 (23.85, 24.25) &                   $\leq~$20.61 &         14.25 &                   5.93 &             49.73 &               49.16 &         $\leq~$45.70 \\
O4.5 III & 45.0 (43.0, 47.0) &      3.9 (3.7, 4.1) &        24.57 (24.47, 24.64) &          23.97 (23.79, 24.10) &                   $\leq~$20.18 &         13.93 &                   5.86 &             49.64 &               49.05 &         $\leq~$45.25 \\
  O5 III & 43.5 (42.0, 45.5) &      3.8 (3.7, 4.1) &        24.49 (24.42, 24.57) &          23.87 (23.75, 24.01) &                   $\leq~$20.07 &         13.73 &                   5.79 &             49.55 &               48.93 &         $\leq~$45.13 \\
O5.5 III & 42.0 (40.0, 44.5) &      3.8 (3.6, 4.1) &        24.41 (24.30, 24.51) &          23.74 (23.53, 23.94) &                   $\leq~$19.65 &         13.62 &                   5.72 &             49.46 &               48.79 &         $\leq~$44.70 \\
  O6 III & 41.0 (39.0, 43.0) &      3.8 (3.6, 4.0) &        24.33 (24.24, 24.43) &          23.63 (23.48, 23.82) &                   $\leq~$19.39 &         13.47 &                   5.66 &             49.37 &               48.68 &         $\leq~$44.43 \\
O6.5 III & 39.0 (36.5, 41.5) &      3.7 (3.5, 3.9) &        24.20 (24.05, 24.35) &          23.43 (23.19, 23.69) &                   $\leq~$18.98 &         13.27 &                   5.56 &             49.23 &               48.46 &         $\leq~$44.01 \\
  O7 III & 37.5 (35.5, 39.0) &      3.7 (3.5, 3.8) &        24.07 (23.95, 24.18) &          23.23 (23.02, 23.44) &                   $\leq~$18.11 &         13.06 &                   5.48 &             49.09 &               48.24 &         $\leq~$43.12 \\
O7.5 III & 35.5 (34.5, 36.5) &      3.6 (3.5, 3.7) &        23.93 (23.84, 24.01) &          22.95 (22.80, 23.10) &                   $\leq~$17.16 &         12.91 &                   5.39 &             48.94 &               47.96 &         $\leq~$42.17 \\
  O8 III & 34.5 (33.0, 35.5) &      3.5 (3.4, 3.7) &        23.79 (23.67, 23.87) &          22.63 (22.29, 22.84) &                   $\leq~$16.50 &         12.70 &                   5.30 &             48.78 &               47.62 &         $\leq~$41.49 \\
O8.5 III & 33.0 (32.5, 34.0) &      3.5 (3.4, 3.6) &        23.63 (23.54, 23.72) &          22.20 (21.91, 22.47) &                   $\leq~$15.82 &         12.49 &                   5.23 &             48.61 &               47.17 &         $\leq~$40.80 \\
  O9 III & 32.0 (31.5, 32.5) &      3.5 (3.4, 3.6) &        23.48 (23.41, 23.54) &          21.67 (21.40, 21.96) &                   $\leq~$15.35 &         12.32 &                   5.16 &             48.44 &               46.64 &         $\leq~$40.32 \\
O9.5 III & 31.0 (30.0, 31.5) &      3.4 (3.3, 3.5) &        23.26 (23.16, 23.36) &          20.98 (20.70, 21.23) &                   $\leq~$14.67 &         12.03 &                   5.07 &             48.21 &               45.92 &         $\leq~$39.62 \\
O9.7 III & 29.5 (28.5, 30.5) &      3.4 (3.3, 3.5) &        22.94 (22.73, 23.12) &          20.28 (19.80, 20.68) &                   $\leq~$12.83 &         11.97 &                   4.99 &             47.89 &               45.22 &         $\leq~$37.77 \\
\toprule
    O4 I & 43.5 (41.5, 45.0) &      3.7 (3.6, 3.9) &        24.51 (24.43, 24.59) &          23.88 (23.75, 24.00) &                   $\leq~$19.80 &         17.02 &                   5.97 &             49.76 &               49.13 &         $\leq~$45.05 \\
  O4.5 I & 42.0 (40.5, 43.0) &      3.7 (3.6, 3.9) &        24.42 (24.36, 24.46) &          23.75 (23.66, 23.85) &                   $\leq~$19.17 &         17.27 &                   5.91 &             49.68 &               49.01 &         $\leq~$44.43 \\
    O5 I & 41.0 (39.0, 42.0) &      3.6 (3.5, 3.8) &        24.37 (24.28, 24.44) &          23.67 (23.53, 23.75) &                   $\leq~$19.17 &         17.53 &                   5.89 &             49.64 &               48.94 &         $\leq~$44.44 \\
  O5.5 I & 40.0 (39.0, 41.0) &      3.6 (3.5, 3.6) &        24.32 (24.27, 24.38) &          23.59 (23.50, 23.68) &                   $\leq~$18.59 &         17.93 &                   5.87 &             49.61 &               48.88 &         $\leq~$43.88 \\
    O6 I & 38.5 (37.5, 39.5) &      3.5 (3.4, 3.6) &        24.25 (24.18, 24.30) &          23.47 (23.37, 23.55) &                   $\leq~$18.22 &         18.30 &                   5.83 &             49.55 &               48.78 &         $\leq~$43.52 \\
  O6.5 I & 37.0 (36.5, 38.5) &      3.4 (3.4, 3.5) &        24.14 (24.06, 24.22) &          23.30 (23.17, 23.44) &                   $\leq~$18.05 &         18.61 &                   5.78 &             49.46 &               48.62 &         $\leq~$43.37 \\
    O7 I & 35.5 (34.5, 36.0) &      3.3 (3.3, 3.4) &        23.99 (23.89, 24.05) &          23.05 (22.87, 23.18) &                   $\leq~$17.19 &         19.03 &                   5.70 &             49.33 &               48.39 &         $\leq~$42.53 \\
  O7.5 I & 34.0 (33.0, 35.0) &      3.3 (3.2, 3.4) &        23.85 (23.77, 23.94) &          22.79 (22.61, 22.97) &                   $\leq~$16.71 &         19.52 &                   5.66 &             49.22 &               48.16 &         $\leq~$42.08 \\
    O8 I & 32.5 (31.5, 33.5) &      3.3 (3.1, 3.4) &        23.71 (23.64, 23.81) &          22.48 (22.23, 22.70) &                   $\leq~$15.84 &         19.83 &                   5.60 &             49.09 &               47.86 &         $\leq~$41.22 \\
  O8.5 I & 31.5 (30.5, 32.5) &      3.2 (3.1, 3.4) &        23.57 (23.47, 23.67) &          22.06 (21.65, 22.41) &                   $\leq~$15.29 &         19.98 &                   5.54 &             48.95 &               47.45 &         $\leq~$40.68 \\
    O9 I & 30.5 (29.5, 31.5) &      3.2 (3.0, 3.4) &        23.42 (23.37, 23.51) &          21.55 (21.26, 21.87) &                   $\leq~$15.05 &         20.34 &                   5.51 &             48.82 &               46.95 &         $\leq~$40.45 \\
  O9.5 I & 29.0 (27.5, 31.0) &      3.1 (2.9, 3.4) &        23.23 (23.08, 23.36) &          20.89 (20.47, 21.42) &                   $\leq~$13.99 &         20.80 &                   5.45 &             48.65 &               46.31 &         $\leq~$39.41 \\
  O9.7 I & 28.0 (26.5, 29.5) &      3.1 (2.8, 3.3) &        22.88 (22.42, 23.06) &          20.03 (19.35, 20.40) &                   $\leq~$11.46 &         20.84 &                   5.38 &             48.31 &               45.45 &         $\leq~$36.88 \\
    B0 I & 26.5 (25.0, 28.0) &      3.0 (2.7, 3.2) &        22.64 (22.42, 22.95) &          19.48 (19.05, 19.92) &                   $\leq~$10.37 &         30.95 &                   5.64 &             48.41 &               45.24 &         $\leq~$36.14 \\
  B0.5 I & 25.0 (23.5, 27.0) &      2.9 (2.6, 3.1) &        22.04 (21.81, 22.47) &          18.63 (18.13, 19.40) &                    $\leq~$9.10 &         35.72 &                   5.66 &             47.93 &               44.52 &         $\leq~$34.99 \\
    B1 I & 24.0 (22.0, 25.5) &      2.8 (2.5, 3.1) &        21.40 (20.96, 21.81) &          17.83 (17.17, 18.42) &                    $\leq~$7.87 &         40.49 &                   5.67 &             47.40 &               43.83 &         $\leq~$33.87 \\
  B1.5 I & 21.0 (18.5, 23.5) &      2.6 (2.2, 3.0) &        20.51 (19.71, 21.15) &          16.56 (15.06, 17.57) &                    $\leq~$6.38 &         45.26 &                   5.56 &             46.61 &               42.66 &         $\leq~$32.47 \\
    B2 I & 18.5 (16.5, 20.5) &      2.5 (2.0, 2.9) &        19.66 (18.98, 20.34) &          14.98 (11.97, 16.21) &                    $\leq~$4.49 &         50.03 &                   5.44 &             45.84 &               41.17 &         $\leq~$30.67 \\
  B2.5 I & 17.0 (15.5, 18.5) &      2.4 (2.0, 2.7) &        19.09 (18.58, 19.64) &          13.70 (11.76, 15.10) &                    $\leq~$2.63 &         54.80 &                   5.36 &             45.35 &               39.96 &         $\leq~$28.89 \\
    B3 I & 15.5 (15.0, 17.0) &      2.3 (2.0, 2.6) &        18.66 (18.36, 19.14) &          12.31 (10.52, 14.37) &                    $\leq~$1.07 &         59.57 &                   5.29 &             44.99 &               38.64 &         $\leq~$27.40 \\
\end{longtable}
\end{landscape}
\clearpage
\twocolumn

\subsection{\Teff~scale}
\label{sec:TEFF_scale}

Accurately characterizing the stellar effective temperature
scale of OB stars is fundamental to linking spectral types with physical parameters. Several works have calculated the \Teff~scale in the MW and MCs \citep[e.g.,][]{Crowther2002, Martins2005, Massey2005b, Trundle2007, Massey2009, McEvoy2015, Markova2018}. The works by \citet{GarciaHerrero2013}, \citet{Tramper2014} and \citet{Camacho2016} opened the way in the sub-SMC metallicity regime but an extensive calibration does not yet exist.


\begin{figure*}
    \includegraphics[width=\hsize]{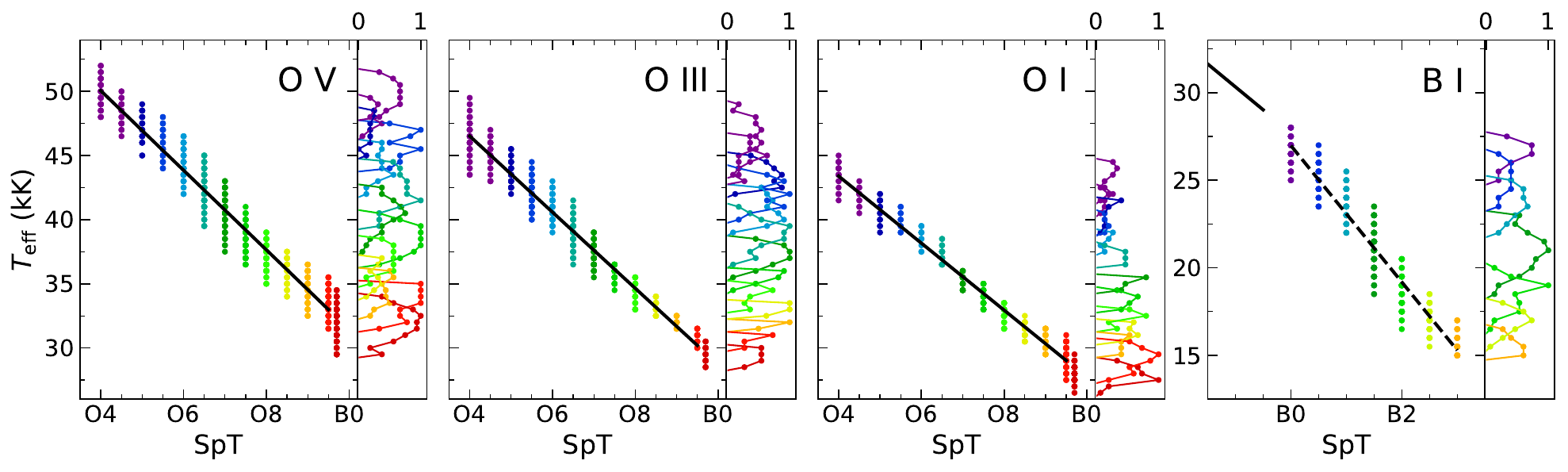}
    \caption{Effective temperature of all the models as a function of their spectral subtype for different luminosity classes. The black solid lines are the linear fits for O stars excluding the O9.7 subtype. The dashed black line is the linear fit for the early-B supergiants.
    We present a histogram on the right side of each panel. Each point represents the number of models with a given effective temperature that are compatible with the same colour-coded spectral subtype, normalized to the maximum of all distributions for visualization purposes.
    }
    \label{fig:TEFFscale}
\end{figure*}

Figure~\ref{fig:TEFFscale} shows the effective temperature of all our 0.10~\Zsun~\textsc{fastwind} models as a function of spectral type for different luminosity classes.
For a given spectral subtype, the models exhibit a range of \Teff~with some overlap between adjacent subtypes (as reflected by Table~\ref{tab:StellarParams}). 
As explained earlier, various parameter combinations can replicate the same spectral morphology, leading to dispersion in the \Teff~scale. 
This is illustrated in the histograms of the right subpanels; they show the number of models with a given \Teff~for each subtype, normalized by the maximum of all distributions.
These distributions demonstrate a continuous variation of \Teff~within spectral type, and smooth transitions between subtypes. Assessing these histograms and minimizing numerical asymmetries drove us to use a fine grid of models with \Teff-steps of 0.5~kK. However, we note that we do not expect uncertainties below 1~kK even in high-precision spectroscopic analyses \citep[e.g.,][]{Holgado2018}.

The distributions are roughly centred around the \Teff~assigned to each subtype. 
In addition, some subtypes are less populated than others, which reflects they occupy a smaller fraction of the considered parameter space.  We note that these distributions do not represent the frequency of the subtypes in nature.

We built the \Teff~scale by fitting the points corresponding to all the models using a least squares method, providing separate \Teff~scales for O and B stars, and for the different luminosity classes.
In this linear fit, we did not consider subtype O9.7 because, as discussed in \citet{Martins2005},
the increment of this spectral subtype is not proportional to the corresponding \HeI~4471/\HeII~4541 increment.
While the difference between O9.5 and O9.7 is only 0.2 in subtype, the line ratio increment is larger than the one between O8.5 and O9.
Therefore, including O9.7 in the \Teff~scale would lead to artificially lower effective temperatures.
For this subtype, we refer the reader to Table~\ref{tab:StellarParams}.

The linear fits for the \Teff~scales are shown in Figure~\ref{fig:TEFFscale} and are expressed as follows,

\begin{equation}
\Teff~(\mathrm{kK}) = \begin{cases}
62.45\pm2.05~-3.10~\times~\text{SpT} & \quad \text{(O4 - O9.5~V)} \\
58.38\pm1.62~-2.97~\times~\text{SpT} & \quad \text{(O4 - O9.5~III)} \\
53.85\pm1.25~-2.61~\times~\text{SpT} & \quad \text{(O4 - O9.5~I)} \\
65.92\pm1.69~-3.89~\times~\text{SpT} & \quad \text{(B0 - B3~I)} \\
\end{cases}
.
\end{equation}

In these and the following relations, the spectral subtype is represented by a number. O stars range from 4 to 9.5 according to their subtype, while B stars span from 10 to 13. 
We included uncertainties in the intercept in all equations. They quantify the dispersion of \Teff~within subtypes and luminosity classes, and express the distributions presented in Fig.~\ref{fig:TEFFscale}, and the minimum and maximum values included in Table~\ref{tab:StellarParams}. 
The provided uncertainties encompass
90\% of the models of the grid.

As expected and reflected by other calibrations, given a spectral subtype dwarfs have higher \Teff~than giants, and giants are hotter than supergiants. 
In addition, we observe that the \Teff~dispersion for O dwarfs is larger than for other classes. This is due to the wider range of \logg~compatible with this luminosity class.
The $\sim$2~kK dispersion of our calibration for dwarfs is in agreement with the scatter seen in observational \Teff~scales of different environments \citep[e.g., ][]{Simon-Diaz2014, Sabin-Sanjulian2017}. The $\sim$1~kK dispersion for giants and supergiants is also consistent with observations \citep[e.g.,][]{Ramirez-Agudelo2017}.

\begin{figure}
    \includegraphics[width=0.9\hsize]{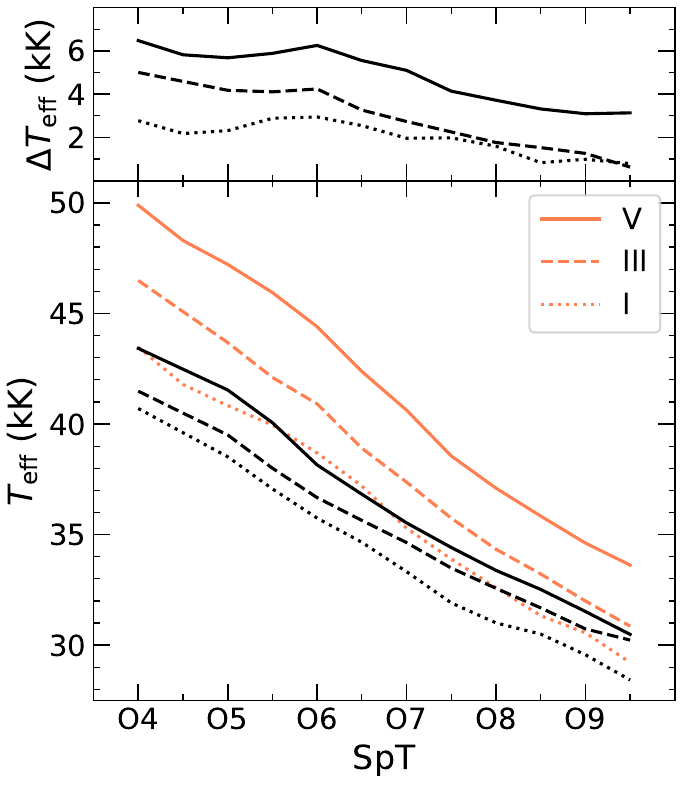}
    \caption{Comparison between \citet{Martins2005}'s calibration for Galactic O~stars (black) and this work's \Teff~scale (orange) for the three luminosity classes indicated in the legend. The upper panel shows the differences between the two scales. 
    }
    \label{fig:TEFFscale_Martins}
\end{figure}

In Figure \ref{fig:TEFFscale_Martins}, we compare the theoretical calibration of \citet{Martins2005} for Galactic O stars with this work.
As expected for the lower metallicity of our models, our \Teff~scales are hotter than the calibrations of \citet{Martins2005}. The difference ranges from 1-2~kK for supergiants and up to 6~kK for dwarfs. 

\begin{figure}
    \includegraphics[width=\hsize]{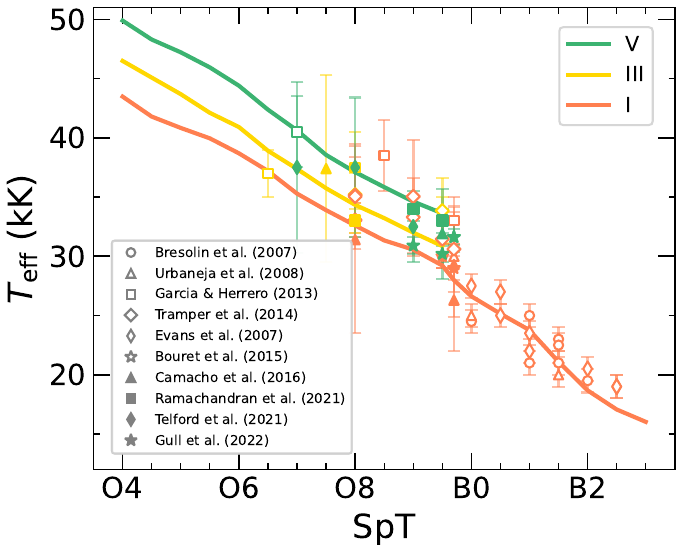}
    \caption{Effective temperature as a function of spectral type, comparing this work's calibration with observations of OB-type stars in environments with sub-SMC oxygen and iron content (filled symbols): 
    \citet{Telford2021b} in Leo~P (0.03~\Zsun),
    \citet{Gull2022} in Leo~A (0.05~\Zsun), \citet{Camacho2016} and \citet{Telford2021b} in Sextans~A (0.10~\Zsun), and 
    \citet{Ramachandran2021} in the Magellanic Bridge (0.20-0.02~\Zsun). 
    We also included \Teff~estimates of massive stars with oxygen abundances smaller than the SMC's, but with iron content similar to or higher than this galaxy (empty symbols):
    \citet{Tramper2014} and \citet{Evans2007} in NGC~3109 (0.12~\Osun), 
    \citet{Urbaneja2008}, \citet{Tramper2014}, \citet{Bouret2015} and \citet{Telford2021b} in WLM (0.14~\Osun),
    and \citet{Bresolin2007}, \citet{GarciaHerrero2013}, \citet{Tramper2014} and \citet{Bouret2015} in IC~1613 (0.16~\Osun). Colours indicate different luminosity classes.}
    \label{fig:TEFFscale_Obs}
\end{figure}

In Figure~\ref{fig:TEFFscale_Obs}, we compare our \Teff~scale with temperatures derived for OB-type stars 
in galaxies with sub-SMC oxygen abundance (empty symbols) and in galaxies with both iron and oxygen content lower than the SMC (filled symbols).
This Figure illustrates the scarcity of quantitative analyses of massive stars in very metal-poor environments, which prompted the construction of our calibration from models.

The \Teff~of O stars is not generally well constrained by these works, as evinced by the large error bars.
There are three O stars whose derived \Teff~deviate significantly from our scale.
First, the O8.5 supergiant identified by \citet{GarciaHerrero2013} in IC~1613 (ID 67684) is much hotter than predicted for class I. 
The star also departs from the \Teff~scale built by the same authors. 
They noted that the normalization of its spectrum was problematic and could have led to spectral misclassification, which may explain the discrepancy.
On the other hand, stars K1~(O9.5~V) and K7 (O9.7~V) in Leo~A \citep{Gull2022}  are 4~kK cooler than predicted for their classification, being more compatible with the Galactic scale for class V. 
When error bars are considered, our \Teff~scale agrees with observations of very metal-poor massive stars. This particularly applies to B supergiants, whose brightness allows for higher S/N observations and smaller error bars. Nonetheless, better quality spectra of a larger sample are needed to thoroughly test consistency, especially for O stars.

\subsection{Ionizing photons}
\label{sec:QH_scale}

Massive stars are key sources of ionization within star-forming regions. Quantifying their production of ionizing photons not only enables studying their feedback but can also be used to calculate the fraction of ionizing photons that escape from a region or galaxy. This quantity is especially interesting in extremely low-$Z$ environments as it helps us understand the role of massive stars in the epoch of reionization.

The ionizing flux is calculated by integrating the SED up to the ionization energy,
\begin{equation}
q_{\mathrm{x}} = \int_0^{\lambda_x} \frac{\lambda F(\lambda)}{hc} d\lambda,
\end{equation}
where $F(\lambda)$ is the stellar flux at one stellar radius and $\lambda_{\mathrm{x}}$ is the ionization wavelength of the atomic species x.

We calculated the ionizing flux of H~\textsc{i} \mbox{($\lambda_{\mathrm{HI}}$ = 911 \AA{})} and \HeI~\mbox{($\lambda_{\mathrm{He I}}$ = 504 \AA{})}
for each model of our grids and provided the average (plus minimum and maximum) for each subtype in Table~\ref{tab:StellarParams}, estimated from all the models with the same assigned spectral subtype and luminosity class (as in Section~\ref{sec:TEFF_scale}).
We computed the ionizing flux instead of the total photon production rate to avoid scaling by the radius.
Nonetheless, in Table~\ref{tab:StellarParams}, we also provide the average of the photon production rate ($Q_{\mathrm{x}}$) using calibrated radii (O stars) or radii derived from analogue stars in previous works (B stars). The ionizing flux of \HeII~\mbox{($\lambda_{\mathrm{He II}}$ = 228 \AA{})} will be discussed separately in the next section.

\begin{figure*}
    \includegraphics[width=\hsize]{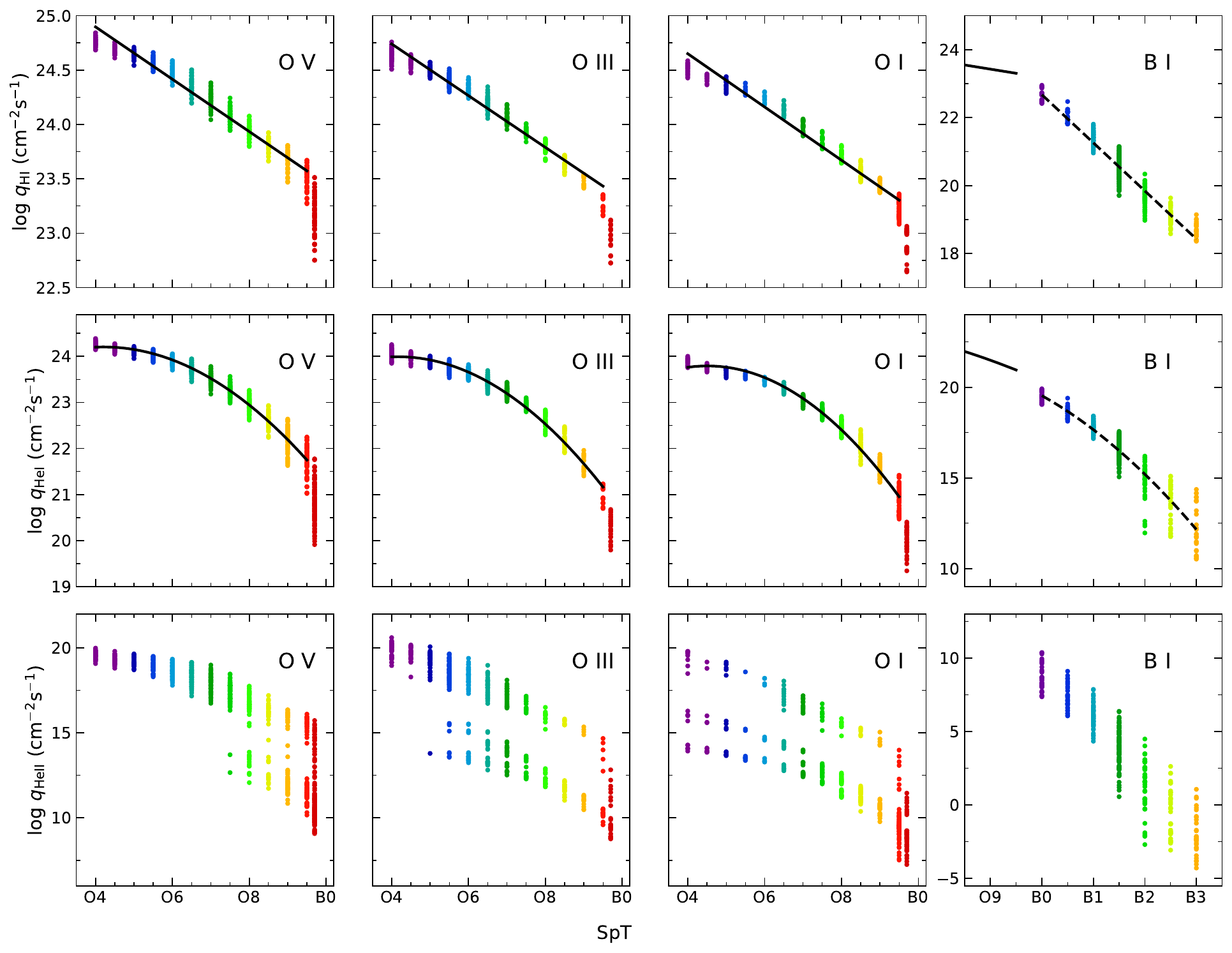}
    \caption{
    H~\textsc{i}-, \HeI- and \HeII-ionizing photon flux (in logarithm) of the grid models as a function of spectral type for different luminosity classes. The black solid lines are the least squared fits to O star models, excluding the O9.7 subtype. The black dashed lines are the fits to the early-B supergiant models. The bimodal distribution of \logqHeII~is discussed in Sect.~\ref{sec:qHeII}.
    }
    \label{fig:logQscale}
\end{figure*}

In Figure~\ref{fig:logQscale}, we display the ionizing flux of the models as a function of spectral type and luminosity class for the three ions considered. 
We fitted the trends of \logqH~and \logqHeI~using a least squares method for the O stars and the B supergiants separately and discarding the models classified as O9.7. 

The logarithm of H~\textsc{i}-ionizing photon flux (in units of cm$^{-2}$ s$^{-1}$) can be expressed as a linear function of the spectral subtype,

\begin{equation}
\log q_{\mathrm{HI}} = 
\begin{cases}
25.86\pm0.16~-0.24~\times~\text{SpT} & \quad \text{(O4 - O9.5~V)} \\
25.69\pm0.14~-0.24~\times~\text{SpT} & \quad \text{(O4 - O9.5~III)} \\
25.63\pm0.15~-0.24~\times~\text{SpT} & \quad \text{(O4 - O9.5~I)} \\
36.83\pm0.50~-1.42~\times~\text{SpT} & \quad \text{(B0 - B3~I)} \\
\end{cases}
\end{equation}

while \logqHeI~is better reproduced by a second-degree polynomial,
\newline\newline
\noindent \logqHeI~(cm$^{-2}$ s$^{-1}$) =
\begin{equation}
\quad 
\begin{cases}
22.65\pm0.29~+~0.74~\times~\text{SpT}~-0.09~\times~\text{SpT}^2~~~\text{(O4 - O9.5~V)} \\
22.27\pm0.22~+~0.83~\times~\text{SpT}~-0.10~\times~\text{SpT}^2~~~\text{(O4 - O9.5~III)} \\
21.47\pm0.31~+~1.03~\times~\text{SpT}~-0.11~\times~\text{SpT}^2~~~\text{(O4 - O9.5~I)} \\
\hspace{1.5mm}6.09\pm1.23~+~4.27~\times~\text{SpT}~-0.29~\times~\text{SpT}^2~~~\text{(B0 - B3~I)} \\
\end{cases}
\end{equation}

\begin{figure}
    \includegraphics[width=0.9\hsize]{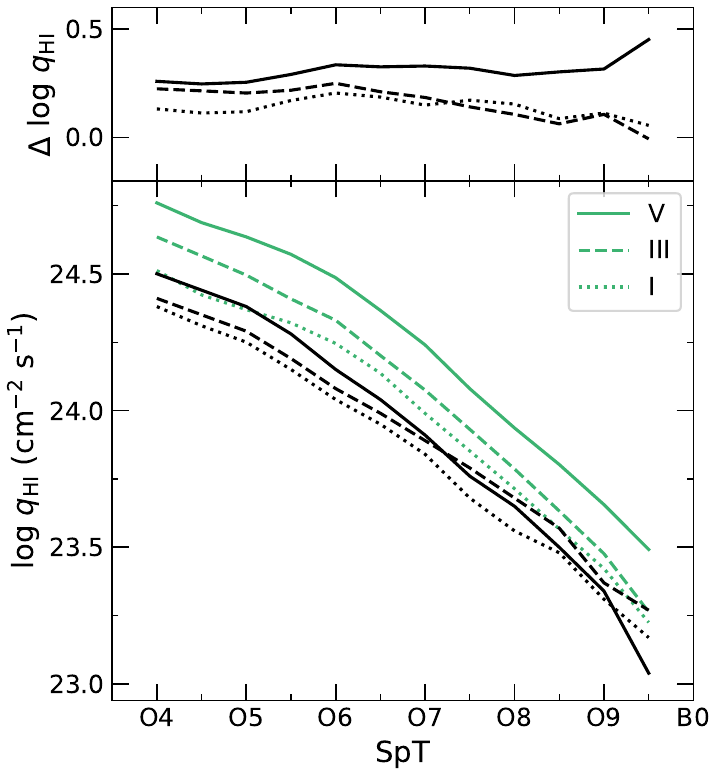}
    \caption{Our derived \logqH~scale (green) compared against \citet{Martins2005}'s (black),  for all luminosity classes. The difference between the two scales is plotted in the upper panel.
    }
    \label{fig:qHscale_Martins}
\end{figure}

\begin{figure}
    \includegraphics[width=0.9\hsize]{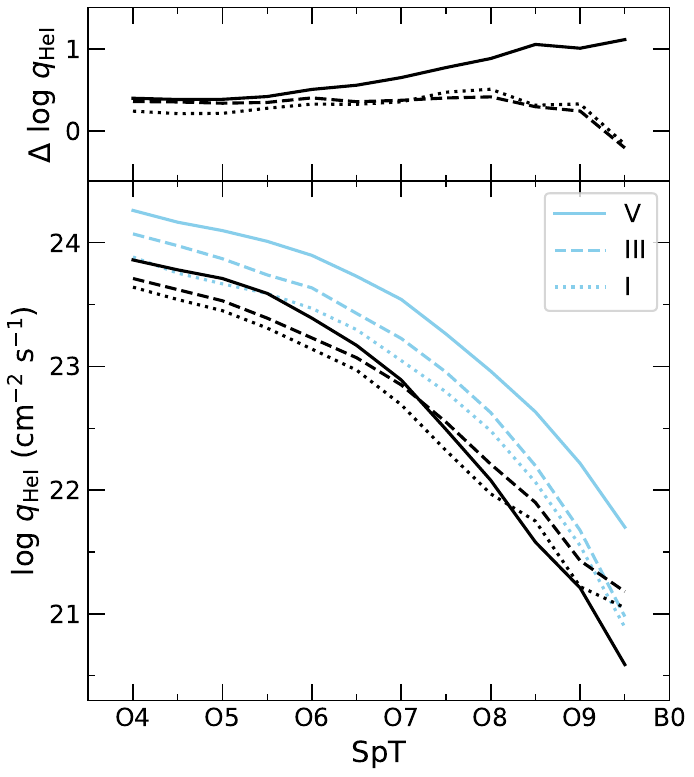}
    \caption{Same as Fig.~\ref{fig:qHscale_Martins}, now showing the \logqHeI~scale.
    }
    \label{fig:qHeIscale_Martins}
\end{figure}

\noindent We again included uncertainty in the intercept of the scales as a way to express the typical dispersion of H~\textsc{i}- and \HeI-ionizing flux compatible with a spectral type.


\citet{Martins2005} estimated the ionizing fluxes of H~\textsc{i} and \HeI~as a function of spectral type for Galactic O stars. In Figures \ref{fig:qHscale_Martins} and \ref{fig:qHeIscale_Martins}, we compare their theoretical calibrations with our own to study the effect of metallicity on these quantities. 
Both \logqH~and \logqHeI~increase as $Z$ decreases, with the largest differences seen in O dwarfs that exceed the Galactic calibration by 0.25~dex on average. 

\begin{figure}
    \includegraphics[width=0.9\hsize]{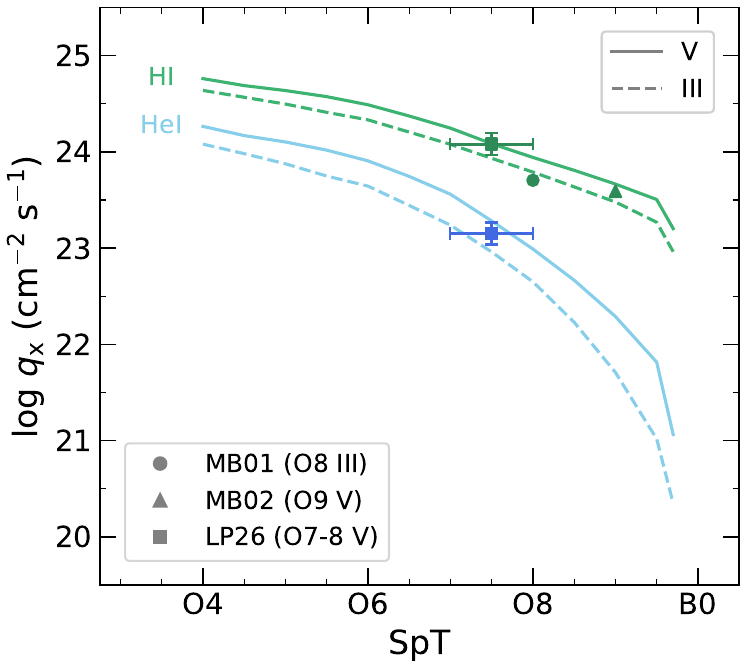}
    \caption{The \logqH~(green) and \logqHeI~(blue) scales obtained in this work are compared against measurements of the ionizing flux of XMP O stars: MB01 and MB02 in the Magellanic Bridge \citep{Ramachandran2021} and LP26 in Leo~P \citep{Telford2023}. We note that \citet{Ramachandran2021} did not provide error bars for \logQH.
    }
    \label{fig:logQ_Obs}
\end{figure}

There are few direct estimates of the ionizing photon production of massive stars at extremely low metallicities. \citet{Ramachandran2021} determined \logQH~for two metal-poor O stars in the Magellanic Bridge (0.20-0.02~\Zsun) from the quantitative analysis of their spectra. \citet{Telford2023} recently calculated the H~\textsc{i}- and He-ionizing photon yields of an O7-8~V star in Leo~P (0.03~\Zsun) from the nebular emission lines of the H~{\sc ii} region powered by the star, assuming that no photons were escaping. 
In Figure \ref{fig:logQ_Obs}, we compare our ionizing flux scales with their results. 
For the objects in \citet{Ramachandran2021}, we used the radii they derived for each star to convert the given photon production rate to ionizing photon flux. 
\citet{Telford2023} did not estimate the radius of the O7-8~V star, thus we used the average of our calibrated radii for O7 and O8~dwarfs and propagated the radius uncertainty to the error bars of the ionizing flux of this star.  
Our calibration is in agreement with both studies. 
For the particular case of the Leo~P star, our results support \citet{Telford2023}’s hypothesis that no ionizing photons are leaking from the nebula.

\begin{figure}
    \includegraphics[width=\hsize]{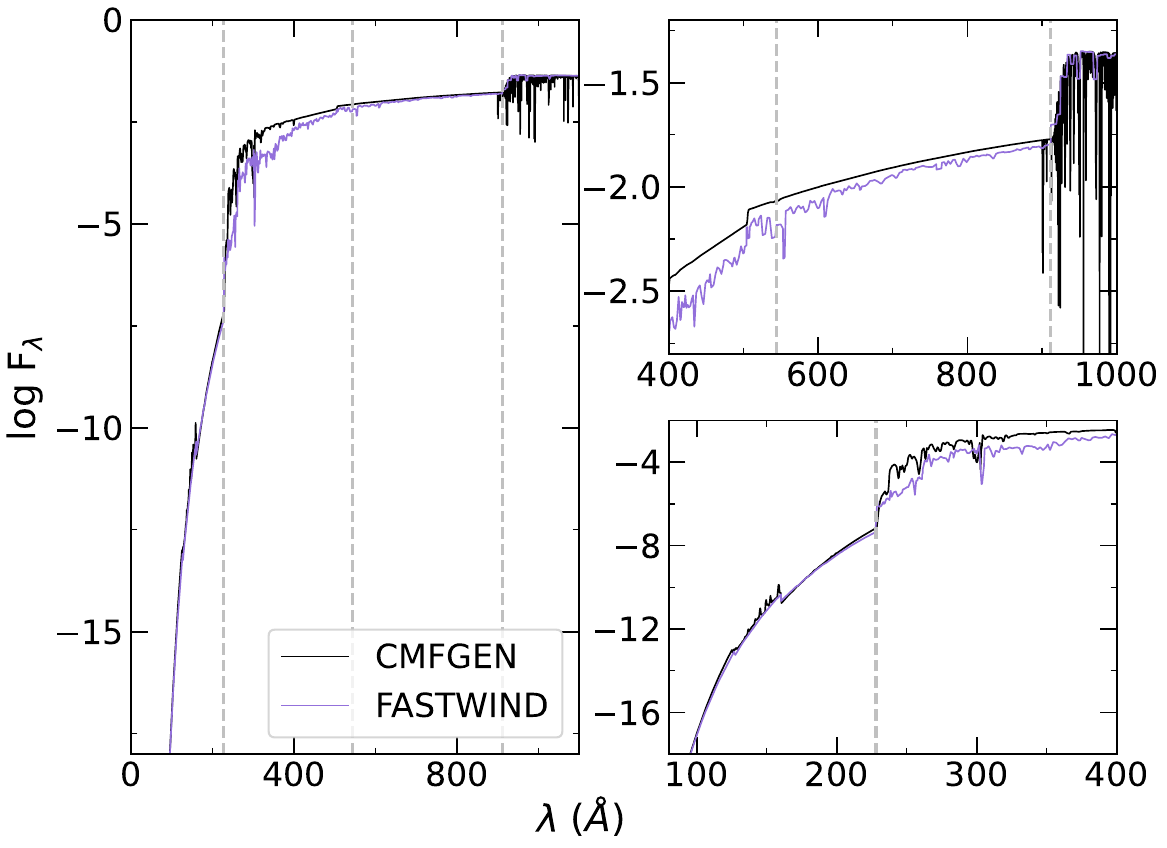}
    \caption{
    Spectral energy distributions of \Teff~=~38~kK, \logg~=~4.0~dex and \mbox{\logQ~=~-14~dex} models calculated with \textsc{fastwind} (purple) and \textsc{cmfgen} (black). The right panels zoom into the ionization edges of H, \HeI~and \HeII. The ionization wavelengths of these atomic species are indicated by grey dashed lines for reference.
    }
    \label{fig:QH_CMFGEN}
\end{figure}

Finally, we discuss whether the estimated ionizing fluxes depend on the stellar atmosphere code used. 
Although \textsc{fastwind} does not explicitly include the detailed atomic models of some species, as for instance \textsc{cmfgen} does  \citep{HillierMiller1998}, it does provide an approximate treatment of the full comprehensive line list inherited from WM-basic in the extreme ultraviolet (EUV) and shorter wavelengths \citep{Pauldrach2001}. 
Because of the severe absorption by interstellar H photons at $\lambda <$ 912~\AA\, except in very low H~{\sc i} column density sight-lines, 
only a few stars have been observed in the EUV \citep[e.g.,][]{Hoare1993, Cassinelli1994} and 
there are scarce observational constraints 
on this part of the SED.
However, we estimated that the impact on the resulting ionizing fluxes is not significant. In Figure~\ref{fig:QH_CMFGEN}, we compare SEDs computed with \textsc{fastwind} and \textsc{cmfgen}, for models with \Teff~=~38~kK, \logg~=~4.0~dex and \logQ~=~-14~dex.
The small discrepancies between the SEDs do not result in significant differences in the ionizing fluxes: 0.10~dex for \logqH~and 0.20~dex for \logqHeI~and \logqHeII. 
If we now compare a \textsc{fastwind} model with negligible winds against a \textsc{tlusty} model 
\citep[OSTAR2002 grid,][]{LanzHubeny2003} with \Teff~=~37.5~kK and  \logg~=~4.0~dex, we find 
only small differences ($\Delta$\logqH~=~0.05~dex, $\Delta$\logqHeI~=~0.15~dex $\Delta$\logqHeII~=~0.06~dex)
despite \textsc{tlusty}'s more sophisticated treatment of line blanketing by metals.
The differences may increase at lower effective temperatures \citep[e.g.][]{Simon-diazStasinska2008}.


\subsubsection{The \HeII~ionizing flux}
\label{sec:qHeII}

The \logqHeII~scale of O stars exhibits a bimodal distribution in the mid- and late-O types (see Fig.~\ref{fig:logQscale}). 
In this section, we discuss that this behaviour is a consequence of known the dependence of \qHeII~on \Teff~and \Mdot~\citep[e.g.,][]{SchmutzHamann1986, Gabler1989, SchaererdeKoter1997}.

The flux capable of ionizing \HeII~depends on the available photons (\Teff, \logL) and on the population of the ground state of this ion. 
The more populated the ground state is, the higher the continuum opacity at wavelengths shorter than~$\lambda_{\mathrm{He II}}$, resulting in a reduction of \qHeII. The population of the ground state is primarily determined by \Teff, wind density (\Mdot)~and \logg.

For a negligible wind outflow, equivalent to a hydrostatic atmosphere, the radiation field drives the population of the ground state. Thus, the behaviour of \qHeII~as a function of \Teff~or spectral type is similar to \qH~and \qHeI.
As \Mdot~increases, the wind velocity field desaturates the resonance lines, which enhances the excitation of electrons to the upper levels and depopulates the ground state \citep{Gabler1989}.
At a certain \Mdot, however, the wind is so strong that all the resonance lines become saturated and cannot absorb more photons. If \Mdot~continues increasing, the ground state will become overpopulated by the enhanced recombination rate, significantly increasing the opacity of the continuum and leading to a strong reduction of the \HeII-ionizing flux \citep{SchaererdeKoter1997}. The same process also impacts the \HeI-ionizing continuum of early B~giants \citep[][]{Najarro1996}.

The \Mdot-threshold that defines the transition between resonance lines and recombinations as dominant drivers of the ground-state \HeII~population depends mainly on the effective temperature of the star \citep{SchmutzHamann1986}. 
At higher \Teff, the \HeII~is almost fully ionized into He~\textsc{iii}, hence strong \Mdot~as high as measured in Wolf-Rayet stars would be needed to repopulate the ground state of \HeII. Conversely, the number of available ionizing photons is significantly reduced at lower \Teff~(Wien side of the SED) and even with weak winds the lines will saturate and the \HeII~ground state will become overpopulated.

\begin{figure*}
    \includegraphics[width=\hsize]{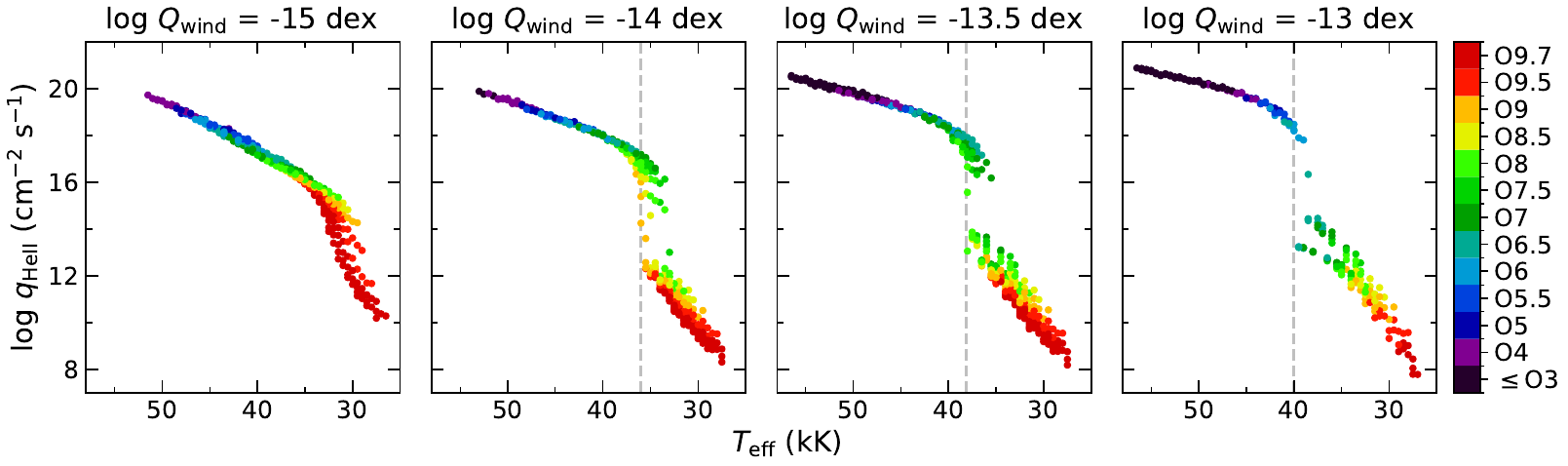}
    \caption{\HeII-ionizing flux of all the models classified as O stars, regardless their luminosity class, as a function of \Teff~for different values of the wind strength parameter. The coloured sidebar indicates the assigned spectral subtypes. The Figure illustrates how the production of \HeII-ionizing photons experiences a bimodal behaviour that depends on \Teff~and \logQ. The sharp decrease of \qHeII~(dashed line) occurs at higher \Teff~as the wind strength increases.
    }
    \label{fig:logQHeII_TEFF_SPT}
\end{figure*}

Figure~\ref{fig:logQHeII_TEFF_SPT} illustrates this effect. We show the \HeII-ionizing flux as a function of \Teff~for different values of \logQ~in our grid.
XMP early-O stars are always in the regime dominated by resonance lines. Thus, the \HeII-ionizing flux is slightly higher as the wind is stronger, with differences of less than 1~dex.
The transition between a high and low production of \HeII-ionizing photons occurs at the mid- and late-types, which may lead to differences of \logqHeII~of up to 4~dex between the same spectral subtype.
The dependence of \qHeII~on the wind strength explains the bimodal distribution seen in Fig.~\ref{fig:logQscale}, since for a given spectral type and luminosity class our collection includes both models with negligible wind outflows (\mbox{\logQ = -15~dex}) and with weak and moderate winds (\mbox{\logQ $\geq$ -14~dex}).

In Figure~\ref{fig:logQHeII_TEFF_SPT}, the transition between a high and low production of \HeII-ionizing photons shifts to higher temperatures as $Q_{\mathrm{wind}}$ increases.
An analogue behaviour was detected by \citet{MartinsPalacios2021} when they compared the \QHeII/\QH~ratio for different metallicities. They found that the transition \Teff~between high and low \QHeII/\QH~increases with increasing metallicity. This behaviour actually reflects the dependence of \QHeII~on the wind strength, which the authors had parameterized as a function of metallicity.  

The dependence of \qHeII~on \Mdot~imprints large uncertainties on the estimated production of \HeII-ionizing photons by individual massive stars. If we examine, e.g., an O8~dwarf in Fig.~\ref{fig:logQscale}, the star could lie at either side of the bimodal distribution of \logqHeII~depending on its stellar parameters. 
We now know that winds as weak as \mbox{\logQ~=~-14~dex} already impact \qHeII~(Fig.~\ref{fig:logQHeII_TEFF_SPT}), hence \Mdot~must be accurately determined to estimate \qHeII~from the SED of the best-fitting model. 
However, the wind diagnostics in the optical range (Balmer lines and \HeII~4686) are insensitive to these weak wind values (e.g., Lorenzo et al. in prep.). Therefore, additional diagnostics in the UV \citep[e.g.,][]{Garcia2014,Telford2021b} or in the NIR \citep[e.g.,][]{Najarro2011} are critical to constrain the wind, and consequently \qHeII.
Due to the strong dispersion of \logqHeII~within the spectral types, we provide only the maximum values in Table~\ref{tab:StellarParams}.


\begin{figure}
    \includegraphics[width=\hsize]{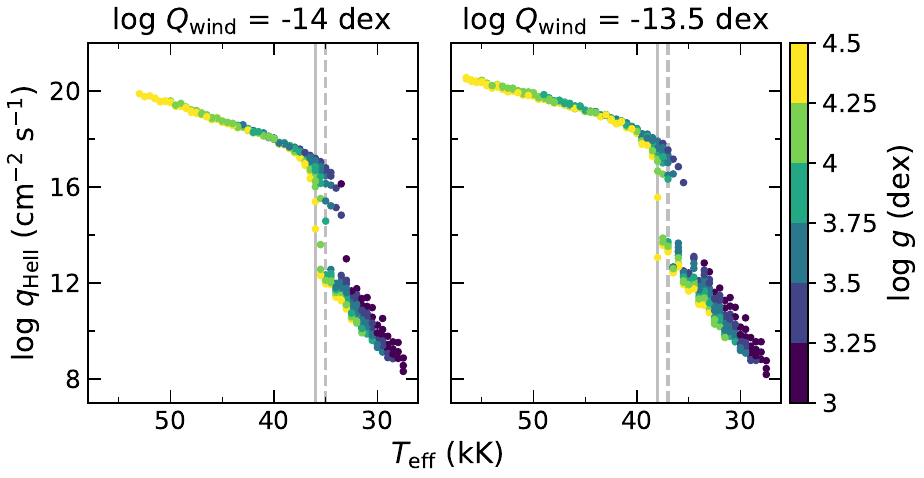}
    \caption{Impact of \logg~on \qHeII. All models classified as O stars with \logQ~=~-14~dex and -13.5~dex are shown, independently of their spectral type and luminosity class. The sidebar illustrates the colour code according to \logg. We mark the transition \Teff~for models with \logg~=~4.5~dex (solid line) and \logg~=~4.0~dex (dashed line).
    }
    \label{fig:logQHeII_TEFF_logg}
\end{figure}

Finally, surface gravity also impacts \qHeII. For constant \Mdot~and \Teff, lower \logg~will result in a larger scale-height, enhancing the escape probability of the ionizing photons and therefore increasing the probability of recombination and impacting the \HeII~continuum. Consequently, the transition between high/low \qHeII~will occur at higher \Teff~when \logg~increases.
This is in apparent contradiction with what we observe in Figure~\ref{fig:logQscale}, where the bimodal distribution of class III starts at earlier types than in class V.
We note, however, that the impact of \logg~is weaker than the impact of \Mdot~and \Teff, and therefore Fig.~\ref{fig:logQscale} rather reflects the different \Teff~between classes~III and V. 

Figure~\ref{fig:logQHeII_TEFF_logg} illustrates that \logg~has a second-order effect compared to the role of \Teff~and \Mdot.
If the wind strength parameter is enhanced by 0.5~dex, the transition \Teff~shifts by +2~kK. An increase of 0.5~dex in \logg~only shifts the position of the transition \Teff~by +1~kK.
We note, however, that outside the \logg~ranges considered in this work, that correspond to $\Gamma_{\mathrm{e}} \leq$~0.52 for the O stars and $\Gamma_{\mathrm{e}} \leq$~0.55 for the B stars, this parameter could have a stronger impact.

\begin{figure}
    \includegraphics[width=\hsize]{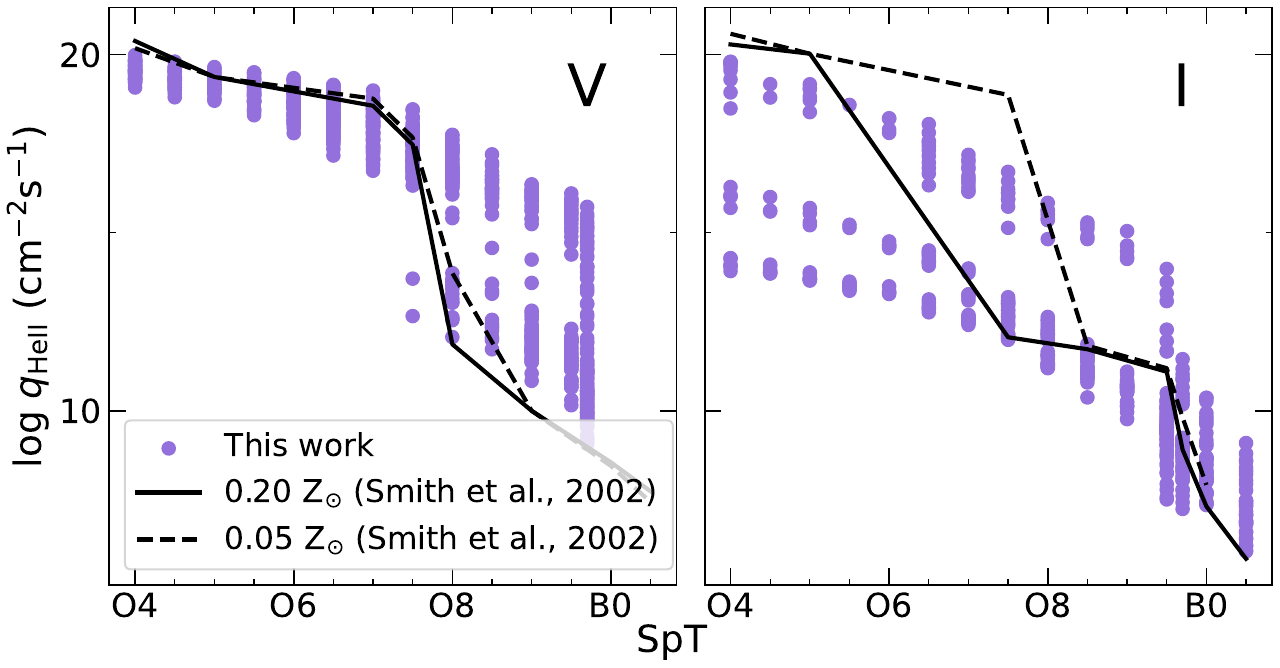}
    \caption{The \HeII-ionizing flux estimated in this work for O dwarfs and supergiants (points) compared against \citet{Smith2002}'s scales for metallicities 0.20~\Zsun~(solid lines) and 0.05~\Zsun~(dashed lines).
    }
    \label{fig:Smith2002}
\end{figure}

In Figure~\ref{fig:Smith2002}, we compare our derived \logqHeII~for O dwarfs and supergiants with the scales calculated by \citet{Smith2002} for metallicity 0.20~\Zsun~and 0.05~\Zsun, which are used in population synthesis codes, such as \textsc{starburst99} \citep{ Leitherer1999}, and in analyses of single-star \HII~regions.
\citet{Smith2002} estimated ionizing fluxes by integrating the SED of a grid of models computed with \textsc{wm-basic} \citep{Pauldrach2001}. 
In contrast to our grid, \citet{Smith2002} assigned one single set of parameters (\Teff, \logg, \Mdot, v$_{\infty}$) for each spectral type and luminosity class at a given metallicity.
Consequently, as shown in Figure~\ref{fig:Smith2002}, their scales do not reflect the bimodal behaviour of the \HeII-ionizing flux. 
Their calibration may overestimate \qHeII~from early- and mid-O supergiants.
On the other hand, it underestimates the production of \HeII-ionizing photons by the more numerous late O~dwarfs,
which may also contribute to power nebular \HeII~4686 and \HeII~1640 emission \citep[e.g.,][]{Kehrig2015, Wofford2021, Senchyna2021}.



There are three key takeaway messages for this section. Firstly, in order to quantify the \HeII-ionizing flux of O stars all stellar parameters need to be constrained, especially \Teff~and \Mdot, and calibrations are not reliable. Secondly, this implies that in the case of XMP O stars, observations outside the optical range are needed to reach diagnostics that are sensitive to the expected low \Mdot.
Finally, the stellar grids to be used by population synthesis codes cannot assume one single value of \Mdot~because the winds can significantly impact \qHeII~even if they are weak.



\section{Photometric calibration} 
\label{sec:phot_results}

The grid of \textsc{fastwind} models and their SEDs also enabled us to compute calibrations of magnitudes, colours and bolometric corrections for XMP O stars and B supergiants. 
Because we usually draw our candidate XMP OB~stars from \citet{Massey2007}’s catalogue \citep[e.g.,][\citetalias{Lorenzo2022}]{Camacho2016} we firstly computed synthetic photometry in the same photometric system: the $UBVRI$ filters of the instrument MOSAIC installed at the 4-m Mayall telescope on Kitt Peak National Observatory (KPNO).
In addition, we calculated synthetic photometry in other frequently used filters: the \textit{UBVRIJHK} generic Bessell bands, the \textit{ugri} bands of the Sloan Digital Sky Survey (SLOAN-SDSS) and a set of optical and UV filters of the Wide-Field Planetary Camera~2 (WFPC2) and the Wide Field Camera 3 (WFC3) onboard the Hubble Space Telescope (HST).

From the emergent SEDs at the stellar radius [$F(\lambda)$] of each \textsc{fastwind} model, we calculated $F_{\mathrm{T}}$,
\begin{equation}
F_{\mathrm{T}} = \frac{\int \lambda F(\lambda) T(\lambda) \mathrm{d}\lambda}{\int   \lambda T(\lambda) \mathrm{d}\lambda},
\label{Eq:ref_pho}
\end{equation}
where $T(\lambda)$ is the transmission curve of filter T\footnote{The transmission response curves of all filters were downloaded from the Spanish Virtual Observatory \citep[][]{Rodrigo2012, Rodrigo2020, Rodrigo2024}.}.
Then, we computed the absolute magnitude by applying the corresponding zero points ($ZP_{\mathrm{T}}$) and accounting for the radii adopted for each model and a distance of 10~pc,
\begin{equation}
    M_{\mathrm{T}} = -2.5\, \log F_{\mathrm{T}} + 2.5\, \log ZP_{\mathrm{T}} - 5 \log R_* + 5 \log 10~\mathrm{pc}.
\end{equation}

Zero points were calculated with Eq.~\ref{Eq:ref_pho} using the spectrum of Vega ($\alpha$ Lyrae)\footnote{\href{http://svo2.cab.inta-csic.es/theory/fps/morefiles/vega.dat}{http://svo2.cab.inta-csic.es/theory/fps/morefiles/vega.dat}}.
For the SLOAN-SDSS bands, which are usually in the AB system, we used a reference spectrum of constant flux equal to 3631~Jy.

The bolometric correction in the corresponding band was calculated following the expression,
\begin{equation}
BC_{\mathrm{T}} = M_{\odot}^{\mathrm{bol}} - M_{\mathrm{T}} - 2.5 \times \log \frac{L}{L_{\odot}},
\end{equation}
where $M_{\odot}^{\mathrm{bol}}$ is the solar bolometric magnitude in filter T.

We provide the calibrated photometric colours, absolute $V$-magnitude and bolometric corrections for the KPNO-MOSAIC system in Table~\ref{tab:Phot_KPNO_Mosaic}, as a function of spectral type and luminosity class. 
As we did with stellar parameters, we computed the average, minimum and maximum value of colours and bolometric corrections by considering all \textsc{fastwind} models compatible with each spectral subtype and luminosity class.
The calibrations for other photometric systems are provided in Appendix~\ref{appx:PhotCal}.
We remind the reader that, while the calibrated colours and bolometric corrections are radius-independent, absolute magnitudes may need revision once extensive information on XMP OB star radii is available.

\clearpage
\onecolumn
\setlength\LTleft{-15pt}
\setlength\LTright{-20pt}
\setlength\LTcapwidth{\linewidth}
\captionsetup[longtable]{labelfont=bf, font = footnotesize, labelsep=period}
\begin{longtable}{lccccc|c}
\caption{Calibrated photometric colours, bolometric corrections and absolute visual magnitudes of O stars and early-B supergiants in the KPNO-MOSAIC system. We include the average (av.), minimum (min.) and maximum (max.) values calculated with the SED of all grid models classified with the corresponding subtypes, except for $M_{\mathrm{V}}$, for which we only provide the average. The latter parameter was calculated with the calibrated (O stars) and observed (B stars) radii provided in Table~\ref{tab:StellarParams}.}
\label{tab:Phot_KPNO_Mosaic}\\
\toprule
\toprule
     SpT &            $(U-B)_0$ &            $(B-V)_0$ &            $(V-R)_0$ &            $(R-I)_0$ &             $BC_{V}$ & $M_{\mathrm{V}}$ \\
 & av. (min., max.) & av. (min., max.) & av. (min., max.) & av. (min., max.) & av. (min., max.) & av. \\
\midrule
\endfirsthead
\caption[]{continued.} \\
\toprule
\toprule
     SpT &            $(U-B)_0$ &            $(B-V)_0$ &            $(V-R)_0$ &            $(R-I)_0$ &             $BC_{V}$ & $M_{\mathrm{V}}$ \\
 & av. (min., max.) & av. (min., max.) & av. (min., max.) & av. (min., max.) & av. (min., max.) & av. \\
\midrule
\endhead
\midrule
\multicolumn{7}{r}{{\textit{continued on next page}}} \\

\endfoot

\bottomrule
\endlastfoot
    O4 V & -1.25 (-1.26, -1.23) & -0.30 (-0.31, -0.30) & -0.16 (-0.16, -0.15) & -0.19 (-0.20, -0.19) & -4.32 (-4.46, -4.19) &            -5.21 \\
  O4.5 V & -1.24 (-1.25, -1.23) & -0.30 (-0.31, -0.30) & -0.16 (-0.16, -0.15) & -0.19 (-0.19, -0.19) & -4.22 (-4.33, -4.10) &            -5.06 \\
    O5 V & -1.24 (-1.25, -1.22) & -0.30 (-0.31, -0.29) & -0.15 (-0.16, -0.15) & -0.19 (-0.20, -0.19) & -4.15 (-4.27, -4.01) &            -4.91 \\
  O5.5 V & -1.23 (-1.25, -1.22) & -0.30 (-0.31, -0.29) & -0.15 (-0.16, -0.15) & -0.19 (-0.20, -0.19) & -4.07 (-4.20, -3.93) &            -4.78 \\
    O6 V & -1.23 (-1.25, -1.22) & -0.30 (-0.30, -0.29) & -0.15 (-0.15, -0.15) & -0.19 (-0.19, -0.19) & -3.96 (-4.11, -3.79) &            -4.66 \\
  O6.5 V & -1.22 (-1.24, -1.20) & -0.29 (-0.30, -0.28) & -0.15 (-0.15, -0.14) & -0.19 (-0.20, -0.19) & -3.82 (-4.00, -3.61) &            -4.50 \\
    O7 V & -1.22 (-1.23, -1.19) & -0.29 (-0.30, -0.28) & -0.15 (-0.15, -0.14) & -0.19 (-0.20, -0.18) & -3.69 (-3.86, -3.45) &            -4.36 \\
  O7.5 V & -1.20 (-1.22, -1.18) & -0.28 (-0.29, -0.28) & -0.14 (-0.15, -0.14) & -0.19 (-0.19, -0.18) & -3.53 (-3.71, -3.37) &            -4.19 \\
    O8 V & -1.20 (-1.21, -1.17) & -0.28 (-0.29, -0.27) & -0.14 (-0.15, -0.14) & -0.19 (-0.20, -0.18) & -3.41 (-3.56, -3.24) &            -4.04 \\
  O8.5 V & -1.19 (-1.21, -1.17) & -0.28 (-0.29, -0.27) & -0.14 (-0.15, -0.14) & -0.19 (-0.20, -0.18) & -3.31 (-3.44, -3.16) &            -3.88 \\
    O9 V & -1.18 (-1.20, -1.16) & -0.28 (-0.29, -0.27) & -0.14 (-0.15, -0.14) & -0.19 (-0.20, -0.18) & -3.22 (-3.37, -3.05) &            -3.71 \\
  O9.5 V & -1.17 (-1.19, -1.15) & -0.28 (-0.28, -0.27) & -0.14 (-0.15, -0.14) & -0.19 (-0.20, -0.18) & -3.15 (-3.30, -2.98) &            -3.56 \\
  O9.7 V & -1.15 (-1.18, -1.12) & -0.27 (-0.28, -0.26) & -0.14 (-0.15, -0.13) & -0.19 (-0.20, -0.18) & -3.04 (-3.23, -2.83) &            -3.31 \\
\toprule
  O4 III & -1.22 (-1.24, -1.20) & -0.29 (-0.30, -0.28) & -0.15 (-0.16, -0.14) & -0.19 (-0.19, -0.18) & -4.14 (-4.33, -3.93) &            -5.89 \\
O4.5 III & -1.22 (-1.24, -1.20) & -0.29 (-0.30, -0.28) & -0.15 (-0.15, -0.14) & -0.19 (-0.19, -0.18) & -4.05 (-4.17, -3.88) &            -5.81 \\
  O5 III & -1.21 (-1.23, -1.20) & -0.29 (-0.30, -0.28) & -0.15 (-0.15, -0.14) & -0.19 (-0.19, -0.18) & -3.95 (-4.08, -3.84) &            -5.74 \\
O5.5 III & -1.21 (-1.23, -1.19) & -0.28 (-0.29, -0.27) & -0.14 (-0.15, -0.13) & -0.19 (-0.19, -0.18) & -3.83 (-4.01, -3.66) &            -5.68 \\
  O6 III & -1.20 (-1.22, -1.19) & -0.28 (-0.29, -0.27) & -0.14 (-0.15, -0.13) & -0.19 (-0.19, -0.18) & -3.74 (-3.90, -3.62) &            -5.62 \\
O6.5 III & -1.19 (-1.21, -1.18) & -0.28 (-0.29, -0.27) & -0.14 (-0.15, -0.13) & -0.18 (-0.19, -0.18) & -3.59 (-3.79, -3.42) &            -5.52 \\
  O7 III & -1.19 (-1.20, -1.17) & -0.28 (-0.28, -0.27) & -0.14 (-0.14, -0.13) & -0.18 (-0.19, -0.18) & -3.46 (-3.60, -3.34) &            -5.44 \\
O7.5 III & -1.18 (-1.18, -1.17) & -0.27 (-0.28, -0.26) & -0.14 (-0.14, -0.13) & -0.18 (-0.19, -0.18) & -3.33 (-3.39, -3.25) &            -5.36 \\
  O8 III & -1.17 (-1.18, -1.15) & -0.27 (-0.28, -0.26) & -0.13 (-0.14, -0.13) & -0.18 (-0.19, -0.17) & -3.21 (-3.31, -3.09) &            -5.27 \\
O8.5 III & -1.16 (-1.17, -1.15) & -0.27 (-0.27, -0.26) & -0.13 (-0.14, -0.13) & -0.18 (-0.19, -0.18) & -3.11 (-3.19, -3.05) &            -5.18 \\
  O9 III & -1.15 (-1.16, -1.14) & -0.26 (-0.27, -0.26) & -0.13 (-0.14, -0.13) & -0.18 (-0.19, -0.18) & -3.02 (-3.07, -2.97) &            -5.09 \\
O9.5 III & -1.14 (-1.15, -1.13) & -0.26 (-0.27, -0.25) & -0.13 (-0.14, -0.13) & -0.18 (-0.18, -0.17) & -2.93 (-3.00, -2.84) &            -4.97 \\
O9.7 III & -1.13 (-1.14, -1.11) & -0.26 (-0.26, -0.25) & -0.13 (-0.14, -0.13) & -0.18 (-0.18, -0.17) & -2.82 (-2.91, -2.73) &            -4.87 \\
\toprule
    O4 I & -1.21 (-1.22, -1.19) & -0.28 (-0.29, -0.27) & -0.14 (-0.14, -0.13) & -0.18 (-0.18, -0.17) & -3.97 (-4.08, -3.85) &            -6.17 \\
  O4.5 I & -1.20 (-1.22, -1.19) & -0.28 (-0.28, -0.27) & -0.14 (-0.14, -0.13) & -0.18 (-0.18, -0.17) & -3.85 (-3.93, -3.77) &            -6.15 \\
    O5 I & -1.20 (-1.21, -1.19) & -0.27 (-0.28, -0.26) & -0.14 (-0.14, -0.13) & -0.18 (-0.18, -0.17) & -3.78 (-3.85, -3.66) &            -6.16 \\
  O5.5 I & -1.19 (-1.20, -1.19) & -0.27 (-0.28, -0.27) & -0.13 (-0.14, -0.13) & -0.18 (-0.18, -0.17) & -3.71 (-3.79, -3.64) &            -6.18 \\
    O6 I & -1.18 (-1.19, -1.18) & -0.27 (-0.27, -0.26) & -0.13 (-0.13, -0.12) & -0.18 (-0.18, -0.17) & -3.61 (-3.68, -3.53) &            -6.18 \\
  O6.5 I & -1.17 (-1.18, -1.17) & -0.26 (-0.27, -0.26) & -0.13 (-0.13, -0.12) & -0.17 (-0.18, -0.17) & -3.48 (-3.58, -3.40) &            -6.18 \\
    O7 I & -1.16 (-1.17, -1.16) & -0.26 (-0.26, -0.25) & -0.12 (-0.13, -0.12) & -0.17 (-0.17, -0.16) & -3.32 (-3.40, -3.24) &            -6.16 \\
  O7.5 I & -1.16 (-1.17, -1.15) & -0.26 (-0.26, -0.24) & -0.12 (-0.13, -0.11) & -0.17 (-0.18, -0.16) & -3.20 (-3.29, -3.12) &            -6.16 \\
    O8 I & -1.15 (-1.16, -1.14) & -0.25 (-0.26, -0.24) & -0.12 (-0.13, -0.11) & -0.17 (-0.18, -0.15) & -3.09 (-3.17, -3.00) &            -6.14 \\
  O8.5 I & -1.14 (-1.16, -1.13) & -0.25 (-0.26, -0.24) & -0.12 (-0.13, -0.11) & -0.17 (-0.18, -0.16) & -2.98 (-3.09, -2.90) &            -6.09 \\
    O9 I & -1.14 (-1.15, -1.12) & -0.25 (-0.26, -0.24) & -0.12 (-0.14, -0.11) & -0.17 (-0.18, -0.15) & -2.91 (-3.00, -2.82) &            -6.09 \\
  O9.5 I & -1.13 (-1.15, -1.11) & -0.24 (-0.26, -0.23) & -0.12 (-0.13, -0.10) & -0.17 (-0.18, -0.15) & -2.80 (-2.96, -2.66) &            -6.06 \\
  O9.7 I & -1.12 (-1.13, -1.10) & -0.24 (-0.26, -0.22) & -0.12 (-0.13, -0.10) & -0.17 (-0.18, -0.15) & -2.69 (-2.84, -2.55) &            -5.99 \\
    B0 I & -1.10 (-1.11, -1.09) & -0.23 (-0.25, -0.21) & -0.11 (-0.13, -0.10) & -0.16 (-0.17, -0.14) & -2.57 (-2.71, -2.42) &            -6.74 \\
  B0.5 I & -1.08 (-1.10, -1.07) & -0.23 (-0.24, -0.20) & -0.11 (-0.12, -0.09) & -0.15 (-0.17, -0.13) & -2.44 (-2.63, -2.28) &            -6.94 \\
    B1 I & -1.06 (-1.07, -1.03) & -0.22 (-0.24, -0.19) & -0.10 (-0.12, -0.08) & -0.15 (-0.16, -0.13) & -2.32 (-2.49, -2.13) &            -7.09 \\
  B1.5 I & -1.01 (-1.04, -0.95) & -0.20 (-0.22, -0.16) & -0.09 (-0.11, -0.06) & -0.13 (-0.15, -0.10) & -2.03 (-2.31, -1.73) &            -7.10 \\
    B2 I & -0.93 (-0.96, -0.89) & -0.17 (-0.20, -0.13) & -0.07 (-0.09, -0.04) & -0.11 (-0.13, -0.08) & -1.75 (-1.99, -1.47) &            -7.08 \\
  B2.5 I & -0.86 (-0.90, -0.82) & -0.16 (-0.18, -0.13) & -0.06 (-0.08, -0.04) & -0.10 (-0.11, -0.08) & -1.54 (-1.73, -1.32) &            -7.09 \\
    B3 I & -0.79 (-0.83, -0.76) & -0.15 (-0.16, -0.14) & -0.06 (-0.07, -0.05) & -0.09 (-0.10, -0.08) & -1.39 (-1.54, -1.31) &            -7.15 \\
\end{longtable}
\begin{tablenotes}
\item \textbf{Notes.} The filter transmission responses used to calculate the synthetic photometry were taken from the SVO Carlos Rodrigo filter profile service (\href{http://svo2.cab.inta-csic.es/theory/fps/index.php?mode=browse&gname=KPNO&gname2=Mosaic&asttype=}{http://svo2.cab.inta-csic.es/theory/fps/index.php?mode=browse\&gname=KPNO\&gname2=Mosaic\&asttype=}).
\end{tablenotes}
\clearpage
\twocolumn

\subsection{Accounting for the higher hydrogen series}

\begin{figure}
    \includegraphics[width=\hsize]{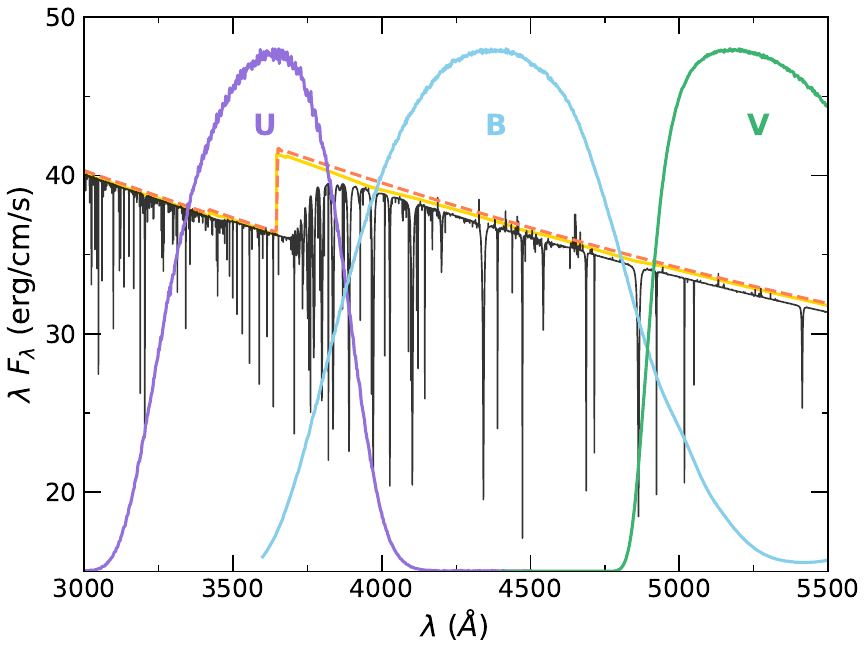}
    \caption{SEDs delivered by different atmosphere codes: \textsc{fastwind} (yellow solid line), \textsc{cmfgen} (black solid line) and \textsc{cmfgen}  with lines and level dissolution de-activated in the formal solution (red dashed line).
    All models have \Teff~=~33~kK and \logg~=~3.4~dex. As a reference, we show the transmission curves of KPNO-MOSAIC $UBV$ filters.
    }
    \label{fig:SEDjump}
\end{figure}

Our photometric calibrations were computed from the SEDs delivered by \textsc{fastwind}, which do not include spectral lines or the continuum decrease due to the higher lines of the hydrogen series (level dissolution). 
This can affect the synthetic photometry, especially in the bands covering the hydrogen jumps (see e.g., the Balmer jump in Fig.~\ref{fig:SEDjump}). 
We used \textsc{cmfgen} models to quantify this effect and apply corrections to our derived photometry when needed.

\begin{figure}
    \includegraphics[width=\hsize]{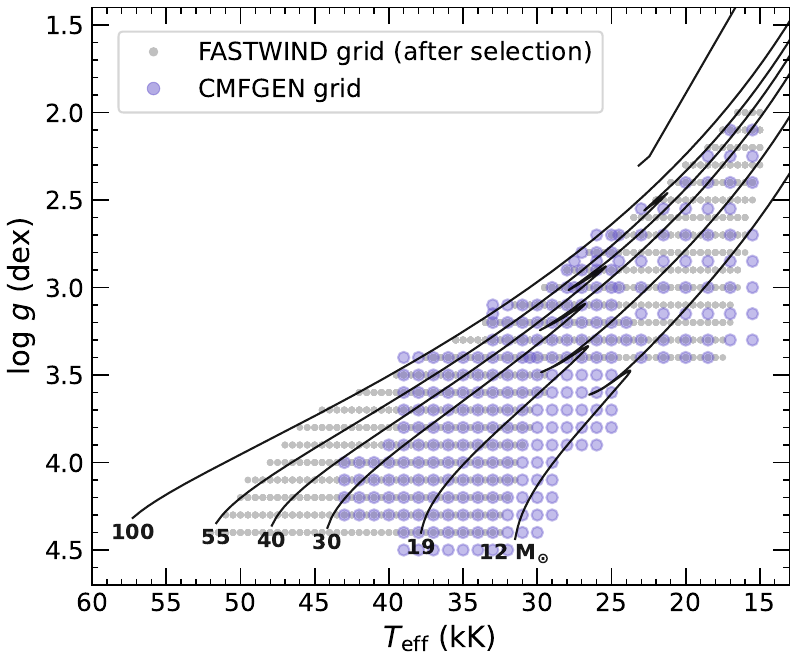}
    \caption{Parameter space covered by the \textsc{cmfgen} (purple) and the \textsc{fastwind} grids of models (grey). Only the subset of \textsc{fastwind} models used for the calibrations is shown.
    }
    \label{fig:CMFGENgrid}
\end{figure}

\textsc{cmfgen} \citep{Hillier1990,HillierMiller1998} 
delivers SEDs including the full set of spectral lines and level dissolution, but it is also possible to de-activate these options, thus delivering SED equivalent to \textsc{fastwind}’s.
We built a grid of \textsc{cmfgen} models covering the \Teff--\logg~space of our \textsc{fastwind} grid (see Fig.~\ref{fig:CMFGENgrid}) at the \Teff~ranges where we expect the strongest hydrogen jump (cooler \Teff), hence a stronger impact in the calculated synthetic photometry.
For each pair of \Teff~and \logg, we considered only one value of mass-loss rate as we measured that the effect of this parameter is small (of the order of 0.001~mag).

\begin{figure}
    \includegraphics[width=\hsize]{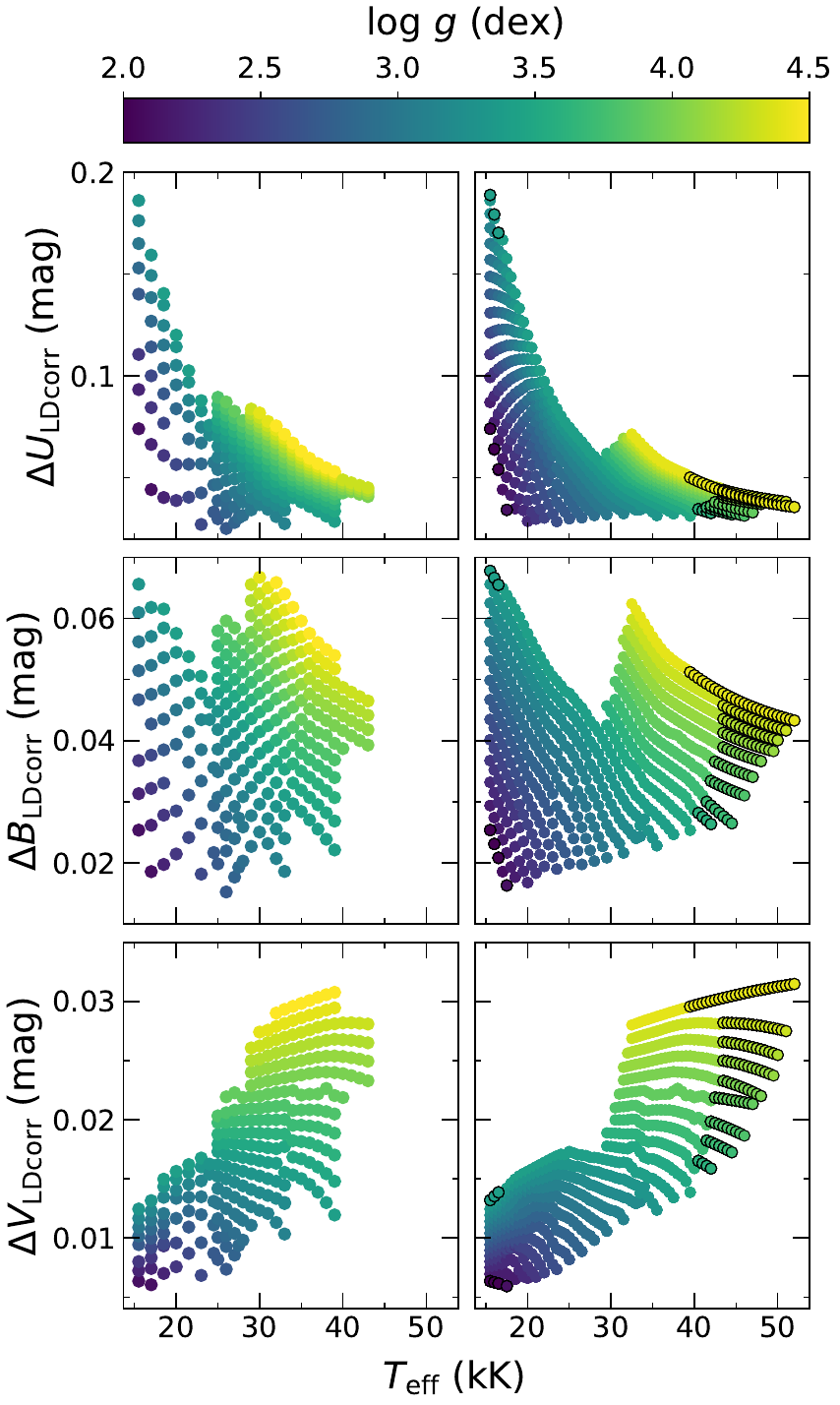}
    \caption{
    Left panels: Differences between the KPNO-MOSAIC $UBV$ magnitudes calculated from the SED of \textsc{cmfgen} models with and without including spectral lines and the continuum dissolution by the higher levels of the Balmer series ($\Delta M_{\mathrm{T, LDcorr}}$). These effects have the largest impact on the $U$-band, which overlaps with the Balmer jump. Increments are plotted as a function of \Teff~and colour-coded by \logg. 
    Right panels: $\Delta M_{\mathrm{T, LDcorr}}$ values are interpolated to the \Teff--\logg~sampling of the \textsc{fastwind} grid. Circled symbols mark extrapolated values. These and analogue corrections were applied to the photometric calibrations provided in Tables~\ref{tab:Phot_KPNO_Mosaic} and \ref{tab:Phot_Generic_Bessell}--\ref{tab:Phot_HST}.
    }
    \label{fig:LDcorr}
\end{figure}

For each model and filter, we computed the absolute magnitudes with the full \textsc{cmfgen} SED and also without considering lines and level dissolution, and calculated the difference between the two of them ($\Delta M_{\mathrm{T, LDcorr}}$).
This quantity thus represents the correction to be applied to the magnitudes calculated from \textsc{fastwind} SEDs in order to account for the not included lines and level dissolution.
The left panels of Figure~\ref{fig:LDcorr} show $\Delta M_{\mathrm{T, LDcorr}}$ computed for the $UBV$ bands of KPNO-MOSAIC. As expected, the effect of lines and level dissolution is larger at lower \Teff~and higher \logg, and it is more pronounced in the $U$-band.

We then interpolated the \textsc{cmfgen} $\Delta M_{\mathrm{T, LDcorr}}$ values to the \Teff--\logg~values of the \textsc{fastwind} grid, using a cubic spline in the \Teff--\logg~space. 
For the \Teff--\logg~values outside the \textsc{cmfgen} grid coverage, we extrapolated the correction as follows.
We considered separately each \logg~value and fitted $\Delta M_{\mathrm{T, LDcorr}}$ with respect to \Teff~using a least squares method.
For the bands most affected by the level dissolution, those that overlap with the hydrogen jumps (such as $U, B, I$, etc.), we used an exponential function to fit the points. For others mostly affected by spectral lines (such as $V, R, J$, etc.), we used a second-degree polynomial. 
In the right panels of Fig.~\ref{fig:LDcorr}, we show the interpolated $\Delta M_{\mathrm{T, LDcorr}}$ for the \textsc{fastwind} grid and the extrapolated values circled in black.

These and analogue corrections were applied to all the synthetic photometric calibrations provided in Tables~\ref{tab:Phot_KPNO_Mosaic} and \ref{tab:Phot_Generic_Bessell}--\ref{tab:Phot_HST}.

\subsection{Comparison with existing calibrations}

\begin{figure*}
\centering
    \includegraphics[width=\hsize]{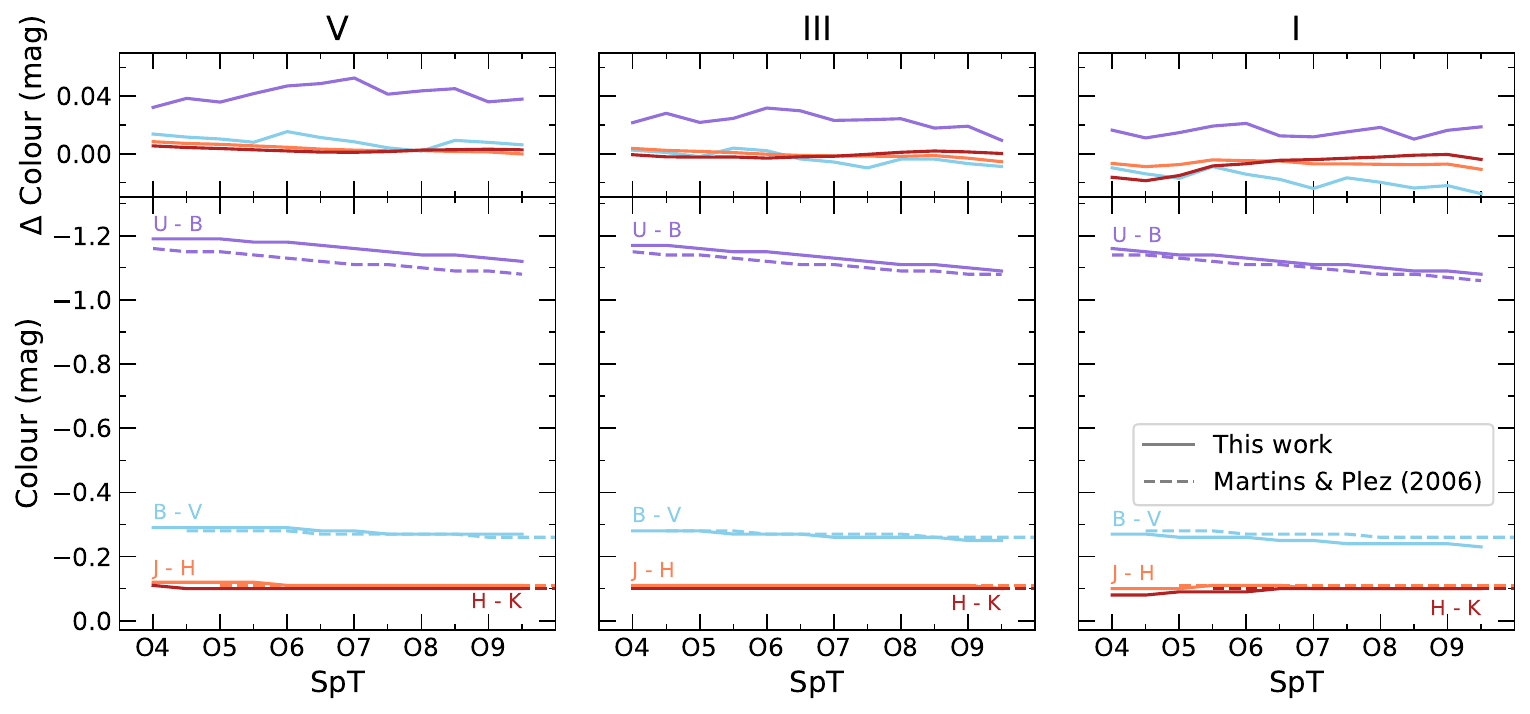}
    \caption{
    \citet{MartinsPlez2006}'s calibration for the Bessell filters is compared against this work. We show the difference between the two systems in the upper panels. Colours are indicated in the legend.
    }
    \label{fig:PHOT_Martins}
\end{figure*}

In Figure~\ref{fig:PHOT_Martins}, we compare our calibration for the Bessell bands with the photometric calibration of \citet{MartinsPlez2006} for massive stars in the MW.
The two calibrations are similar for all colours except for $(U-B)_{\mathrm{0}}$ where XMP O stars are up to 0.04~mag bluer, as expected because of their higher \Teff.

XMP O dwarfs and giants exhibit up to 0.01~mag bluer $(B-V)_{\mathrm{0}}$, but the supergiants show redder $(B-V)_{\mathrm{0}}$ than the stars with \Zsun. 
This apparent contradiction with expectations may be due to the different treatment of gravity in the two calibrations.

Surface gravity has an impact on $(B-V)_{\mathrm{0}}$: the lower the gravity, the redder the SED of O stars \citep{AbbottHummer1985}. 
\citet{MartinsPlez2006} calculated photometry for \citet{Martins2005}’s grid of models, who assigned one value of \logg~per subtype and luminosity class.
In contrast, we left this parameter free and obtained a range of \logg~compatible with each pair of (SpT, LC). Our treatment resulted in higher mean \logg~values for dwarfs and early giants, and slightly lower (on average) for supergiants, which could explain the small differences seen in Figure~\ref{fig:PHOT_Martins}.


\subsection{The reddening-free pseudo-colour $\mathbf{Q_{\mathrm{phot}}}$}

\citet{JohnsonMorgan1953} introduced the pseudo-colour \Qphot~[\mbox{\Qphot~$= (U-B) - 0.72 \times (B-V)$}], a parameter independent of reddening as long as the total to selective extinction ratio is \mbox{$R_{\mathrm{v}}$ = 3.1}. 
This parameter holds a biunivocal relation with spectral type from O stars to B0 types and
can be used as an effective criterion to select OB~star candidates \citep[e.g.,][]{Garcia2009}.

\citet{GarciaHerrero2013} used the $(U-B)$ vs. \Qphot~diagram from a photometric catalogue to search for O stars in the metal-poor galaxy IC~1613. They set \mbox{\Qphot~$<$~-0.8~mag} as their target selection criterion, which resulted in the positive confirmation of 70~per cent of their candidates as OB-type stars.
Using the same selection criterion, \citet{Camacho2016} and \citetalias{Lorenzo2022} achieved success rates of 50~per cent and 66~per cent in Sextans~A, respectively. 
In addition, \citetalias{Lorenzo2022} used their large sample and found that by using \mbox{\Qphot~$<$~-1.0~mag}, the success rate could increase to 84~per cent.

\begin{figure*}
    \includegraphics[width=\hsize]{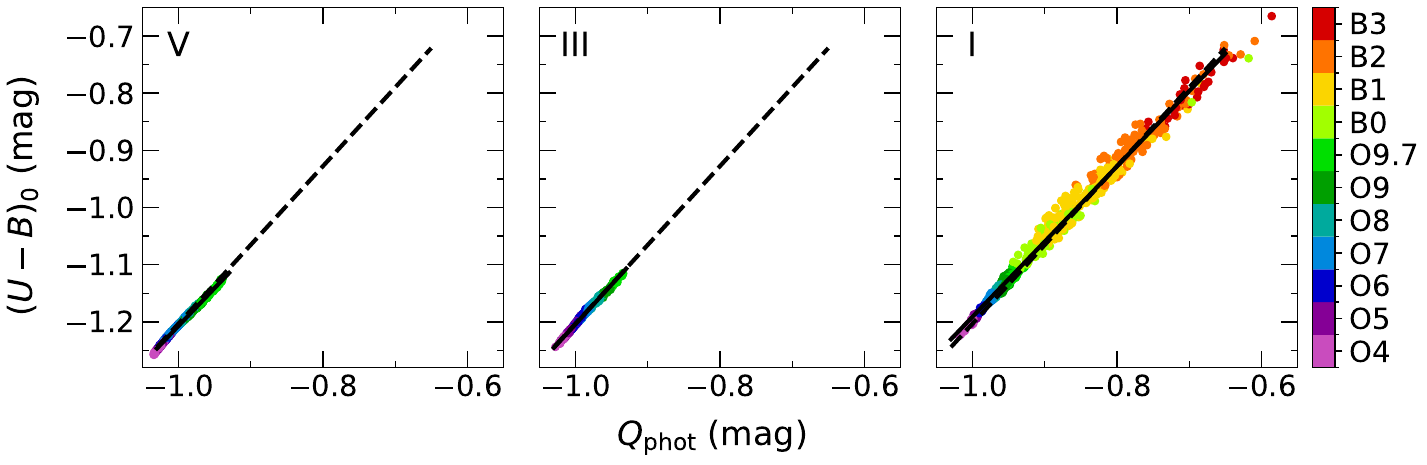}
    \caption{\Qphot~vs. $(U-B)_{\mathrm{0}}$~calculated for our grid models, separated by luminosity class and colour-coded according to spectral type. The solid lines are the linear fits to the points for each luminosity class.
    The black dashed line represents the fit to all models regardless their luminosity class (Eq.~\ref{eq:UB0vsQ_all}). We note the overlap between the two fits.}
    \label{fig:UB0vsQ}
\end{figure*}

Figure~\ref{fig:UB0vsQ} shows \Qphot~vs. $(U-B)_{\mathrm{0}}$ as a function of spectral subtype and luminosity class. 
\Qphot~and $(U-B)_{\mathrm{0}}$ hold a linear relation 
with small dispersion in the earliest subtypes.
Furthermore, Fig.~\ref{fig:UB0vsQ} shows the strong monotonic relation of \Qphot~and $(U-B)_{\mathrm{0}}$ with spectral type, supporting the use of the $(U-B)$ vs. \Qphot~diagram to unveil OB~stars. All models classified as O stars are found at \mbox{\Qphot~$<$~-0.90~mag}, in agreement with the results of \citetalias{Lorenzo2022}.

In order to provide a relation between \Qphot~and the intrinsic $(U-B)_{\mathrm{0}}$ that can be used to identify OB-type candidates and to estimate extinction, we fitted the calibrated intrinsic colours of all our \textsc{fastwind} models. The least squares fit returns the following relations (solid black lines in Fig.~\ref{fig:UB0vsQ}): 

\begin{equation}
(U-B)_{\mathrm{0}} = 
\begin{cases}
0.116\pm0.003~+~1.323~\times~Q_{\mathrm{phot}} &~\text{(O4 - O9.7 V)} \\
0.179\pm0.004~+~1.383~\times~Q_{\mathrm{phot}} &~\text{(O4 - O9.7 III)} \\
0.125\pm0.020~+~1.316~\times~Q_{\mathrm{phot}} &~\text{(O4 - B3 I)} \\
\end{cases}
.
\label{eq:UB0vsQ}
\end{equation}

The \Qphot~vs. $(U-B)_{\mathrm{0}}$~relation is approximately the same for all luminosity classes, and only the intercept varies.
The value of the intercept is smaller (bluer) for dwarfs than for giants and supergiants. This implies that the X-axis dispersion in the observed \Qphot~vs. $(U-B)$ diagram of massive star populations is partly intrinsic, and not only due to extinction.

Because of the difficulty of obtaining high-quality photometry in the $U$-band, reddening is most often estimated with the $(B-V)$ colour for which \Qphot~is also a good proxy.
Using Kurucz's ATLAS9 models with $Z$~=~0.80~\Zsun,
\citet{Massey2000} parameterized the relation between $(B-V)_{\mathrm{0}}$ and \Qphot~as 
$(B-V)_{\mathrm{0}} = -0.005 + 0.317 \times Q_{\mathrm{phot}} $. This relation
does not take into account the luminosity class of the stars, which may have an impact as we show below.

\begin{figure*}
    \includegraphics[width=\hsize]{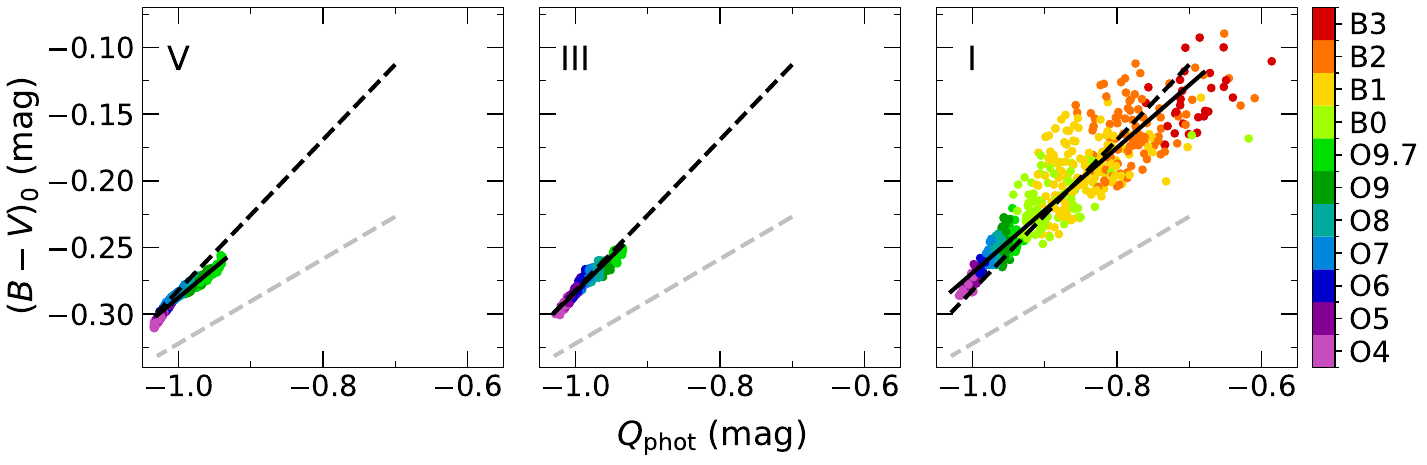}
    \caption{
    \Qphot~vs. $(B-V)_{\mathrm{0}}$~calculated for our grid models, separated by luminosity class and colour-coded according to spectral type. The solid lines are the linear fits to the points for each luminosity class. The black dashed line represents the fit to all models regardless their luminosity class (Eq.~\ref{eq:BV0vsQ_all}) and the grey dashed line represents \citet{Massey2000}'s relation for $Z$~=~0.80~\Zsun.
    }
    \label{fig:BV0vsQ}
\end{figure*}

Figure~\ref{fig:BV0vsQ} shows \Qphot~vs $(B-V)_{\mathrm{0}}$ for all our \textsc{fastwind} models, as a function of their spectral type and luminosity class.
For comparison, we included \citet{Massey2000}'s relation. 
Unlike $(U-B)_{\mathrm{0}}$ vs. \Qphot, the $(B-V)_{\mathrm{0}}$ vs. \Qphot~diagram shows larger scatter, especially for B stars. Hence, the $(U-B)$ colour excess is more reliable for quantifying extinction when $U$-band photometry is available.
Figure~\ref{fig:BV0vsQ} also shows that our models are 0.07~mag redder on average than \citet{Massey2000}'s relation.

We fitted the intrinsic colors of the \textsc{fastwind} models with a least square method, and obtained the following relations:

\begin{equation}
(B-V)_{\mathrm{0}} = 
\begin{cases}
0.161\pm0.005~+~0.448~\times~Q_{\mathrm{phot}} &~\text{(O4 - O9.7 V)} \\
0.249\pm0.006~+~0.532~\times~Q_{\mathrm{phot}} &~\text{(O4 - O9.7 III)} \\
0.200\pm0.029~+~0.469~\times~Q_{\mathrm{phot}} &~\text{(O4 - B3 I)} \\
\end{cases}
.
\label{eq:BV0vsQ}
\end{equation}
\noindent The error bars given in eqs.~\ref{eq:UB0vsQ} and ~\ref{eq:BV0vsQ} represent the dispersion of the intrinsic colours in our grid, calculated so that the defined ranges encompass 90\%~of the models.

The largest advantage of using \Qphot, however, relies on providing a good approximation to the intrinsic colours of sources only from photometry, lacking any information on their spectral classification. To enable exploiting this advantage, we also computed the $(B-V)_0$ vs. \Qphot~and $(U-B)_0$ vs. \Qphot~relations using all the grid models, without separating them into luminosity class:

\begin{equation}
(B-V)_{\mathrm{0}} = 
0.282\pm0.021~+~0.564~\times~Q_{\mathrm{phot}}
.
\label{eq:BV0vsQ_all}
\end{equation}
\begin{equation}
(U-B)_{\mathrm{0}} = 
0.174\pm0.015~+~1.377~\times~Q_{\mathrm{phot}}
.
\label{eq:UB0vsQ_all}
\end{equation}


\section{A case example: mapping the extinction of Sextans A}
\label{sec:ExtMap}

Dwarf irregular galaxies are often assumed to experience negligible and uniform extinction. 
However, massive stars in these galaxies enable studying extinction in resolved sightlines and results are challenging this working hypothesis. For instance, \citet{Garcia2019} measured a colour excess of up to \EBV~=~0.250~mag towards massive stars in the South of the 0.10~\Zsun~dwarf irregular galaxy Sextans~A, a significantly larger value than its foreground extinction \citep[\EBV$_\mathrm{fg}$ = 0.044~mag,][]{Tammann2011}.
These findings suggest that the local reddening in Sextans~A could be non-negligible and uneven.
With a larger sample covering the whole galaxy, \citetalias{Lorenzo2022} showed that the locus of the OB-type stars in the colour-magnitude diagram (CMD) did not match the expected location for their spectral classification, in line with \citet{Garcia2019}'s results.

Characterizing and mapping the internal extinction of Sextans~A is not only useful to search for its massive stellar population or to estimate the completeness of \citetalias{Lorenzo2022}'s catalogue, but it could also aid tracing the molecular gas content of this XMP galaxy.

In the Milky Way and similarly metal-rich environments, the molecular gas is easily traced by CO.
However, the detection of this molecule is challenging in XMP environments, and inferring the molecular gas content from dust emission (from e.g. Herschel observations) is also problematic as the gas-to-dust ratio is poorly calibrated at very low $Z$.

There is only one weak detection of  CO (J~=~1$\rightarrow$0) in Sextans~A with IRAM’s 30-m telescope \citep{Shi2015} that has not been confirmed with interferometric observations taken with ALMA or CARMA \citep{Warren2015}. This apparent paucity of molecular gas is at odds with the young population of massive stars in the galaxy, although a number of possibilities could explain the non-detection of CO. In this particular example, the CO emission could be extended and therefore be missed by the interferometers.
In general, the CO emission will also be weaker because of the lower metal content of galaxies like Sextans~A. 
On the other hand, the higher porosity of the XMP interstellar medium makes CO more prone to dissociation \citep[e.g.,][]{Ramambason2022}, and may result in a different stratification of the photodissociation region 
\citep[smaller and more compact CO cores surrounded by wider \Htwo~distributions, e.g., ][]{Bolatto2013, Madden2020}. All these factors could make molecular gas in XMP galaxies like Sextans~A CO-dark.

\begin{figure*}
    \centering
    \includegraphics[width=\hsize]{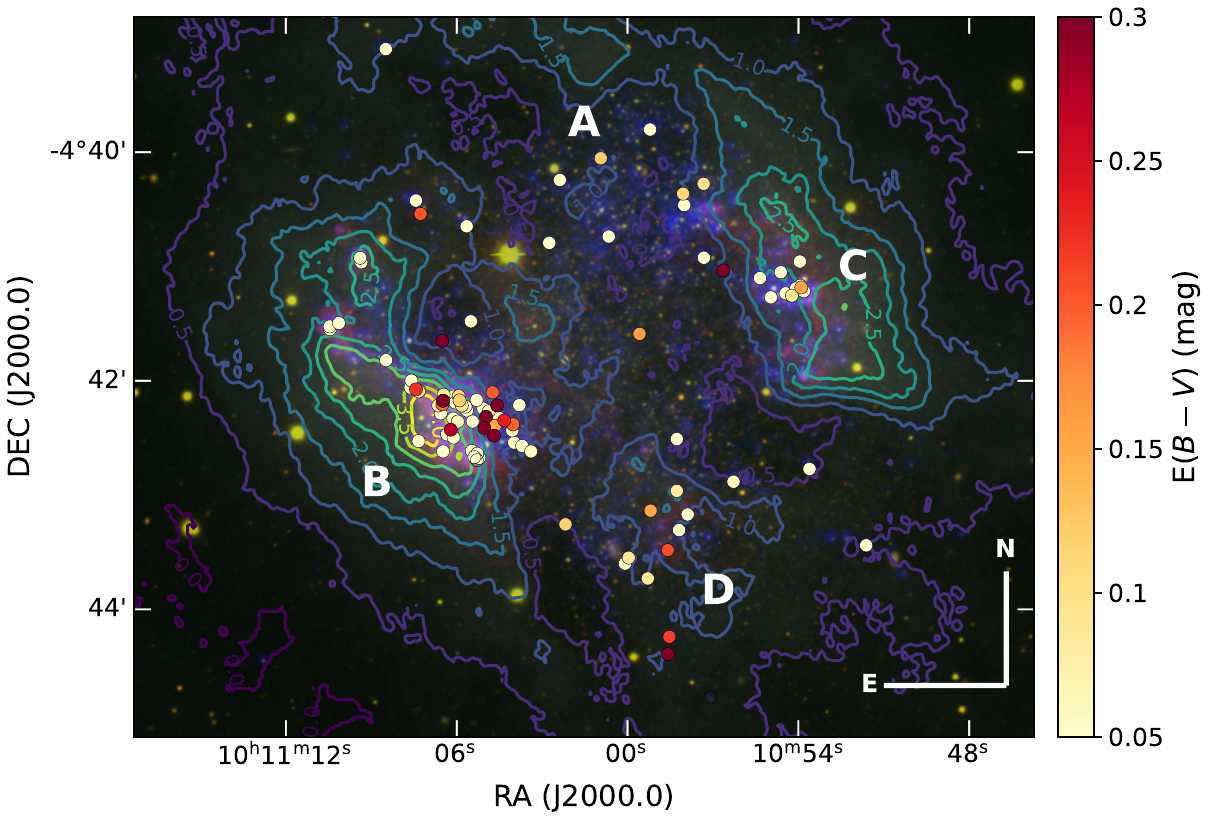}
    \caption{Extinction map of Sextans~A. The background image is an RGB composite of Sextans~A, built with \ha-- (red) and $V$--bands (green) from \citet{Massey2007}, and GALEX FUV (blue) data. We overlay a \textsc{little things} \HI~map \citep{Hunter2012} with contour lines marking regions of constant column density (in 10$^{21}$~cm$^{-2}$ units). We colour-coded the O4-B3~stars of \citetalias{Lorenzo2022}'s catalogue by the colour excess towards the line of sight. The latter was calculated using the calibration of synthetic colours provided in this work and \citet{Massey2007}'s observed photometry. We also mark the A, B, C and D regions with ongoing star formation as a reference.
    }
    \label{fig:EXTmap}
\end{figure*}

Using the intrinsic colours calibrated for 0.10~\Zsun~stars in this work and the observed photometry of the OB stars in \citetalias{Lorenzo2022}'s catalogue, we built a two-dimensional map of colour excess and extinction in Sextans~A.
In Figure~\ref{fig:EXTmap}, we show the colour excess towards the lines of sight of the OB stars in \citetalias{Lorenzo2022}.
We selected stars with subtypes B3 and earlier with an entry in \citet{Massey2007}.
We then calculated \EBV~using the intrinsic colours calibrated in this work for the spectral type and luminosity class of each target.
For those cases where the derived \EBV~was lower than the foreground value, for example, the \textit{blue outliers} of \citetalias{Lorenzo2022} (stars with bluer colours than the ZAMS in the CMD of Sextans~A), we adopted the foreground extinction of the galaxy.
In Fig.~\ref{fig:EXTmap}, we also provide the \textsc{little things} map of atomic hydrogen of Sextans~A \citep{Hunter2012} to better interpret the extinction map.

Region~B of Sextans~A shows the highest concentration of OB-type stars and neutral atomic hydrogen, and we find large variations of \EBV~at very short length scales. Most of the stars in this region have low colour excess consistent with the foreground extinction, however, 10 stars show \EBV~between 
 0.2 and 0.6~mag.
This may be explained by their relative location to the local gas within the line of sight, with stars with larger colour excess being more embedded in the \HI~cloud.
Applying \citet{Bohlin1978}'s relation ($N_{\rm H}/$\EBV~=~5.8~10$^{21}$~cm$^{-2}$mag$^{-1}$, where $N_H = N_{\rm HI} + N_{\rm H_2}$) to our derived \EBV~and the measured $N_{\rm HI}$ column densities, we estimated that the column density of molecular hydrogen could reach up to $N_{\mathrm{H_2}}$~$\sim$~0.80~10$^{21}$~cm$^{-2}$.

In region C, however, we do not find significant \EBV~variations. Unlike region B, this complex does not overlap with the western \HI~cloud, but lies on its inner edge. Only s096, the closest star to the \HI~cloud, has a high colour excess of 0.5 mag. Using again \citet{Bohlin1978}'s relation, we find that the density of \Htwo~at this location could be $\sim$~1.00~10$^{21}$~cm$^{-2}$.

Finally, six stars lie in regions of low \HI~column density, between 0.5 and 1~10$^{21}$~cm$^{-2}$, and strikingly suffer from strong colour excess. 
One possible explanation is that they may be surrounded by a colder disk that results in redder colours.
This is the case for star s003, located in region~D. Recent observations with higher resolution and S/N revealed that s003 is an Oe star (M. Garcia priv. comm.), i.e. a star that rotates so fast that has developed a circumstellar decretion disk.

An alternative explanation is that these sources are multiple systems or very compact clusters, for which our calibration would be assigning incorrect intrinsic colours. Only two of these six objects have been proposed as candidate binary systems, s060 and s094.
Finally, the strong extinction could also indicate that these stars are located in compact pockets of gas, detectable only with higher resolution maps of gas and dust or higher sensitivity radio observations. 

This is one example of the many applications our calibrations
offer to study and characterize XMP environments. In this case, our extinction map, further complemented with future observations of additional massive stars in Sextans~A, may guide the search of molecular gas within the galaxy.


\section{Summary and concluding remarks}
\label{sec:conclusions}

In this work, we present calibrations and material to analyze future observations of $Z$~=~0.10~\Zsun~OB stars.

We processed a grid of \textsc{fastwind} synthetic spectra to match the observations of \citetalias{Sota2011}'s standards, classified the models, 
and removed those with unrealistic stellar parameters. 
We used the distilled subset to build the first calibration of stellar parameters and synthetic photometry of 0.10~\Zsun~massive stars as a function of spectral type.
Because we left all stellar parameters free to allow for different combinations to yield the same spectral morphology, we obtained a range of parameters for each (SpT, LC) pair, more realistically accounting for the scatter seen in nature.



Our derived scales of effective temperature and H- and \HeI-ionizing flux are in agreement with results for OB stars and \HII~regions in sub-SMC metallicity environments. 0.10~\Zsun~OB stars are 1-6~kK hotter (depending on luminosity class) and produce higher H- and \HeI-ionizing fluxes than their Galactic analogues.
As reported in previous works, our \textsc{fastwind} grid showed that the \HeII-ionizing flux depends on the wind density, even if winds are weak (\logQ~$\sim$~-14~dex). This implies that the \HeII-ionizing photon production of XMP massive stars is unconstrained unless their \Teff~and \Mdot~are determined. Moreover, in order to study the \HeII-ionizing photon budget of XMP galaxies, population synthesis codes should be provided with stellar libraries that vary mass-loss rates within the HRD. The calibrations used in some stellar population synthesis codes may be underestimating the production of \HeII~ionizing photons by the more numerous O dwarfs.

This work also delivers calibrated intrinsic colours, magnitudes and bolometric corrections for different photometric systems. The calibrated photometry accounts for spectral lines in the considered ranges and the level dissolution caused by the high hydrogen series thanks to correction factors calculated with \textsc{cmfgen} models.
We calculated the relation between $(B-V)_{\mathrm{0}}$ and \Qphot~that can be used to estimate extinction and to identify 0.10~\Zsun~ OB-star candidates.
For a given \Qphot~pseudo-colour, our calibration returns redder $(B-V)_{\mathrm{0}}$ than \citet{Massey2000}'s relation by 0.07~mag on average.


Finally, as an example of the possible application of our theoretical calibrations, we mapped the internal extinction of the 0.10~\Zsun~dwarf irregular galaxy Sextans~A. We used \citetalias{Lorenzo2022}'s spectroscopic catalogue, their spectral types and our calibrated intrinsic colours to estimate colour excess. The map shows Sextans~A suffers from significant and uneven interstellar extinction, especially in region~B that overlaps with the eastern \HI~cloud, but also in other areas with low \HI~column density.

To summarize, this paper unites the tools needed to search and provide a preliminary analysis of OB-type stars in XMP galaxies. Two of the most advanced atmosphere codes, \textsc{fastwind} and \textsc{cmfgen}, were used in our calculations to provide long-lasting materials. All tables will be available online so they may serve as references for future studies of XMP OB~stars with a variety of instruments.

\section*{Acknowledgements}

M. Lorenzo, M. Garcia and F. Najarro gratefully acknowledge support by grants PID2019-105552RB-C41 and PID2022-137779OB-C41, and M. Garcia further acknowledges grant PID2022-140483NB-C22, funded by the Spanish Ministry of Science, Innovation and Universities/State Agency of Research MICIU/AEI/10.13039/501100011033 and by “ERDF
A way of making Europe”.
M. Lorenzo also acknowledges funding from grant PRE2019-087988 under project MDM-2017-0737-19-3 Unidad de Excelencia "María de Maeztu"-Centro de Astrobiología (INTA-CSIC).
M. Cerviño acknowledges financial support from the Spanish MICIU through grants PID2019-107408GB-C41 and PID2022-136598NB-C33.
N. Castro gratefully acknowledges funding from the Deutsche Forschungsgemeinschaft \mbox{(DFG) – CA 2551/1-1,2}.
A. Herrero and S. Simón-Díaz acknowledge support by the Spanish MICIU through grant PID2021-122397NB-C21 and the Severo Ochoa Program \mbox{CEX2019-000920-S}.
The authors would also like to thank D. Lennon for his valuable comments and discussion on the classification of B-type stars.
We would like to thank our three anonymous referees for their reviews.

This research made use of the Spanish Virtual Observatory (https://svo.cab.inta-csic.es) project funded by the Spanish MCI through grant PID2020-112949GB-I00. It also used the IAC HTCondor facility (\href{http://research.cs.wisc.edu/htcondor/}{http://research.cs.wisc.edu/htcondor/}), partly financed by the Ministry of Economy and Competitiveness with FEDER funds, code IACA13-3E-2493. 

\section*{Data Availability}

All data are incorporated into the article and its online supplementary material.


\bibliographystyle{mnras}
\bibliography{AA_article.bib}


\newpage
\newpage
\newpage
\pagebreak[3]

\appendix
\section*{Appendix}

\section{The effect of microturbulence}
\label{sec:Appx_micro}

Measuring the microturbulence in O-type stars is challenging. Unlike early-B supergiants, the number of metal lines (best diagnostics for calculating $\xi$) available in the spectra of O stars is scarce. This situation is even more pronounced when we consider the XMP~regime. In this work, we set $\xi$~=~5~\kms~for the models with \Teff~typical of O stars. However, \citet{Holgado2018} measured a higher $\xi$~=~15~\kms~value for Galactic O stars at \mbox{$\mathcal{L} >$ 3.8 dex}.
In this section, we study how the classification assigned to the \textsc{fastwind} models changes if we increase $\xi$ in these regions of the sHRD.

\begin{figure}
    \includegraphics[width=0.9\hsize]{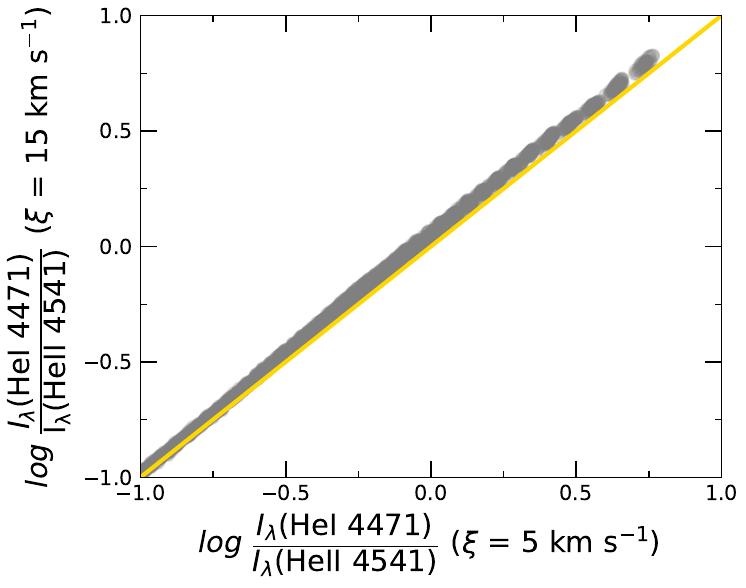}
    \caption{\HeI~4471/\HeII~4541 line intensity ratio for models with $\xi$~=~5~\kms~(adopted in this work) versus $\xi$~=~15~\kms. The yellow line represents the 1:1 relation.}
    \label{fig:MICRO_ratio}
\end{figure}

We followed the same approach described in Section~\ref{sec:GRID_proc}, but this time using $\xi = 15$~\kms~for the models with \mbox{$\mathcal{L} >$ 3.8 dex}. 
In Figure~\ref{fig:MICRO_ratio}, we compare the \HeI~4471/\HeII~4541 ratio of these models with the two values of microturbulence considered. 
We observe that $\xi = $15~\kms~returns higher values than $\xi$~=~5~\kms~by roughly 0.1~dex, although in later types the $\xi = $15~\kms~models exhibit even larger \HeI/\HeII~ratios. Consequently, some models would have been classified with later types than with $\xi$~=~5~\kms. However, this only affects about 20\%  of the models in the range considered, and the changes never exceed 0.5~subtypes.

Therefore, we conclude that varying the microturbulence in our models does not significantly affect the results of our work. Moreover, since this parameter is predicted to decrease with metallicity, we do not expect strong $\xi$ variations for XMP massive stars across the sHRD.

\section{Photometric calibration in additional photometric systems}
\label{appx:PhotCal}

This section collects calibrated colours and bolometric corrections of O~stars and early-B~supergiants in the Bessell (Table~\ref{tab:Phot_Generic_Bessell}) and the SLOAN-SDSS (Table~\ref{tab:Phot_SLOAN_SDSS}) systems, and for a subset of filters of the WFPC2 and WFC3 of the HST (Table~\ref{tab:Phot_HST}). Because of space issues, only absolute magnitudes are provided for the HST filters. 

\onecolumn
\begin{landscape}
\setlength\LTleft{-15pt}
\setlength\LTright{-20pt}
\setlength\LTcapwidth{\linewidth}
\captionsetup[longtable]{labelfont=bf, font = footnotesize, labelsep=period}
\begin{longtable}{lccccccc|c}
\caption{Calibrated photometric colours, bolometric corrections and absolute visual magnitudes of O stars and early-B supergiants in the Bessell system. We include the average (av.), minimum (min.) and maximum (max.) values calculated with the SED of all grid models classified over the corresponding subtypes, except for $M_{\mathrm{V}}$, for which we only provide the average. The latter parameter was calculated with the calibrated (O stars) and observed (B stars) radii provided in Table~\ref{tab:StellarParams}.}
\label{tab:Phot_Generic_Bessell}\\
\toprule
\toprule
     SpT &            $(U-B)_0$ &            $(B-V)_0$ &            $(V-R)_0$ &            $(R-I)_0$ &            $(J-H)_0$ &            $(H-K)_0$ &             $BC_{V}$ & $M_{\mathrm{V}}$ \\
 & av. (min., max.) & av. (min., max.) & av. (min., max.) & av. (min., max.) & av. (min., max.) & av. \\
\midrule
\endfirsthead
\caption[]{continued.} \\
\toprule
\toprule
     SpT &            $(U-B)_0$ &            $(B-V)_0$ &            $(V-R)_0$ &            $(R-I)_0$ &            $(J-H)_0$ &            $(H-K)_0$ &             $BC_{V}$ & $M_{\mathrm{V}}$ \\
 & av. (min., max.) & av. (min., max.) & av. (min., max.) & av. (min., max.) & av. (min., max.) & av. \\
\midrule
\endhead
\midrule
\multicolumn{9}{r}{{\textit{continued on next page}}} \\

\endfoot

\bottomrule
\endlastfoot
    O4 V & -1.19 (-1.20, -1.18) & -0.29 (-0.30, -0.29) & -0.14 (-0.14, -0.13) & -0.18 (-0.18, -0.17) & -0.12 (-0.12, -0.12) & -0.11 (-0.11, -0.10) & -4.33 (-4.47, -4.21) &            -5.19 \\
  O4.5 V & -1.19 (-1.20, -1.18) & -0.29 (-0.30, -0.29) & -0.13 (-0.14, -0.13) & -0.18 (-0.18, -0.17) & -0.12 (-0.12, -0.11) & -0.10 (-0.11, -0.10) & -4.23 (-4.34, -4.11) &            -5.04 \\
    O5 V & -1.19 (-1.20, -1.17) & -0.29 (-0.30, -0.28) & -0.13 (-0.14, -0.13) & -0.18 (-0.18, -0.17) & -0.12 (-0.12, -0.11) & -0.10 (-0.11, -0.10) & -4.16 (-4.28, -4.02) &            -4.89 \\
  O5.5 V & -1.18 (-1.20, -1.17) & -0.29 (-0.30, -0.28) & -0.13 (-0.14, -0.13) & -0.18 (-0.18, -0.17) & -0.12 (-0.12, -0.11) & -0.10 (-0.11, -0.10) & -4.08 (-4.22, -3.95) &            -4.77 \\
    O6 V & -1.18 (-1.19, -1.16) & -0.29 (-0.29, -0.28) & -0.13 (-0.13, -0.13) & -0.18 (-0.18, -0.17) & -0.11 (-0.12, -0.11) & -0.10 (-0.10, -0.10) & -3.97 (-4.12, -3.80) &            -4.64 \\
  O6.5 V & -1.17 (-1.19, -1.15) & -0.28 (-0.29, -0.27) & -0.13 (-0.13, -0.12) & -0.17 (-0.18, -0.17) & -0.11 (-0.12, -0.11) & -0.10 (-0.10, -0.10) & -3.83 (-4.02, -3.62) &            -4.49 \\
    O7 V & -1.16 (-1.18, -1.14) & -0.28 (-0.29, -0.27) & -0.13 (-0.13, -0.12) & -0.17 (-0.18, -0.17) & -0.11 (-0.11, -0.11) & -0.10 (-0.10, -0.10) & -3.70 (-3.87, -3.46) &            -4.34 \\
  O7.5 V & -1.15 (-1.17, -1.13) & -0.27 (-0.28, -0.26) & -0.13 (-0.13, -0.12) & -0.17 (-0.18, -0.17) & -0.11 (-0.11, -0.11) & -0.10 (-0.10, -0.10) & -3.54 (-3.73, -3.38) &            -4.18 \\
    O8 V & -1.14 (-1.16, -1.12) & -0.27 (-0.28, -0.26) & -0.13 (-0.13, -0.12) & -0.17 (-0.18, -0.17) & -0.11 (-0.12, -0.11) & -0.10 (-0.11, -0.10) & -3.42 (-3.57, -3.26) &            -4.02 \\
  O8.5 V & -1.14 (-1.15, -1.12) & -0.27 (-0.28, -0.26) & -0.13 (-0.13, -0.12) & -0.17 (-0.18, -0.17) & -0.11 (-0.12, -0.11) & -0.10 (-0.11, -0.10) & -3.33 (-3.46, -3.18) &            -3.86 \\
    O9 V & -1.13 (-1.15, -1.10) & -0.27 (-0.28, -0.26) & -0.13 (-0.13, -0.12) & -0.17 (-0.18, -0.17) & -0.11 (-0.12, -0.11) & -0.10 (-0.11, -0.10) & -3.24 (-3.38, -3.06) &            -3.70 \\
  O9.5 V & -1.12 (-1.14, -1.09) & -0.27 (-0.27, -0.26) & -0.13 (-0.13, -0.12) & -0.17 (-0.18, -0.17) & -0.11 (-0.12, -0.10) & -0.10 (-0.11, -0.10) & -3.17 (-3.31, -2.99) &            -3.55 \\
  O9.7 V & -1.10 (-1.13, -1.07) & -0.26 (-0.27, -0.25) & -0.12 (-0.13, -0.11) & -0.17 (-0.18, -0.16) & -0.11 (-0.11, -0.10) & -0.10 (-0.11, -0.10) & -3.05 (-3.24, -2.84) &            -3.30 \\
\toprule
  O4 III & -1.17 (-1.19, -1.15) & -0.28 (-0.29, -0.27) & -0.13 (-0.14, -0.12) & -0.17 (-0.18, -0.16) & -0.11 (-0.12, -0.11) & -0.10 (-0.10, -0.09) & -4.15 (-4.34, -3.94) &            -5.88 \\
O4.5 III & -1.17 (-1.18, -1.15) & -0.28 (-0.29, -0.27) & -0.13 (-0.13, -0.12) & -0.17 (-0.18, -0.16) & -0.11 (-0.12, -0.11) & -0.10 (-0.10, -0.09) & -4.06 (-4.19, -3.89) &            -5.79 \\
  O5 III & -1.16 (-1.18, -1.15) & -0.28 (-0.29, -0.27) & -0.13 (-0.13, -0.12) & -0.17 (-0.18, -0.16) & -0.11 (-0.12, -0.11) & -0.10 (-0.10, -0.09) & -3.96 (-4.09, -3.85) &            -5.73 \\
O5.5 III & -1.15 (-1.18, -1.14) & -0.27 (-0.28, -0.26) & -0.12 (-0.13, -0.12) & -0.17 (-0.17, -0.16) & -0.11 (-0.11, -0.11) & -0.10 (-0.10, -0.09) & -3.84 (-4.02, -3.68) &            -5.67 \\
  O6 III & -1.15 (-1.17, -1.14) & -0.27 (-0.28, -0.26) & -0.12 (-0.13, -0.12) & -0.17 (-0.18, -0.16) & -0.11 (-0.11, -0.11) & -0.10 (-0.10, -0.09) & -3.76 (-3.91, -3.64) &            -5.60 \\
O6.5 III & -1.14 (-1.16, -1.12) & -0.27 (-0.28, -0.26) & -0.12 (-0.13, -0.11) & -0.17 (-0.17, -0.16) & -0.11 (-0.11, -0.11) & -0.10 (-0.10, -0.09) & -3.60 (-3.80, -3.43) &            -5.51 \\
  O7 III & -1.13 (-1.15, -1.12) & -0.26 (-0.27, -0.25) & -0.12 (-0.12, -0.11) & -0.17 (-0.17, -0.16) & -0.11 (-0.11, -0.10) & -0.10 (-0.10, -0.09) & -3.48 (-3.61, -3.35) &            -5.43 \\
O7.5 III & -1.12 (-1.13, -1.12) & -0.26 (-0.27, -0.25) & -0.12 (-0.12, -0.11) & -0.17 (-0.17, -0.16) & -0.11 (-0.11, -0.11) & -0.10 (-0.10, -0.10) & -3.34 (-3.41, -3.26) &            -5.35 \\
  O8 III & -1.11 (-1.13, -1.10) & -0.26 (-0.27, -0.25) & -0.12 (-0.12, -0.11) & -0.16 (-0.17, -0.16) & -0.11 (-0.11, -0.11) & -0.10 (-0.10, -0.10) & -3.22 (-3.33, -3.10) &            -5.26 \\
O8.5 III & -1.11 (-1.12, -1.10) & -0.26 (-0.26, -0.25) & -0.12 (-0.12, -0.11) & -0.17 (-0.17, -0.16) & -0.11 (-0.11, -0.10) & -0.10 (-0.10, -0.10) & -3.13 (-3.21, -3.06) &            -5.17 \\
  O9 III & -1.10 (-1.11, -1.09) & -0.25 (-0.26, -0.25) & -0.12 (-0.12, -0.11) & -0.17 (-0.17, -0.16) & -0.11 (-0.11, -0.10) & -0.10 (-0.10, -0.10) & -3.03 (-3.09, -2.98) &            -5.08 \\
O9.5 III & -1.09 (-1.10, -1.08) & -0.25 (-0.26, -0.24) & -0.12 (-0.12, -0.11) & -0.16 (-0.17, -0.16) & -0.10 (-0.11, -0.10) & -0.10 (-0.10, -0.10) & -2.94 (-3.01, -2.86) &            -4.96 \\
O9.7 III & -1.08 (-1.09, -1.06) & -0.25 (-0.25, -0.24) & -0.11 (-0.12, -0.11) & -0.16 (-0.17, -0.16) & -0.10 (-0.11, -0.10) & -0.10 (-0.10, -0.10) & -2.83 (-2.92, -2.74) &            -4.86 \\
\toprule
    O4 I & -1.16 (-1.17, -1.14) & -0.27 (-0.28, -0.26) & -0.12 (-0.13, -0.11) & -0.16 (-0.17, -0.16) & -0.10 (-0.11, -0.10) & -0.08 (-0.10, -0.07) & -3.98 (-4.09, -3.86) &            -6.15 \\
  O4.5 I & -1.15 (-1.17, -1.14) & -0.27 (-0.27, -0.26) & -0.12 (-0.12, -0.12) & -0.16 (-0.17, -0.16) & -0.10 (-0.11, -0.10) & -0.08 (-0.10, -0.07) & -3.86 (-3.95, -3.78) &            -6.13 \\
    O5 I & -1.14 (-1.16, -1.13) & -0.26 (-0.27, -0.25) & -0.12 (-0.12, -0.11) & -0.16 (-0.17, -0.16) & -0.10 (-0.11, -0.09) & -0.09 (-0.10, -0.07) & -3.79 (-3.86, -3.68) &            -6.14 \\
  O5.5 I & -1.14 (-1.14, -1.13) & -0.26 (-0.27, -0.25) & -0.12 (-0.12, -0.11) & -0.16 (-0.17, -0.16) & -0.11 (-0.11, -0.10) & -0.09 (-0.10, -0.09) & -3.72 (-3.80, -3.65) &            -6.17 \\
    O6 I & -1.13 (-1.14, -1.12) & -0.26 (-0.26, -0.25) & -0.11 (-0.12, -0.11) & -0.16 (-0.16, -0.15) & -0.11 (-0.11, -0.10) & -0.09 (-0.10, -0.09) & -3.62 (-3.69, -3.55) &            -6.17 \\
  O6.5 I & -1.12 (-1.13, -1.12) & -0.25 (-0.26, -0.25) & -0.11 (-0.12, -0.11) & -0.16 (-0.16, -0.15) & -0.11 (-0.11, -0.10) & -0.10 (-0.10, -0.09) & -3.50 (-3.60, -3.42) &            -6.17 \\
    O7 I & -1.11 (-1.12, -1.10) & -0.25 (-0.25, -0.24) & -0.11 (-0.11, -0.10) & -0.15 (-0.16, -0.15) & -0.10 (-0.11, -0.10) & -0.10 (-0.10, -0.09) & -3.34 (-3.41, -3.25) &            -6.15 \\
  O7.5 I & -1.11 (-1.11, -1.09) & -0.24 (-0.25, -0.23) & -0.11 (-0.11, -0.10) & -0.15 (-0.16, -0.14) & -0.10 (-0.11, -0.10) & -0.10 (-0.10, -0.09) & -3.21 (-3.31, -3.13) &            -6.15 \\
    O8 I & -1.10 (-1.11, -1.09) & -0.24 (-0.25, -0.23) & -0.11 (-0.11, -0.09) & -0.15 (-0.16, -0.14) & -0.10 (-0.11, -0.10) & -0.10 (-0.10, -0.09) & -3.10 (-3.18, -3.01) &            -6.13 \\
  O8.5 I & -1.09 (-1.10, -1.08) & -0.24 (-0.25, -0.23) & -0.10 (-0.12, -0.10) & -0.15 (-0.16, -0.14) & -0.10 (-0.11, -0.10) & -0.10 (-0.10, -0.09) & -2.99 (-3.10, -2.91) &            -6.08 \\
    O9 I & -1.09 (-1.10, -1.07) & -0.24 (-0.25, -0.22) & -0.11 (-0.12, -0.09) & -0.15 (-0.17, -0.14) & -0.10 (-0.11, -0.09) & -0.10 (-0.10, -0.09) & -2.92 (-3.02, -2.83) &            -6.08 \\
  O9.5 I & -1.08 (-1.10, -1.06) & -0.23 (-0.25, -0.21) & -0.10 (-0.12, -0.09) & -0.15 (-0.17, -0.14) & -0.10 (-0.11, -0.09) & -0.10 (-0.10, -0.08) & -2.81 (-2.98, -2.67) &            -6.04 \\
  O9.7 I & -1.07 (-1.08, -1.05) & -0.23 (-0.25, -0.21) & -0.10 (-0.12, -0.09) & -0.15 (-0.16, -0.13) & -0.10 (-0.10, -0.09) & -0.10 (-0.10, -0.09) & -2.70 (-2.85, -2.56) &            -5.98 \\
    B0 I & -1.05 (-1.06, -1.04) & -0.22 (-0.24, -0.20) & -0.10 (-0.11, -0.08) & -0.14 (-0.16, -0.13) & -0.09 (-0.10, -0.08) & -0.09 (-0.10, -0.09) & -2.58 (-2.73, -2.43) &            -6.73 \\
  B0.5 I & -1.03 (-1.05, -1.02) & -0.21 (-0.23, -0.19) & -0.09 (-0.11, -0.08) & -0.14 (-0.15, -0.12) & -0.09 (-0.10, -0.07) & -0.09 (-0.10, -0.08) & -2.45 (-2.64, -2.28) &            -6.93 \\
    B1 I & -1.01 (-1.02, -0.98) & -0.21 (-0.23, -0.18) & -0.09 (-0.10, -0.07) & -0.13 (-0.15, -0.11) & -0.08 (-0.09, -0.07) & -0.09 (-0.09, -0.08) & -2.33 (-2.51, -2.13) &            -7.08 \\
  B1.5 I & -0.96 (-0.99, -0.91) & -0.18 (-0.21, -0.14) & -0.07 (-0.09, -0.05) & -0.11 (-0.14, -0.09) & -0.07 (-0.08, -0.05) & -0.08 (-0.09, -0.07) & -2.04 (-2.32, -1.73) &            -7.09 \\
    B2 I & -0.89 (-0.92, -0.85) & -0.16 (-0.19, -0.12) & -0.06 (-0.08, -0.04) & -0.10 (-0.12, -0.07) & -0.05 (-0.07, -0.04) & -0.07 (-0.08, -0.06) & -1.75 (-2.00, -1.47) &            -7.07 \\
  B2.5 I & -0.82 (-0.86, -0.78) & -0.15 (-0.17, -0.11) & -0.05 (-0.07, -0.04) & -0.09 (-0.10, -0.07) & -0.04 (-0.05, -0.03) & -0.07 (-0.07, -0.06) & -1.54 (-1.74, -1.33) &            -7.09 \\
    B3 I & -0.75 (-0.79, -0.72) & -0.14 (-0.15, -0.12) & -0.05 (-0.06, -0.04) & -0.08 (-0.09, -0.07) & -0.04 (-0.04, -0.03) & -0.06 (-0.07, -0.06) & -1.40 (-1.55, -1.32) &            -7.14 \\
\end{longtable}
\begin{tablenotes}
\item \textbf{Notes.} The filter transmission responses used to calculate the synthetic photometry were obtained from the SVO Carlos Rodrigo filter profile service (\href{http://svo2.cab.inta-csic.es/theory/fps/index.php?mode=browse&gname=Generic&asttype=}{http://svo2.cab.inta-csic.es/theory/fps/index.php?mode=browse\&gname=Generic\&asttype=}).
\end{tablenotes}
\end{landscape}
\setlength\LTleft{-15pt}
\setlength\LTright{-20pt}
\setlength\LTcapwidth{\linewidth}
\captionsetup[longtable]{labelfont=bf, font = footnotesize, labelsep=period}
\begin{longtable}{lcccc|c}
\caption{Calibrated photometric colours, bolometric corrections and absolute visual magnitudes of O stars and early-B supergiants in the SLOAN-SDSS system. The table follows a similar format as Table~\ref{tab:Phot_Generic_Bessell}. We note that the SLOAN-SDSS magnitudes are provided in the AB system.}
\label{tab:Phot_SLOAN_SDSS}\\
\toprule
\toprule
     SpT &            $(u-g)_0$ &            $(g-r)_0$ &            $(r-i)_0$ &                $BC_{g}$ & $M_{\mathrm{g}}$ \\
 & av. (min., max.) & av. (min., max.) & av. (min., max.) & av. (min., max.) & av. \\
\midrule
\endfirsthead
\caption[]{continued.} \\
\toprule
\toprule
     SpT &            $(u-g)_0$ &            $(g-r)_0$ &            $(r-i)_0$ &                $BC_{g}$ & $M_{\mathrm{g}}$ \\
 & av. (min., max.) & av. (min., max.) & av. (min., max.) & av. (min., max.) & av. \\
\midrule
\endhead
\midrule
\multicolumn{6}{r}{{\textit{continued on next page}}} \\

\endfoot

\bottomrule
\endlastfoot
    O4 V & -1.34 (-1.35, -1.33) & -1.43 (-1.44, -1.42) & -1.01 (-1.02, -1.01) & -44.06 (-44.20, -43.94) &            34.54 \\
  O4.5 V & -1.34 (-1.35, -1.32) & -1.43 (-1.44, -1.42) & -1.01 (-1.01, -1.01) & -43.96 (-44.08, -43.85) &            34.70 \\
    O5 V & -1.33 (-1.35, -1.32) & -1.43 (-1.43, -1.42) & -1.01 (-1.01, -1.01) & -43.90 (-44.01, -43.76) &            34.84 \\
  O5.5 V & -1.33 (-1.34, -1.31) & -1.42 (-1.43, -1.42) & -1.01 (-1.01, -1.00) & -43.81 (-43.95, -43.68) &            34.97 \\
    O6 V & -1.32 (-1.34, -1.31) & -1.42 (-1.43, -1.41) & -1.01 (-1.01, -1.00) & -43.71 (-43.85, -43.54) &            35.10 \\
  O6.5 V & -1.31 (-1.33, -1.29) & -1.42 (-1.43, -1.40) & -1.01 (-1.01, -1.00) & -43.57 (-43.75, -43.36) &            35.26 \\
    O7 V & -1.31 (-1.33, -1.28) & -1.41 (-1.42, -1.40) & -1.00 (-1.01, -1.00) & -43.43 (-43.60, -43.21) &            35.40 \\
  O7.5 V & -1.29 (-1.32, -1.27) & -1.41 (-1.42, -1.40) & -1.00 (-1.01, -1.00) & -43.27 (-43.46, -43.12) &            35.57 \\
    O8 V & -1.28 (-1.31, -1.26) & -1.41 (-1.41, -1.39) & -1.00 (-1.01, -1.00) & -43.16 (-43.30, -43.00) &            35.73 \\
  O8.5 V & -1.27 (-1.30, -1.25) & -1.40 (-1.41, -1.39) & -1.00 (-1.01, -1.00) & -43.06 (-43.19, -42.92) &            35.89 \\
    O9 V & -1.26 (-1.29, -1.24) & -1.40 (-1.41, -1.39) & -1.00 (-1.01, -1.00) & -42.97 (-43.11, -42.81) &            36.05 \\
  O9.5 V & -1.25 (-1.28, -1.23) & -1.40 (-1.41, -1.39) & -1.00 (-1.01, -1.00) & -42.90 (-43.04, -42.73) &            36.21 \\
  O9.7 V & -1.23 (-1.27, -1.20) & -1.40 (-1.41, -1.38) & -1.00 (-1.01, -0.99) & -42.79 (-42.97, -42.59) &            36.46 \\
\toprule
  O4 III & -1.32 (-1.34, -1.29) & -1.42 (-1.43, -1.40) & -1.01 (-1.01, -1.00) & -43.89 (-44.07, -43.69) &            33.86 \\
O4.5 III & -1.31 (-1.33, -1.29) & -1.42 (-1.43, -1.40) & -1.00 (-1.01, -1.00) & -43.80 (-43.92, -43.64) &            33.95 \\
  O5 III & -1.30 (-1.33, -1.29) & -1.41 (-1.42, -1.40) & -1.00 (-1.01, -1.00) & -43.70 (-43.83, -43.60) &            34.02 \\
O5.5 III & -1.30 (-1.32, -1.27) & -1.41 (-1.42, -1.39) & -1.00 (-1.01, -0.99) & -43.59 (-43.76, -43.43) &            34.09 \\
  O6 III & -1.29 (-1.31, -1.27) & -1.40 (-1.41, -1.39) & -1.00 (-1.01, -0.99) & -43.50 (-43.65, -43.39) &            34.15 \\
O6.5 III & -1.28 (-1.30, -1.26) & -1.40 (-1.41, -1.38) & -1.00 (-1.00, -0.99) & -43.35 (-43.54, -43.19) &            34.25 \\
  O7 III & -1.27 (-1.29, -1.25) & -1.40 (-1.41, -1.38) & -1.00 (-1.00, -0.99) & -43.22 (-43.35, -43.10) &            34.33 \\
O7.5 III & -1.26 (-1.27, -1.25) & -1.39 (-1.40, -1.38) & -1.00 (-1.00, -0.99) & -43.09 (-43.15, -43.01) &            34.42 \\
  O8 III & -1.25 (-1.26, -1.23) & -1.39 (-1.40, -1.38) & -1.00 (-1.00, -0.99) & -42.97 (-43.07, -42.86) &            34.51 \\
O8.5 III & -1.24 (-1.25, -1.23) & -1.39 (-1.40, -1.38) & -1.00 (-1.00, -0.99) & -42.88 (-42.95, -42.82) &            34.59 \\
  O9 III & -1.23 (-1.24, -1.22) & -1.39 (-1.40, -1.38) & -1.00 (-1.00, -0.99) & -42.78 (-42.83, -42.73) &            34.69 \\
O9.5 III & -1.22 (-1.23, -1.21) & -1.38 (-1.39, -1.37) & -1.00 (-1.00, -0.99) & -42.69 (-42.75, -42.61) &            34.81 \\
O9.7 III & -1.20 (-1.22, -1.19) & -1.38 (-1.39, -1.37) & -0.99 (-1.00, -0.99) & -42.58 (-42.67, -42.50) &            34.91 \\
\toprule
    O4 I & -1.30 (-1.31, -1.28) & -1.40 (-1.41, -1.39) & -1.00 (-1.00, -0.99) & -43.73 (-43.84, -43.61) &            33.60 \\
  O4.5 I & -1.29 (-1.31, -1.28) & -1.40 (-1.41, -1.39) & -1.00 (-1.00, -0.99) & -43.61 (-43.69, -43.54) &            33.62 \\
    O5 I & -1.28 (-1.30, -1.27) & -1.39 (-1.40, -1.38) & -0.99 (-1.00, -0.99) & -43.54 (-43.61, -43.44) &            33.62 \\
  O5.5 I & -1.28 (-1.28, -1.27) & -1.39 (-1.40, -1.38) & -0.99 (-1.00, -0.99) & -43.48 (-43.55, -43.41) &            33.60 \\
    O6 I & -1.27 (-1.28, -1.25) & -1.38 (-1.39, -1.37) & -0.99 (-1.00, -0.99) & -43.38 (-43.44, -43.31) &            33.59 \\
  O6.5 I & -1.26 (-1.27, -1.25) & -1.38 (-1.39, -1.37) & -0.99 (-0.99, -0.99) & -43.25 (-43.35, -43.18) &            33.60 \\
    O7 I & -1.24 (-1.25, -1.23) & -1.37 (-1.38, -1.36) & -0.99 (-0.99, -0.98) & -43.10 (-43.17, -43.02) &            33.63 \\
  O7.5 I & -1.23 (-1.25, -1.22) & -1.37 (-1.38, -1.36) & -0.99 (-0.99, -0.98) & -42.98 (-43.06, -42.90) &            33.63 \\
    O8 I & -1.23 (-1.24, -1.21) & -1.37 (-1.38, -1.35) & -0.99 (-1.00, -0.98) & -42.87 (-42.94, -42.79) &            33.65 \\
  O8.5 I & -1.22 (-1.24, -1.20) & -1.36 (-1.38, -1.35) & -0.99 (-1.00, -0.98) & -42.76 (-42.85, -42.68) &            33.70 \\
    O9 I & -1.21 (-1.23, -1.19) & -1.37 (-1.39, -1.35) & -0.99 (-1.00, -0.98) & -42.69 (-42.77, -42.61) &            33.70 \\
  O9.5 I & -1.20 (-1.23, -1.18) & -1.36 (-1.39, -1.34) & -0.99 (-1.00, -0.97) & -42.58 (-42.73, -42.45) &            33.74 \\
  O9.7 I & -1.19 (-1.21, -1.17) & -1.36 (-1.38, -1.33) & -0.99 (-1.00, -0.97) & -42.47 (-42.60, -42.35) &            33.81 \\
    B0 I & -1.17 (-1.19, -1.15) & -1.35 (-1.37, -1.32) & -0.98 (-0.99, -0.97) & -42.36 (-42.48, -42.22) &            33.06 \\
  B0.5 I & -1.15 (-1.17, -1.13) & -1.34 (-1.36, -1.31) & -0.98 (-0.99, -0.96) & -42.23 (-42.40, -42.08) &            32.86 \\
    B1 I & -1.12 (-1.14, -1.09) & -1.33 (-1.35, -1.30) & -0.97 (-0.98, -0.95) & -42.11 (-42.27, -41.94) &            32.72 \\
  B1.5 I & -1.05 (-1.10, -0.99) & -1.30 (-1.34, -1.26) & -0.95 (-0.97, -0.93) & -41.84 (-42.09, -41.56) &            32.73 \\
    B2 I & -0.97 (-1.00, -0.92) & -1.28 (-1.31, -1.23) & -0.94 (-0.96, -0.92) & -41.57 (-41.78, -41.32) &            32.76 \\
  B2.5 I & -0.89 (-0.93, -0.84) & -1.27 (-1.29, -1.23) & -0.93 (-0.94, -0.92) & -41.37 (-41.54, -41.18) &            32.75 \\
    B3 I & -0.81 (-0.84, -0.77) & -1.26 (-1.28, -1.24) & -0.93 (-0.94, -0.92) & -41.22 (-41.36, -41.15) &            32.70 \\
\end{longtable}
\begin{tablenotes}
\item \textbf{Notes.} The filter transmission responses used to calculate the synthetic photometry were obtained from the SVO Carlos Rodrigo filter profile service (\href{http://svo2.cab.inta-csic.es/theory/fps/index.php?mode=browse&gname=SLOAN&asttype=}{http://svo2.cab.inta-csic.es/theory/fps/index.php?mode=browse\&gname=SLOAN\&asttype=}).
\end{tablenotes}
\setlength\LTleft{-15pt}
\setlength\LTright{-20pt}
\setlength\LTcapwidth{\linewidth}
\captionsetup[longtable]{labelfont=bf, font = footnotesize, labelsep=period}
\begin{longtable}{lcccccccccc}
\caption{Calibrated absolute magnitudes of O~stars and early-B~supergiants in a selection of HST-WFPC2 and WFPC3 filters.  We caution the reader that they were calculated with calibrated radii (O stars) or radii derived from analogue stars in previous works (B stars).}
\label{tab:Phot_HST}\\
\toprule
\toprule
     SpT & $M_{F170W}$ & $M_{F225W}$ & $M_{F255W}$ & $M_{F275W}$ & $M_{F336W}$ & $M_{F475W}$ & $M_{F555W}$ & $M_{F606W}$ & $M_{F675W}$ & $M_{F814W}$ \\
\midrule
\endfirsthead
\caption[]{continued.} \\
\toprule
\toprule
     SpT & $M_{F170W}$ & $M_{F225W}$ & $M_{F255W}$ & $M_{F275W}$ & $M_{F336W}$ & $M_{F475W}$ & $M_{F555W}$ & $M_{F606W}$ & $M_{F675W}$ & $M_{F814W}$ \\
\midrule
\endhead
\midrule
\multicolumn{11}{r}{{\textit{continued on next page}}} \\

\endfoot

\bottomrule
\endlastfoot
    O4 V &       -6.35 &       -7.89 &       -7.90 &       -7.87 &       -7.21 &       -5.39 &       -5.27 &       -5.20 &       -5.07 &       -4.92 \\
  O4.5 V &       -6.19 &       -7.73 &       -7.74 &       -7.71 &       -7.05 &       -5.24 &       -5.12 &       -5.05 &       -4.92 &       -4.77 \\
    O5 V &       -6.04 &       -7.58 &       -7.59 &       -7.56 &       -6.90 &       -5.09 &       -4.97 &       -4.90 &       -4.77 &       -4.62 \\
  O5.5 V &       -5.90 &       -7.44 &       -7.45 &       -7.42 &       -6.76 &       -4.96 &       -4.85 &       -4.77 &       -4.64 &       -4.49 \\
    O6 V &       -5.77 &       -7.31 &       -7.32 &       -7.29 &       -6.63 &       -4.84 &       -4.72 &       -4.65 &       -4.52 &       -4.37 \\
  O6.5 V &       -5.60 &       -7.13 &       -7.14 &       -7.11 &       -6.46 &       -4.68 &       -4.57 &       -4.49 &       -4.37 &       -4.22 \\
    O7 V &       -5.44 &       -6.98 &       -6.99 &       -6.95 &       -6.31 &       -4.54 &       -4.42 &       -4.35 &       -4.22 &       -4.08 \\
  O7.5 V &       -5.26 &       -6.78 &       -6.79 &       -6.76 &       -6.12 &       -4.36 &       -4.25 &       -4.18 &       -4.06 &       -3.91 \\
    O8 V &       -5.09 &       -6.61 &       -6.62 &       -6.59 &       -5.95 &       -4.21 &       -4.10 &       -4.02 &       -3.90 &       -3.76 \\
  O8.5 V &       -4.91 &       -6.44 &       -6.45 &       -6.42 &       -5.78 &       -4.05 &       -3.94 &       -3.87 &       -3.74 &       -3.60 \\
    O9 V &       -4.74 &       -6.25 &       -6.27 &       -6.24 &       -5.60 &       -3.88 &       -3.77 &       -3.70 &       -3.58 &       -3.44 \\
  O9.5 V &       -4.57 &       -6.09 &       -6.10 &       -6.07 &       -5.44 &       -3.73 &       -3.62 &       -3.55 &       -3.42 &       -3.28 \\
  O9.7 V &       -4.29 &       -5.80 &       -5.82 &       -5.79 &       -5.17 &       -3.48 &       -3.37 &       -3.30 &       -3.18 &       -3.04 \\
\toprule
  O4 III &       -6.99 &       -8.52 &       -8.53 &       -8.50 &       -7.85 &       -6.06 &       -5.95 &       -5.88 &       -5.75 &       -5.61 \\
O4.5 III &       -6.90 &       -8.43 &       -8.44 &       -8.41 &       -7.76 &       -5.98 &       -5.86 &       -5.79 &       -5.67 &       -5.52 \\
  O5 III &       -6.82 &       -8.35 &       -8.36 &       -8.33 &       -7.68 &       -5.91 &       -5.80 &       -5.72 &       -5.60 &       -5.46 \\
O5.5 III &       -6.74 &       -8.27 &       -8.28 &       -8.25 &       -7.61 &       -5.84 &       -5.73 &       -5.66 &       -5.54 &       -5.40 \\
  O6 III &       -6.67 &       -8.20 &       -8.21 &       -8.18 &       -7.54 &       -5.78 &       -5.67 &       -5.60 &       -5.48 &       -5.34 \\
O6.5 III &       -6.56 &       -8.08 &       -8.09 &       -8.06 &       -7.42 &       -5.68 &       -5.58 &       -5.51 &       -5.39 &       -5.25 \\
  O7 III &       -6.46 &       -7.98 &       -7.99 &       -7.96 &       -7.33 &       -5.60 &       -5.49 &       -5.42 &       -5.31 &       -5.17 \\
O7.5 III &       -6.36 &       -7.88 &       -7.89 &       -7.86 &       -7.23 &       -5.52 &       -5.41 &       -5.34 &       -5.23 &       -5.09 \\
  O8 III &       -6.26 &       -7.77 &       -7.78 &       -7.75 &       -7.13 &       -5.42 &       -5.32 &       -5.25 &       -5.14 &       -5.00 \\
O8.5 III &       -6.16 &       -7.67 &       -7.68 &       -7.65 &       -7.03 &       -5.34 &       -5.23 &       -5.17 &       -5.05 &       -4.92 \\
  O9 III &       -6.05 &       -7.55 &       -7.57 &       -7.54 &       -6.92 &       -5.24 &       -5.14 &       -5.07 &       -4.96 &       -4.82 \\
O9.5 III &       -5.91 &       -7.41 &       -7.43 &       -7.40 &       -6.79 &       -5.12 &       -5.02 &       -4.95 &       -4.84 &       -4.71 \\
O9.7 III &       -5.79 &       -7.28 &       -7.30 &       -7.27 &       -6.67 &       -5.02 &       -4.92 &       -4.86 &       -4.74 &       -4.62 \\
\toprule
    O4 I &       -7.22 &       -8.75 &       -8.76 &       -8.73 &       -8.09 &       -6.33 &       -6.22 &       -6.15 &       -6.04 &       -5.89 \\
  O4.5 I &       -7.19 &       -8.71 &       -8.73 &       -8.70 &       -8.06 &       -6.30 &       -6.20 &       -6.13 &       -6.02 &       -5.88 \\
    O5 I &       -7.19 &       -8.71 &       -8.72 &       -8.69 &       -8.06 &       -6.31 &       -6.21 &       -6.14 &       -6.03 &       -5.89 \\
  O5.5 I &       -7.20 &       -8.72 &       -8.73 &       -8.70 &       -8.07 &       -6.33 &       -6.23 &       -6.16 &       -6.05 &       -5.91 \\
    O6 I &       -7.18 &       -8.70 &       -8.71 &       -8.68 &       -8.06 &       -6.33 &       -6.23 &       -6.17 &       -6.06 &       -5.92 \\
  O6.5 I &       -7.16 &       -8.67 &       -8.69 &       -8.66 &       -8.04 &       -6.32 &       -6.22 &       -6.16 &       -6.05 &       -5.92 \\
    O7 I &       -7.12 &       -8.62 &       -8.64 &       -8.61 &       -8.00 &       -6.30 &       -6.20 &       -6.14 &       -6.04 &       -5.91 \\
  O7.5 I &       -7.11 &       -8.61 &       -8.63 &       -8.60 &       -7.99 &       -6.30 &       -6.20 &       -6.14 &       -6.04 &       -5.91 \\
    O8 I &       -7.07 &       -8.57 &       -8.59 &       -8.56 &       -7.96 &       -6.28 &       -6.18 &       -6.12 &       -6.02 &       -5.89 \\
  O8.5 I &       -7.01 &       -8.50 &       -8.52 &       -8.49 &       -7.90 &       -6.23 &       -6.14 &       -6.08 &       -5.98 &       -5.85 \\
    O9 I &       -7.00 &       -8.50 &       -8.51 &       -8.49 &       -7.89 &       -6.23 &       -6.14 &       -6.08 &       -5.97 &       -5.84 \\
  O9.5 I &       -6.94 &       -8.43 &       -8.45 &       -8.42 &       -7.84 &       -6.19 &       -6.10 &       -6.04 &       -5.94 &       -5.81 \\
  O9.7 I &       -6.85 &       -8.34 &       -8.36 &       -8.33 &       -7.75 &       -6.12 &       -6.03 &       -5.97 &       -5.87 &       -5.74 \\
    B0 I &       -7.57 &       -9.04 &       -9.06 &       -9.04 &       -8.47 &       -6.87 &       -6.78 &       -6.73 &       -6.63 &       -6.52 \\
  B0.5 I &       -7.73 &       -9.19 &       -9.21 &       -9.19 &       -8.63 &       -7.07 &       -6.98 &       -6.93 &       -6.83 &       -6.72 \\
    B1 I &       -7.83 &       -9.27 &       -9.30 &       -9.27 &       -8.74 &       -7.21 &       -7.13 &       -7.08 &       -6.99 &       -6.88 \\
  B1.5 I &       -7.72 &       -9.13 &       -9.16 &       -9.14 &       -8.65 &       -7.20 &       -7.13 &       -7.09 &       -7.01 &       -6.92 \\
    B2 I &       -7.56 &       -8.92 &       -8.96 &       -8.95 &       -8.51 &       -7.17 &       -7.10 &       -7.07 &       -7.01 &       -6.93 \\
  B2.5 I &       -7.46 &       -8.78 &       -8.83 &       -8.81 &       -8.42 &       -7.18 &       -7.12 &       -7.09 &       -7.03 &       -6.97 \\
    B3 I &       -7.41 &       -8.69 &       -8.74 &       -8.73 &       -8.37 &       -7.23 &       -7.17 &       -7.15 &       -7.09 &       -7.04 \\
\end{longtable}
\begin{tablenotes}
\item \textbf{Notes.} The filter transmission responses used to calculate the synthetic photometry were obtained from the SVO Carlos Rodrigo filter profile service (\href{http://svo2.cab.inta-csic.es/theory/fps/index.php?mode=browse&gname=HST&gname2=WFPC2-PC&asttype=}{http://svo2.cab.inta-csic.es/theory/fps/index.php?mode=browse\&gname=HST\&gname2=WFPC2-PC\&asttype=} and \href{http://svo2.cab.inta-csic.es/theory/fps/index.php?mode=browse&gname=HST&gname2=WFC3_UVIS2&asttype=}{http://svo2.cab.inta-csic.es/theory/fps/index.php?mode=browse\&gname=HST\&gname2=WFC3\_UVIS2\&asttype=}).
\end{tablenotes}
\twocolumn


\bsp	
\label{lastpage}

\end{document}